\documentclass[useAMS,usenatbib,usegraphicx]{mn2e}
\usepackage{amsmath,amssymb,epsfig}
\usepackage{breqn}
\usepackage{placeins}
\usepackage{bm}
\usepackage{xcolor}

\begin{document}

\newcommand{\fixme}[1]{{\textbf{Fixme: #1}}}
\newcommand{\detD}{{\det\!\cld}}
\newcommand{\clh}{\mathcal{H}}
\newcommand{\ud}{{\rm d}}
\newcommand{\eprint}[1]{\href{http://arxiv.org/abs/#1}{#1}}
\newcommand{\adsurl}[1]{\href{#1}{ADS}}
\newcommand{\ISBN}[1]{\href{http://cosmologist.info/ISBN/#1}{ISBN: #1}}
\newcommand{\jcap}{J.\ Cosmol.\ Astropart.\ Phys.}
\newcommand{\mnras}{Mon.\ Not.\ R.\ Astron.\ Soc.}
\newcommand{\physrep}{Phys.\ Rep.}
\newcommand{\progress}{Rep.\ Prog.\ Phys.}
\newcommand{\prlett}{Phys.\ Rev.\ Lett.}
\newcommand{\prd}{Phys.\ Rev.\ D}
\newcommand{\apj}{ApJ}

\newcommand{\aap}{A\&A}
\newcommand{\vort}{\varpi}
\newcommand\ba{\begin{eqnarray}}
\newcommand\ea{\end{eqnarray}}
\newcommand\be{\begin{equation}}
\newcommand\ee{\end{equation}}
\newcommand\lagrange{{\cal L}}
\newcommand\cll{{\cal L}}
\newcommand\cln{{\cal N}}
\newcommand\clx{{\cal X}}
\newcommand\clz{{\cal Z}}
\newcommand\clv{{\cal V}}
\newcommand\cld{{\cal D}}
\newcommand\clt{{\cal T}}

\newcommand{\ThreeJSymbol}[6]{\begin{pmatrix}
#1 & #3 & #5 \\
#2 & #4 & #6
 \end{pmatrix}}

\newcommand\clo{{\cal O}}
\newcommand{\cla}{{\cal A}}
\newcommand{\clp}{{\cal P}}
\newcommand{\clr}{{\cal R}}
\newcommand{\uD}{{\mathrm{D}}}
\newcommand{\calE}{{\cal E}}
\newcommand{\calB}{{\cal B}}
\newcommand{\curl}{\,\mbox{curl}\,}
\newcommand\del{\nabla}
\newcommand\Tr{{\rm Tr}}
\newcommand\half{{\frac{1}{2}}}
\newcommand\fourth{{1\over 8}}
\newcommand\bibi{\bibitem}
\newcommand{\kf}{\beta}
\newcommand{\kfprod}{\alpha}
\newcommand\calS{{\cal S}}
\renewcommand\H{{\cal H}}
\newcommand\K{{\rm K}}
\newcommand\mK{{\rm mK}}
\newcommand\km{{\rm km}}
\newcommand\synch{\text{syn}}
\newcommand\opacity{\tau_c^{-1}}

\newcommand{\Psil}{\Psi_l}
\newcommand{\bI}{\bar{I}}
\newcommand{\bq}{\bar{q}}
\newcommand{\bv}{\bar{v}}
\renewcommand\P{{\cal P}}
\newcommand{\numfrac}[2]{{\textstyle \frac{#1}{#2}}}

\newcommand{\Omtot}{\Omega_{\mathrm{tot}}}
\newcommand\xx{\mbox{\boldmath $x$}}
\newcommand{\phpr} {\phi`}
\newcommand{\gam}{\gamma_{ij}}
\newcommand{\sqgam}{\sqrt{\gamma}}
\newcommand{\delk}{\Delta+3{\K}}
\newcommand{\dph}{\delta\phi}
\newcommand{\om} {\Omega}
\newcommand{\dom}{\delta^{(3)}\left(\Omega\right)}
\newcommand{\rar}{\rightarrow}
\newcommand{\Rar}{\Rightarrow}
\newcommand\gsim{ \lower .75ex \hbox{$\sim$} \llap{\raise .27ex \hbox{$>$}} }
\newcommand\lsim{ \lower .75ex \hbox{$\sim$} \llap{\raise .27ex \hbox{$<$}} }
\newcommand\bigdot[1] {\stackrel{\mbox{{\huge .}}}{#1}}
\newcommand\bigddot[1] {\stackrel{\mbox{{\huge ..}}}{#1}}
\newcommand{\Mpc}{\text{Mpc}}
\newcommand{\Al}{{A_l}}
\newcommand{\Bl}{{B_l}}
\newcommand{\eAl}{e^\Al}
\newcommand{\ix}{{(i)}}
\newcommand{\ixp}{{(i+1)}}
\renewcommand{\k}{\beta}
\newcommand{\HD}{\mathrm{D}}

\newcommand{\nonflat}[1]{#1}
\newcommand{\Cgl}{C_{\text{gl}}}
\newcommand{\Cgltwo}{C_{\text{gl},2}}
\newcommand{\He}{{\rm{He}}}
\newcommand{\Mhz}{{\rm MHz}}
\newcommand{\vx}{{\mathbf{x}}}
\newcommand{\ve}{{\mathbf{e}}}
\newcommand{\vv}{{\mathbf{v}}}
\newcommand{\vk}{{\mathbf{k}}}
\newcommand{\vn}{{\mathbf{n}}}

\newcommand{\vnhat}{{\hat{\mathbf{n}}}}
\newcommand{\vkhat}{{\hat{\mathbf{k}}}}
\newcommand{\taueps}{{\tau_\epsilon}}

\newcommand{\vgrad}{{\mathbf{\nabla}}}
\newcommand{\fbarln}{\bar{f}_{,\ln\epsilon}(\epsilon)}


\title[Cosmic shear measurement]{Cosmic shear measurement with maximum likelihood and maximum a posteriori inference}

\author[Alex Hall and Andy Taylor]
{Alex Hall\thanks{ahall@roe.ac.uk} and Andy Taylor\\
Institute for Astronomy, University of Edinburgh, Royal Observatory, Blackford Hill, Edinburgh, EH9 3HJ, U.K.}
\maketitle


\begin{abstract}
We investigate the problem of noise bias in maximum likelihood and maximum a posteriori estimators for cosmic shear. We derive the leading and next-to-leading order biases and compute them in the context of galaxy ellipticity measurements, extending previous work on maximum likelihood inference for weak lensing. We show that a large part of the bias on these point estimators can be removed using information already contained in the likelihood when a galaxy model is specified, without the need for external calibration. We test these bias-corrected estimators on simulated galaxy images similar to those expected from planned space-based weak lensing surveys, with promising results. We find that the introduction of an intrinsic shape prior can help with mitigation of noise bias, such that the maximum a posteriori estimate can be made less biased than the maximum likelihood estimate. Second-order terms offer a check on the convergence of the estimators, but are largely sub-dominant. We show how biases propagate to shear estimates, demonstrating in our simple setup that shear biases can be reduced by orders of magnitude and potentially to within the requirements of planned space-based surveys at mild signal-to-noise. We find that second-order terms can exhibit significant cancellations at low signal-to-noise when Gaussian noise is assumed, which has implications for inferring the performance of shear-measurement algorithms from simplified simulations. We discuss the viability of our point estimators as tools for lensing inference, arguing that they allow for the robust measurement of ellipticity and shear.
\end{abstract}

\maketitle

\section{Introduction}
\label{sec:intro}

Weak gravitational lensing is developing into a powerful observational probe of the dark matter on large scales. Current cosmic shear surveys are able to break parameter degeneracies affecting more established observables~\citep{2012MNRAS.427..146H,2015arXiv150201590P} and current galaxy-galaxy lensing surveys provide competitive constraints on the cosmological parameters~\citep{2013MNRAS.432.1544M}. However, if weak lensing is to fulfil its potential of providing decisive constraints on the geometry of the large-scale Universe, a number of contaminating systematic errors will have to be resolved. 

One particularly troublesome systematic arises when the fundamental lensing observable, the shear field, is estimated. In this work we will be concerned with a bias that arises in the shear estimator due to the pixel noise in observed galaxy images, sometimes referred to as `noise bias'~\citep{2002AJ....123..583B,2003MNRAS.343..459H}. This arises due to the non-linear mapping from the observed raw photon count in CCD pixels to the shear map. The former contains shot noise from the discreteness of the counts, as well as a fluctuating sky background and read-out noise from the CCD itself~\citep{2014ApJS..212....5M}. Point estimators for galaxy shape parameters will necessarily be biased by the non-linear transformation of this noise. This is potentially disastrous if, as is usual practice, shear is estimated by averaging ellipticity estimates over sources~\citep{2012MNRAS.425.1951R}, since the error bar on the shear estimate will shrink with the number of sources whilst the bias will not.

Shear estimation algorithms therefore generally need to be calibrated against image simulations to correct for these biases, which is clearly not an ideal situation since it relies on the existence of  simulations specific to a given survey against which to calibrate~\citep{2012A&A...547A..98B}. These simulations must be accurate enough and large enough to characterise performance of an algorithm to allow interpolation to real data. In addition, if the simulations themselves change the algorithm must be re-calibrated. Of course, if the bias was known with infinite accuracy it could be subtracted off with no loss of precision. The uncertainty enters when simulations are needed for calibration.

It therefore seems desirable that a method is found requiring minimal external calibration for bias. In this work, we investigate how the method of Maximum Likelihood (ML) might be adapted for use as an unbiased shear estimator, by subtracting off from the Maximum-Likelihood Estimate (MLE) a bias predicted from the likelihood, without external calibration. The MLE is a natural choice for estimating galaxy shape parameters, as it is easy to compute and possesses nice asymptotic properties such as consistency, normality, and efficiency. The MLE has already been implemented as a shear estimator with external bias calibration, with good results~\citep{2012arXiv1211.4847G,2013MNRAS.434.1604Z,2015MNRAS.450.2963M, 2015arXiv150705603J}. Since the noise bias scales roughly as the inverse-square of the signal-to-noise ratio ($S/N$), we are forced to consider the next-to-leading-order behaviour of the MLE around its asymptotic distribution, a Gaussian centred on the truth with $\mathcal{O}(\sqrt{n})$ fluctuations, where $n$ is the number of independent data points. It turns out that the MLE is still an efficient estimator at this order in $n$, a property that may be shown to hold at yet higher order still~\citep{rao}. Thus, the MLE corrected for noise bias seems like a promising choice as a point-estimator for shape and shear, both in terms of its bias and mean-squared error. Similar statements may be made regarding the Maximum A Posteriori estimate (MAP), which arises when a prior is multiplied against the likelihood and the mode of the resulting distribution is found. In this work we also investigate this hitherto unexplored shear estimator.

A further advantage of the MLE (or MAP) is that it is naturally complementary to a Bayesian analysis pipeline for weak lensing. The Bayesian approach to inference can avoid the biases inherent to point estimators~\citep{2015ApJ...807...87S}, and is a preferable approach to weak lensing if it can be made computationally feasible. As the likelihood function is a necessary component of a Bayesian hierarchical model, the MLE can be easily incorporated into such a scheme.

This work will deal with the fundamentals of the MLE and the MAP as lensing estimators, using simple models to elucidate the key properties. We will be particularly concerned with how shear estimates respond to ellipticity bias, and how ellipticity bias is affected by the pixellization of a galaxy image. Our results are not specific to any particular lensing survey, but as a reference we present out final bias measurements alongside the total requirements of a fiducial space-based lensing survey. The stable point-spread function (PSF) of space-based lensing surveys such as \emph{Euclid} and WFIRST potentially means that other systematics, including noise bias, might become limiting factors in the inference of lensing power spectra and cosmological parameters. We also expect our results to be of use for current and upcoming ground-based lensing surveys such as the Kilo-Degree Survey~\citep{2015MNRAS.454.3500K}, the Dark Energy Survey~\citep{2015arXiv150705598B}, or the Large Synoptic Survey Telescope~\citep{2013MNRAS.428.2695C}. Looking beyond weak lensing, many of our results are of more general applicability, concerning as they do the properties of bias in the MLE and the MAP. 

In Section~\ref{sec:bias} we review the calculation of the MLE noise bias, before specialising to the case of estimating ellipticity from galaxy images. In Section~\ref{sec:eresults} we test our bias-correction formalism on simulated noisy galaxy images, and in Section~\ref{sec:MAP} introduce the MAP as an alternative point estimator. In Section~\ref{sec:1Dmresults} we consider how the bias propagates to one-dimensional shear estimates, extending this to two dimensions in Section~\ref{sec:2Dmresults}. In Section~\ref{sec:concs} we compare our formalism to other methods in the literature, and conclude.

\section{Noise bias in the MLE}
\label{sec:bias}

In this section we review the derivation of bias in the MLE, following~\citet{McCullagh}. For simplicity we specialise to a one-dimensional parameter space, but the generalisation to arbitrary dimension is straightforward. We do not assume anything about the form of the likelihood function except that it is differentiable in a sufficiently small region around the true parameter value, making this derivation more general than that presented in the context of weak lensing in~\citet{2012MNRAS.425.1951R}.

Let the log-likelihood be $L(\beta ; \mathbf{Y}) \equiv \log{p(\mathbf{Y}|\beta)}$, where $\mathbf{Y}$ is a data vector and $\beta$ is the true (but unknown) value of the parameter. Suppose we make $n$ observations, which result in $n$ realisations of $\mathbf{Y}$, which we label $\mathbf{Y}_i$ with $i=1...n$. We will assume that the $\mathbf{Y}_i$ are independent but not necessarily identically distributed, each drawn from the distribution $p_i(\mathbf{Y}_i|\beta)$. The joint log-likelihood is then
\begin{equation}
L(\beta) \equiv L(\beta;\mathbf{Y}) = \sum_{i=1}^n \log{p_i(\mathbf{Y}_i|\beta)}.
\end{equation}
The MLE (denoted $\hat{\beta}$) is defined implicitly by the equation $L'(\hat{\beta}) = 0$, with a prime denoting differentiation with respect to $\beta$.

Now, recall the Central Limit Theorem (CLT). Suppose that $\mathbf{X}_j$ are a set of independent (but not necessarily identically distributed) vector-valued random variables, with $j$ labelling members of the set, each with finite mean and variance. The central limit theorem then holds that for sufficiently large $n$
\begin{equation}
\frac{1}{\sqrt{n}}\sum_{j=1}^n (X_j^i - \mu_j^i) \sim N \left (0,\frac{1}{n}\sum_{j=1}^n\mathbf{C}_j \right),
\end{equation}
where $\mu_j^i$ is the $i$-th component of the mean of the $j$-th vector, and $\mathbf{C}_j$ is the covariance matrix of the $j$-th vector. Note that we have glossed over some formal regularity conditions.

Now, expand the ML equation around $\beta$:
\begin{equation}
\label{eq:ML}
L'(\hat{\beta}) \approx L'(\beta) + (\hat{\beta} - \beta)L''(\beta) + \frac{1}{2}(\hat{\beta} - \beta)^2L'''(\beta) = 0,
\end{equation}
where we have neglected higher-order terms. Note that this makes sense in the frequentist approach to statistics since $\beta$ is not considered to be a random variable. Our aim is to find the asymptotic form of this equation in the limit of large $n$. We know that MLEs are asymptotically normal from the CLT, such that in the limit of large $n$
\begin{equation}
\hat{\delta} \equiv \sqrt{n}(\hat{\beta} - \beta) \sim N(0,F^{-1}),
\end{equation}
where the total Fisher information, $F$, is given by
\begin{equation}
F = -\sum_{j=1}^n \left \langle \frac{\partial^2}{\partial \beta^2} \log{p_j(\mathbf{Y}_j|\beta)} \right \rangle,
\end{equation}
with angle brackets denoting an average over realisations of $\mathbf{Y}_i$. Now, from the definition of the log-likelihood
\begin{equation}
L'(\beta) = \sum_{j=1}^n \frac{\partial}{\partial \beta} \log{p_j(\mathbf{Y}_j|\beta)}.
\end{equation}
The mean of each term in the above sum is zero, so by the CLT
\begin{equation}
n^{-1/2}L'(\beta) \sim N(0,\bar{\mathbf{C}}),
\end{equation}
where $\bar{\mathbf{C}}$ is the covariance matrix of the data averaged over $n$ realisations. In other words, $L'(\beta)$ is $\mathcal{O}(\sqrt{n})$, since the average covariance matrix is $\mathcal{O}(1)$. The $\mathcal{O}$ notation here refers to the magnitude of typical r.m.s. fluctuations in a quantity. In this notation, cumulants of individual measurements are taken to be $\mathcal{O}(1)$. Thus we can define a scaled quantity $\mathcal{Z}^{(1)}(\beta) = L'(\beta)/\sqrt{n}$, which is $\mathcal{O}(1)$. 

Now consider the second derivative,
\begin{equation}
L''(\beta) = \sum_{j=1}^n \frac{\partial^2}{\partial\beta^2} \log{p_j(\mathbf{Y}_j|\beta)},
\end{equation}
which has mean $-n\bar{F}$ where $\bar{F}$ is the average Fisher information on $\beta$, defined as $\bar{F} = \frac{1}{n}\sum F$. The total Fisher information is $\mathcal{O}(n)$, so we can define an $\mathcal{O}(1)$ standardised quantity appropriate for using the CLT:
\begin{equation}
\mathcal{Z}^{(2)}(\beta) \equiv \frac{1}{\sqrt{n}}(L''(\beta) + n\bar{F}),
\end{equation}
which is asymptotically Gaussian with zero mean. Similarly, define a standardised third derivative by
\begin{equation}
\mathcal{Z}^{(3)}(\beta) = \frac{1}{\sqrt{n}}(L'''(\beta) - n\bar{K}),
\end{equation}
where the $\mathcal{O}(1)$ quantity $\bar{K}$ is given by
\begin{equation}
\bar{K} \equiv \frac{1}{n}\sum_{j=1}^n \left \langle \frac{\partial^3}{\partial \beta^3} \log{p_j(\mathbf{Y}_j|\beta)} \right \rangle.
\end{equation}
Substituting these definitions into Equation~\eqref{eq:ML}, we have
\begin{equation}
\sqrt{n}\mathcal{Z}^{(1)} + \frac{\hat{\delta}}{\sqrt{n}} ( -n\bar{F} + \sqrt{n}\mathcal{Z}^{(2)}) + \frac{\hat{\delta}^2}{2n}(n\bar{K} + \sqrt{n}\mathcal{Z}^{(3)}) = 0,
\end{equation}
or collecting terms of the same order
\begin{equation}
\sqrt{n}(\mathcal{Z}^{(1)} - \bar{F}\hat{\delta}) + (\hat{\delta}\mathcal{Z}^{(2)} + \frac{\hat{\delta}^2}{2}\bar{K}) + \mathcal{O}\left(\frac{1}{\sqrt{n}}\right) = 0.
\end{equation}
Rearranging, we have
\begin{equation}
\hat{\delta} = \frac{\mathcal{Z}^{(1)}}{\bar{F}} + \frac{1}{\bar{F}\sqrt{n}}\left(\hat{\delta}\mathcal{Z}^{(2)} + \frac{\hat{\delta}^2 \bar{K}}{2}\right) + \mathcal{O}\left(\frac{1}{n}\right).
\end{equation}
To leading order, we only need the term in brackets to $\mathcal{O}(1)$, so we replace $\hat{\delta}$ with its zero-order value $\hat{\delta} = \mathcal{Z}^{(1)}/\bar{F}$ to get
\begin{equation}
\label{eq:MLEseries}
\hat{\delta} = \frac{\mathcal{Z}^{(1)}}{\bar{F}} + \frac{1}{\bar{F}\sqrt{n}} \left( \frac{\mathcal{Z}^{(1)} \mathcal{Z}^{(2)}}{\bar{F}} + \frac{\mathcal{Z}^{(1)\,2} \bar{K}}{2 \bar{F}^2}\right) + \mathcal{O}\left(\frac{1}{n}\right).
\end{equation}
This is the MLE at next-to-leading order in $n$. To find its bias, we take the expectation value of $\hat{\delta}$. To compute this, we need the results $\langle \mathcal{Z}^{(1)} \rangle = 0$, $\langle \mathcal{Z}^{(1)\,2} \rangle = \bar{F}$, as well as the definition
\begin{equation}
\label{eq:J}
\bar{J} \equiv \langle \mathcal{Z}^{(1)}\mathcal{Z}^{(2)} \rangle = \frac{1}{n}\sum_{i,j} \left \langle \frac{\partial \log{p_i}}{\partial \beta} \frac{\partial^2 \log{p_j}}{\partial \beta^2} \right \rangle.
\end{equation}
Note that only the diagonal components contribute to the double sum in Equation~\eqref{eq:J}, ensuring that $\bar{J} \sim \mathcal{O}(1)$. Thus we have the bias of the MLE
\begin{equation}
\label{eq:MLbias}
b \equiv \langle \hat{\beta} - \beta \rangle = \frac{1}{2 n \bar{F}^2}(\bar{K} + 2\bar{J}) + \mathcal{O}\left(\frac{1}{n^{3/2}}\right),
\end{equation}
which is the standard result from~\citet{Bartlett} and subsequent works. It may be generalised to the multi-parameter case to recover the result of~\citet{CoxSnell}. For Gaussian noise it may be shown that this is actually correct to $\mathcal{O}(n^{-2})$. We denote the leading order part of the bias as $b^{(1)}$.

It is clear from the form of the terms $\bar{K}$ and $\bar{J}$ that a bias arises from non-Gaussianity in the likelihood, in particular, $\bar{K}$ measures the expected skewness of the likelihood while $\bar{J} = -\bar{K} - \bar{F}'$ measures a combination of skewness and position-dependent curvature of the log-likelihood around the maximum in parameter space.

To link the above expression with previous work on noise bias, we note that the $1\sigma$ uncertainty on $\beta$ is proportional to $F^{-1/2} \sim \mathcal{O}(n^{-1/2})$, whereas the leading order bias is $\mathcal{O}(n^{-1})$, making the ratio of the noise bias to the standard deviation generically smaller than unity. However, if the MLE is subsequently averaged over a population of galaxies (as in the case of ellipticities and shear estimates), the error on that average will eventually become smaller than the noise bias. Either the bias must be subtracted, or the posteriors on $\beta$ from each galaxy must be correctly propagated to ensure the $b/\sigma \sim \mathcal{O}(n^{-1/2})$ behaviour is recovered. We will explore the former technique in this work, noting only that the latter is at the heart of the Bayesian approach to shear estimation.

In order to calculate the bias, we need to evaluate the expectation values in Equation~\eqref{eq:MLbias} at the true parameter value. However, the difference between the MLE and the true value is $\mathcal{O}(n^{-1/2})$, so to the accuracy of Equation~\eqref{eq:MLbias} we can evaluate the right-hand-side with the MLE as a proxy for the true parameter value. Sensitivity to this choice indicates the significance of higher-order terms in the expansion of the ML equation and higher-order terms in $n$.

\subsection{Pixel noise and galaxy model assumptions}

The noise bias expression Equation~\eqref{eq:MLbias} does not assume any particular functional form for the likelihood. In the case of shape estimation from a galaxy image, the major sources of pixel noise are expected to be Poisson noise from discrete photon counts, a Gaussian background from unresolved sources, and a small Gaussian read-out noise contribution from the CCD~\citep{2014ApJS..212....5M}. For the remainder of this work, we make the simplifying assumption that the noise is Gaussian and uncorrelated between pixels with variance $\sigma_n^2$. The log-likelihood is then
\begin{equation}
L(\beta) = -\frac{1}{2\sigma_n^2}\sum_i  [\hat{I}_i - I^M_i(\beta)]^2 + \mathrm{contant},
\end{equation}
where $\hat{I}_i$ is the measured surface brightness in pixel $i$ and $I^M_i(\beta)$ is the model prediction at the point $\beta$ in parameter space. It is straightforward to show in this case that 
\begin{equation}
\bar{K} = -3\bar{J} = \frac{3}{n \sigma_n^2} \sum_i \frac{\partial I^M_i}{\partial \beta} \frac{\partial^2 I^M_i}{\partial \beta^2},
\end{equation}
where the derivatives are evaluated at the true parameter value. This expression makes it clear that the noise bias arises due to the non-linear mapping between the parameter of interest and the data. A useful expression for the $\mathcal{Z}$-matrices introduced earlier in the derivation of the MLE is, for Gaussian noise,
\begin{equation}
\mathcal{Z}^{(n)}(\beta) = \frac{1}{\sigma_n^2 \sqrt{n}} \sum_i [\hat{I}_i - I^M_i(\beta)] \frac{\partial^{(n)} I^M_i}{\partial \beta^{(n)}}.
\end{equation}
The $\mathcal{Z}$-matrices thus represent weighted linear combinations of the data and are thus Gaussian random variables when the noise is Gaussian.

For the galaxy model, we assume a two-dimensional Gaussian surface-brightness profile with total flux $S_T$, and covariance matrix
\begin{equation}
\mathbf{Q} = \frac{r^2}{2}\left( \begin{array}{cc}
1+e_1 & e_2 \\
e_2 & 1-e_1 \end{array} \right).
\label{eq:galcov}
\end{equation}
This galaxy model is not particularly realistic, but will prove sufficient to demonstrate the fundamentals of noise bias and the MLE. Similarly, we do not consider the effects of the PSF on the biases, but note that such effects could easily be incorporated if a model for the PSF were provided. For example, a Gaussian PSF model would introduce an extra additive covariance matrix into Equation~\eqref{eq:galcov}.

The advantage of analytic functional forms for the likelihood and galaxy model is that the derivatives required to compute the bias in Equation~\eqref{eq:MLbias} can be done exactly, without the need for finite-differencing, although numerical derivatives could be used in this formalism if required.

\subsection{Ellipticity biases from noisy images}

To gain further insight into noise bias in the MLE, we pixellize the surface-brightness profile on a coarse grid of 10 $\times$ 10 pixels. The convolution of the surface-brightness profile with the pixel window function is done on an upsampled grid using Fast Fourier Transforms, then downsampled to the observed resolution. The upsampling resolution is adaptively set depending on the typical curvature radius of the input image profile. We have checked that our results are insensitive to changes in the upsampling resolution\footnote{Note that the convolution of a Gaussian and a top-hat can be done analytically when the image is aligned with the grid, providing a useful check for numerical errors.}. The coarse grid allows for rapid calculation of noise biases and likelihood derivatives.

We fix the galaxy scale-length $r$ to 1.2 pixels. A typical pixel size for planned space-based surveys such as \emph{Euclid} is 0.1 arcseconds, corresponding to a galaxy scale-length of 0.12 arcseconds. This is roughly $1\sigma$ smaller than the peak of the intrinsic distribution of scale-lengths used by the CFHTLenS survey~\citep{2013MNRAS.429.2858M} and the 0.3 arcseconds of a typical \emph{Euclid} source~\citep{2012SPIE.8442E..0VC}. Given the steep slope of the mass function for the typical galaxies used in weak lensing surveys, we expect most of the sources in a flux-limited survey to be close to the resolution limit (PSF size) of the telescope, so our assumed value of $r$ does not represent an extreme choice. However it should be borne in mind that this conservative choice implies that our bias results are slightly pessimistic. Combined with our coarse pixellization, this choice of $r$ has the advantage that biases induced by the finite postage-stamp size are avoided. These extra biases, which do not form part of the current study, have recently been speculated as potentially problematic for ML-based methods~\citep[GREAT3;][]{2015MNRAS.450.2963M}.

We set the peak surface brightness value $I_0 = S_T/2\pi\sqrt{|Q|}$ to unity and then set the pixel noise variance $\sigma_n^2$ by the signal-to-noise ($S/N$), which is defined in the same way as~\citet{2014ApJS..212....5M}
\begin{equation}
S/N = \frac{\sqrt{\sum_i [I^M_i(\beta_0)]^2}}{\sigma_n},
\end{equation}
where the reference model $I^M(\beta_0)$ is taken to have an ellipticity of $(e_1,e_2) = (0.3,0.0)$. Note that we will sometimes use $\nu$ to denote $S/N$ in this work. The $S/N$ as defined above clearly scales as $\sqrt{n}$, and it is straightforward to show that the first-order ML bias in Equation~\eqref{eq:MLbias} scales as $\nu^{-2}$, consistent with previous studies of noise bias~\citep[e.g.][]{2004MNRAS.353..529H}.

We compute derivatives by analytically differentiating the Gaussian surface-brightness profile and then pixellizing. For ellipticity biases this requires some care since oscillatory features in the the image domain induced by differentiating require a higher sampling rate in the convolution step.

\begin{figure}
\centering
\includegraphics[width=\columnwidth]{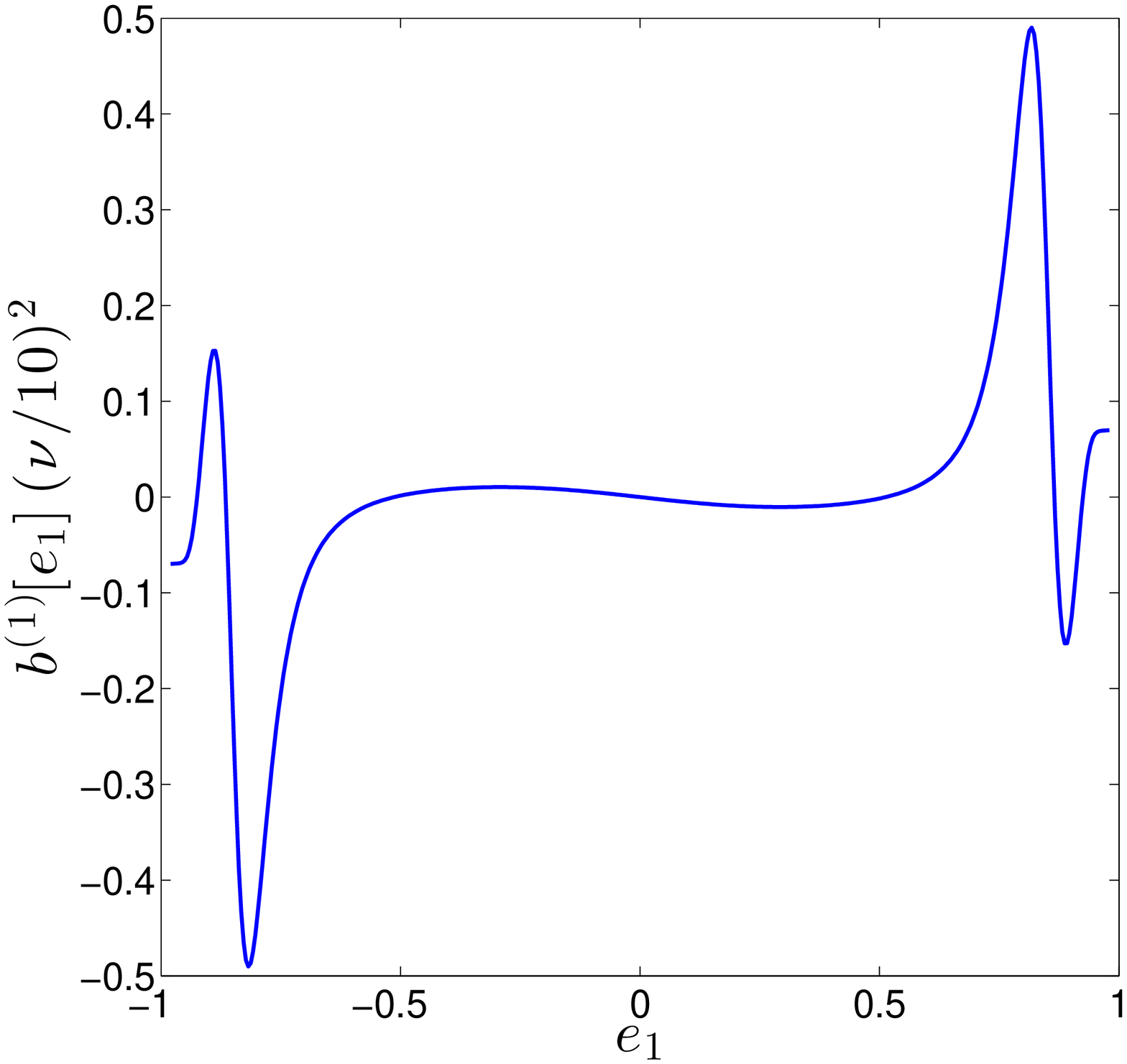}
\includegraphics[width=\columnwidth]{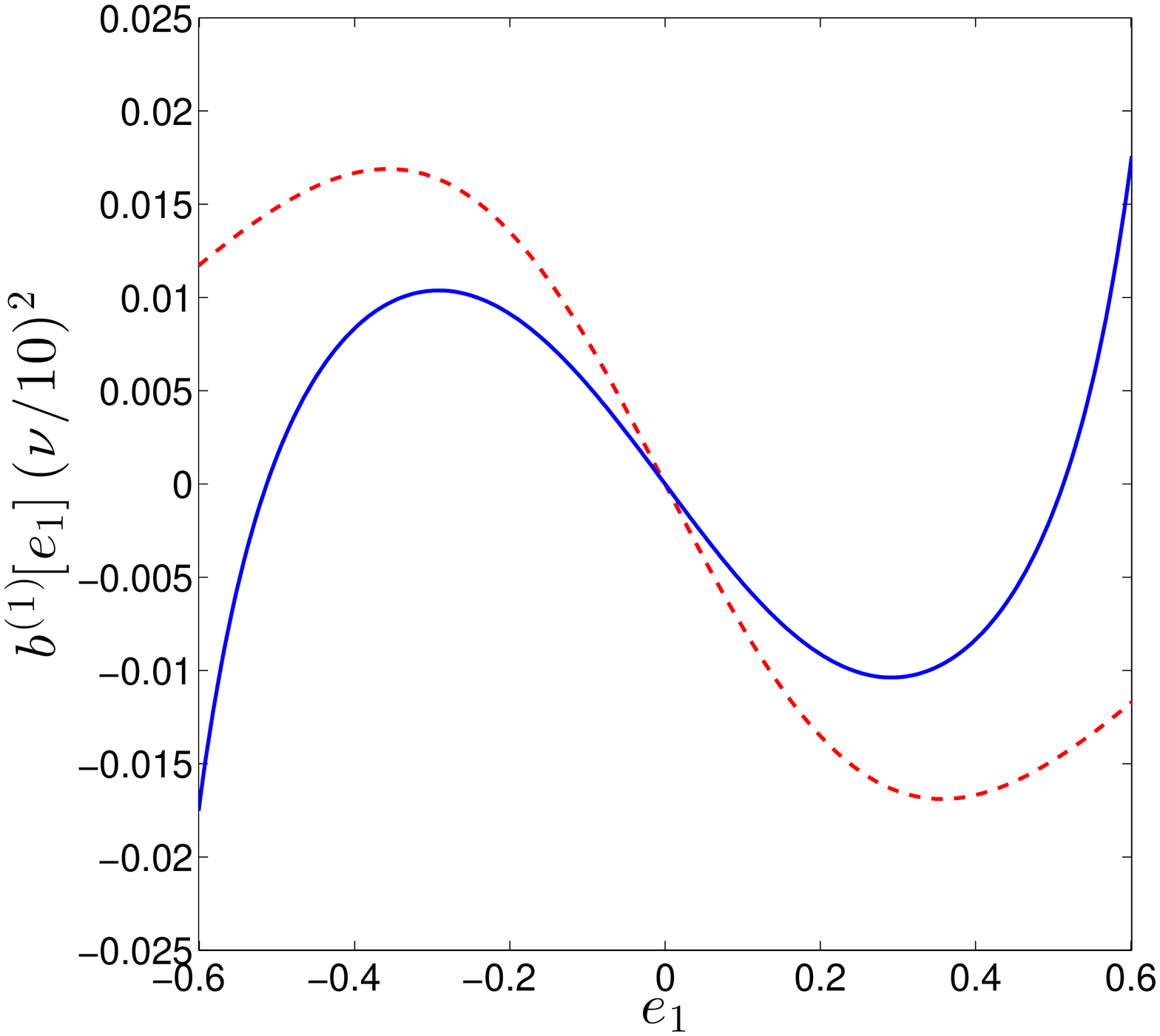}
\includegraphics[width=\columnwidth]{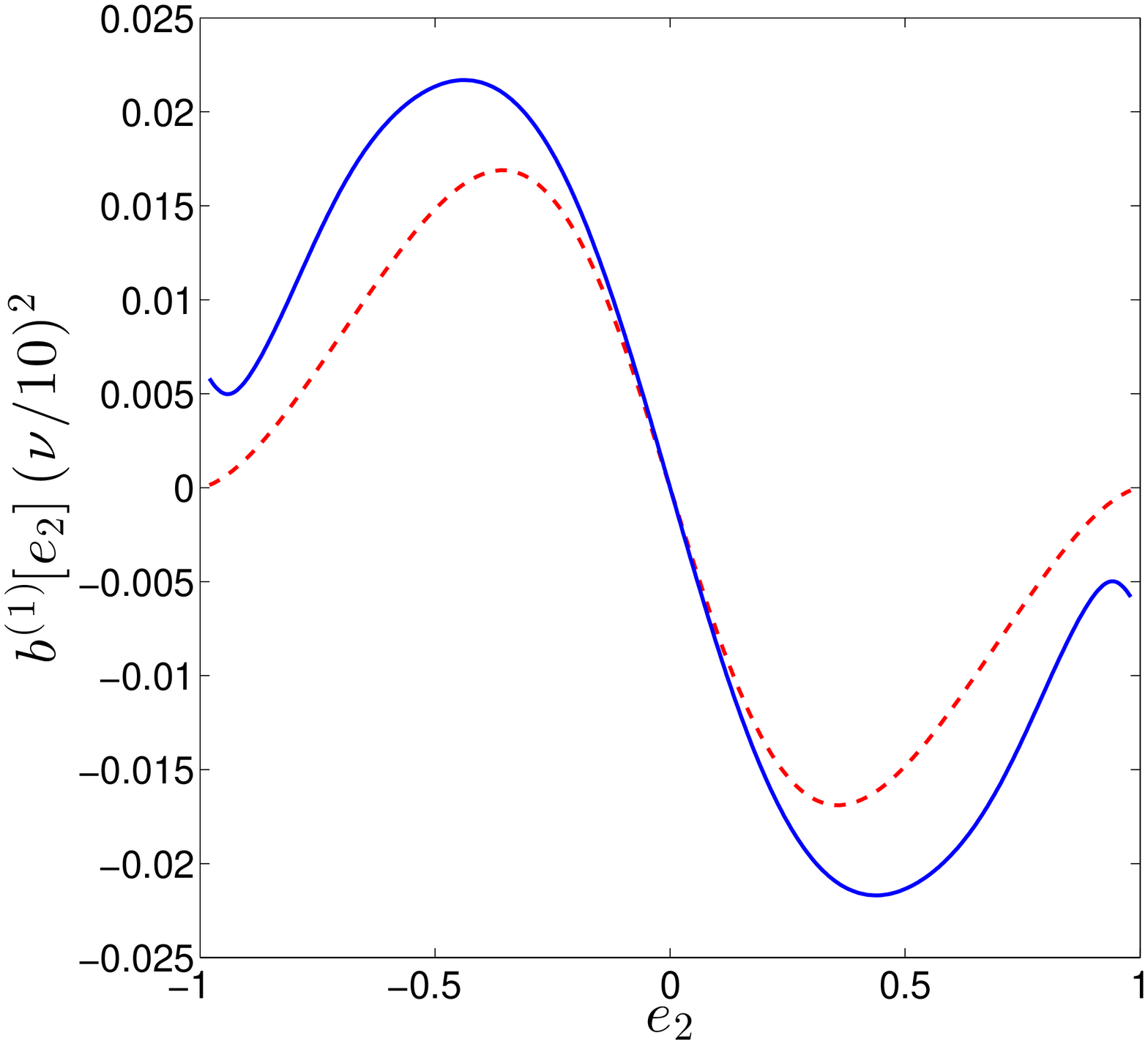}
\caption{\emph{Upper panel:} Normalized first-order MLE bias on $e_1$. \emph{Middle panel:} Normalized first-order MLE bias on $e_1$ (blue solid) and its continuum-limit result (red dashed). Note that the blue curve is the same as that in the upper panel but on a reduced horizontal axis. \emph{Lower panel:} Normalized first-order MLE bias on $e_2$ (blue solid) and its continuum-limit result (red dashed).}
\label{fig:b1np}
\end{figure}

In the top panel of Figure~\ref{fig:b1np} we plot the first-order bias on $e_1$ from Equation~\eqref{eq:MLbias} assuming $e_2 = 0$, as a function of the true value of $e_1$. Given the $S/N$ scaling we choose to plot the $\nu$-independent quantity $b^{(1)}(\nu/10)^2$. The bias is computed over the range $|e_1| \leq 0.98$. Higher values of $|e_1|$ require a prohibitively high sampling rate. Even in this very simplified example a few instructive lessons may be learned. The bias is clearly a highly non-linear function of $e_1$ for $|e_1| \gtrsim 0.1$. We also note that while antisymmetric, the sign of the bias is never guaranteed to be opposite to that of $e_1$, i.e. noise bias does not necessarily isotropize the image.

\begin{figure}
\centering
\includegraphics[width=\columnwidth]{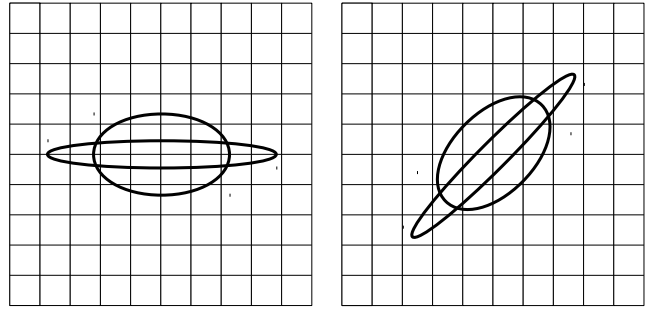}
\caption{\emph{Left panel:} Effect of increasing $e_1$ when $e_2 = 0$ and $r$ is held fixed. Solid lines enclose 1$\sigma$ areas of surface brightness. As $e_1$ grows, all pixel centres eventually sample the tails of the Gaussian surface brightness profile rather than its core. Small changes to $e_1$ can thus gives large changes in surface brightness samples, and hence large noise biases. \emph{Right panel:} The equivalent situation for $e_2$. Now there are always pixel centres sampling the inner regions of the Gaussian. Small changes in $e_2$ thus give smaller changes to the sampled surface brightnesses, and so the bias is smaller.}
\label{fig:grids}
\end{figure}

At large $|e_1|$ the bias grows very large compared to its typical $\mathcal{O}(10^{-2})$ values at small $|e_1|$. This behaviour is due to the finite pixellization scale, and our choice of centroid. Since we have chosen the grid to be centred on the galaxy image with an even number of pixels on each side, for increasing values of $e_1$ the central pixels sample the underlying surface-brightness profile at increasingly higher $\sigma$ away from the image centre. Due to the rapidly decaying tails of the Gaussian galaxy profile, this means that the derivatives of the image profile with respect to $e_1$ become very large at large $|e_1|$. This is illustrated schematically in Figure~\ref{fig:grids}. Had we chosen the grid to have an \emph{odd} number of pixels on each side, we would always sample the central regions of the Gaussian even at large $|e_1|$, and the noise bias would be smaller.

The effects of finite pixellization may be studied by comparing the exact bias of the pixellized image with the exact continuum-limit (i.e. zero pixel-size) result, which are given by the blue solid and red dashed curves in Figure~\ref{fig:b1np} respectively. These two biases agree reasonably well for $|e_1| \lesssim 0.3$. In the middle panel of Figure~\ref{fig:b1np} we show this regime, which represents the typical $1\sigma$ range of observed ellipticities of galaxies in typical weak lensing surveys. Even in this mildly elliptical regime the bias is highly non-linear, a feature which will be important when we try to remove the bias from the MLE. The rough behaviour of the bias with $e_1$ matches the continuum-limit over this range of ellipticity, but the amplitude is reduced even for circular galaxies. Note that the continuum-limit biases are equal for $e_1$ and $e_2$, as required by isotropy.

In the lower panel of Figure~\ref{fig:b1np} we plot the bias on $e_2$ as a function of the true value of $e_2$ assuming $e_1=0$. Unlike for $e_1$, the bias on $e_2$ remains much closer to its continuum-limit form out to high values of $|e_2|$. The reason for this is again due to our assumption of a compact Gaussian functional form for the surface brightness profile coupled with the finite pixellization and our choice for the centroid location. A galaxy with zero $e_1$ lies $45^{\circ}$ to the horizontal axis, and hence when $|e_2|$ is large there are still pixels sampling the core of the Gaussian along its `ridge'. Thus the derivatives with respect to ellipticity are never as large as they are for $e_1$, and the bias is smaller, as illustrated in the right panel of Figure~\ref{fig:grids}.


We emphasise that the pixellization and galaxy size we have chosen for these examples are not extreme choices, and the amplification of noise bias for highly elliptical images could prove a real issue for upcoming weak lensing surveys. One way of mitigating this would be to allow for the centroid to vary and possibly marginalised over, which should go some way to restoring isotropy. Another possibility would be to centre each galaxy image on a grid having an odd number of pixels on each side, as described above. We chose an even grid size to reduce the number of mock images required to detect noise bias - note that even-sized grids are commonly used to test shear measurement methods~\citep[e.g. GREAT10][]{2012MNRAS.423.3163K}, which additionally allows us to compare our results with these studies.

Despite the extreme behaviour of the bias at large $|e_1|$, we will see that the bias can be still be accurately predicted and removed. 

\subsection{The bias-corrected MLE and second-order biases}

The advantage of the analytic form for the noise bias derived above is that the leading-order bias may be subtracted off without external calibration from simulations. We define the first-order bias-corrected MLE as
\begin{equation}
\label{eq:MLEb1}
\hat{\beta}^{(1)} \equiv \hat{\beta} - b^{(1)}(\hat{\beta}),
\end{equation}
where we have used the MLE as a proxy for the true parameter value in the first-order bias function.

If the MLE is used as a proxy for the true parameter value in this way, the sampling distribution of the resulting \emph{bias-corrected} MLE will acquire some non-Gaussianity due to the non-linear dependence of the ML bias on the data. Thus, it may be thought that this procedure for removing bias might spoil the nice asymptotic properties of the MLE, in particular its efficiency over other point estimators. However it may be shown that removing the $\mathcal{O}(n^{-1})$ bias results in a second-order efficient estimator~\citep{rao}. In other words, after correction for the bias, the MLE still has the optimal variance amongst other point estimators not only at leading order in $\mathcal{O}(n^{-1})$ but at next-to-leading order. This suggests that the bias-corrected MLE is the best one can hope to achieve at this order in $\mathcal{O}(n^{-1})$.

The estimator in Equation~\eqref{eq:MLEb1} is unbiased at $\mathcal{O}(n^{-1})$, or equivalently at $\mathcal{O}(\nu^{-2})$. It does however still receive bias both from higher order terms neglected in the perturbative expansion of the ML equation and from the fact that the MLE differs from the true parameter value. We refer to these second-order biases as the intrinsic and the bias-on-bias terms respectively. For Gaussian noise, both are $\mathcal{O}(\nu^{-4})$. The derivation of the intrinsic term follows straightforwardly from the ML equation using the same techniques as above, so we do not reproduce it here. For the one-parameter case the intrinsic term is, for Gaussian noise,
\begin{equation}
b^{(2)} = \frac{1}{n^2 \bar{F}^3}\left[\frac{\bar{K}\bar{P}}{2\bar{F}} + \frac{\bar{K}^3}{24\bar{F}^2} + \frac{\bar{K}\bar{M}}{12\bar{F}} + \frac{\bar{R}}{2} + \frac{\bar{Q}}{40}\right],
\end{equation}
where
\begin{align}
\bar{P} &\equiv \frac{1}{n \sigma_n^2}\sum_i \left(\frac{\partial^2 I^M_i}{\partial \beta^2}\right)^2, \nonumber \\
 \bar{R} &\equiv \frac{1}{n \sigma_n^2} \sum_i \frac{\partial^2 I^M_i}{\partial \beta^2} \frac{\partial^3 I^M_i}{\partial \beta ^3} \nonumber \\
\bar{M} &\equiv \frac{1}{n}\langle L^{\mathrm{IV}} \rangle = -\frac{1}{n \sigma_n^2} \sum_i \left[4\frac{\partial I^M_i}{\partial \beta} \frac{\partial^3 I^M_i}{\partial \beta^3} + 3 \left(\frac{\partial^2 I^M_i}{\partial \beta^2}\right)^2 \right], \nonumber \\
\bar{Q} &\equiv \frac{1}{n}\langle L^{\mathrm{V}} \rangle =  -\frac{5}{n \sigma_n^2} \sum_i \left[\frac{\partial I^M_i}{\partial \beta}\frac{\partial^4 I^M_i}{\partial \beta^4} + 2 \frac{\partial^2 I^M_i}{\partial \beta^2}\frac{\partial^3 I^M_i}{\partial \beta^3} \right].
\end{align}
It is difficult to gain a clear intuition for the various terms in the second-order intrinsic bias, but qualitatively we can say that while the first-order bias arises from skewness in the likelihood, the second-order bias arises from its kurtosis and higher-order non-Gaussian moments.

\begin{figure*}
\centering
\includegraphics[width=0.33\textwidth]{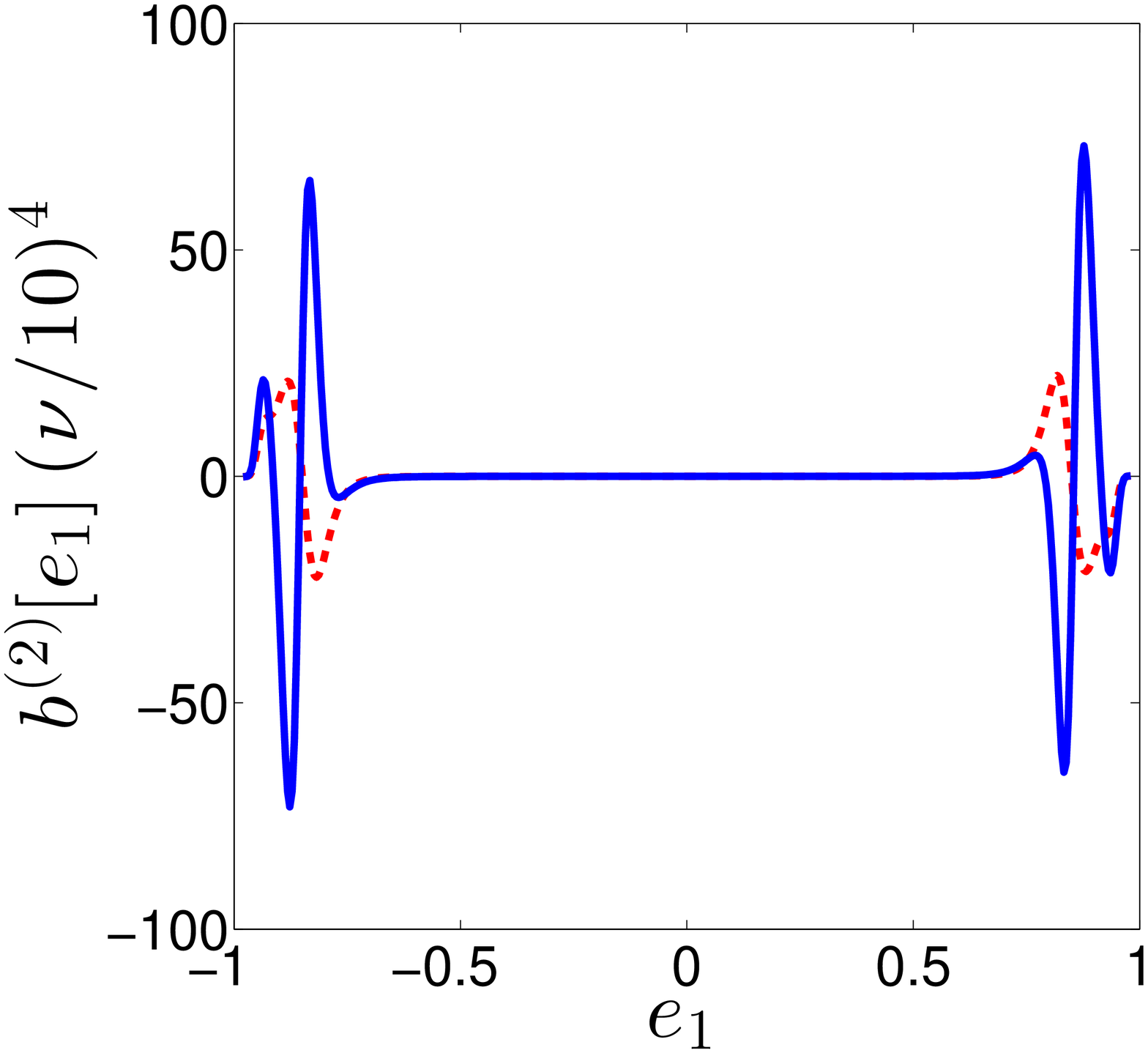}
\includegraphics[width=0.33\textwidth]{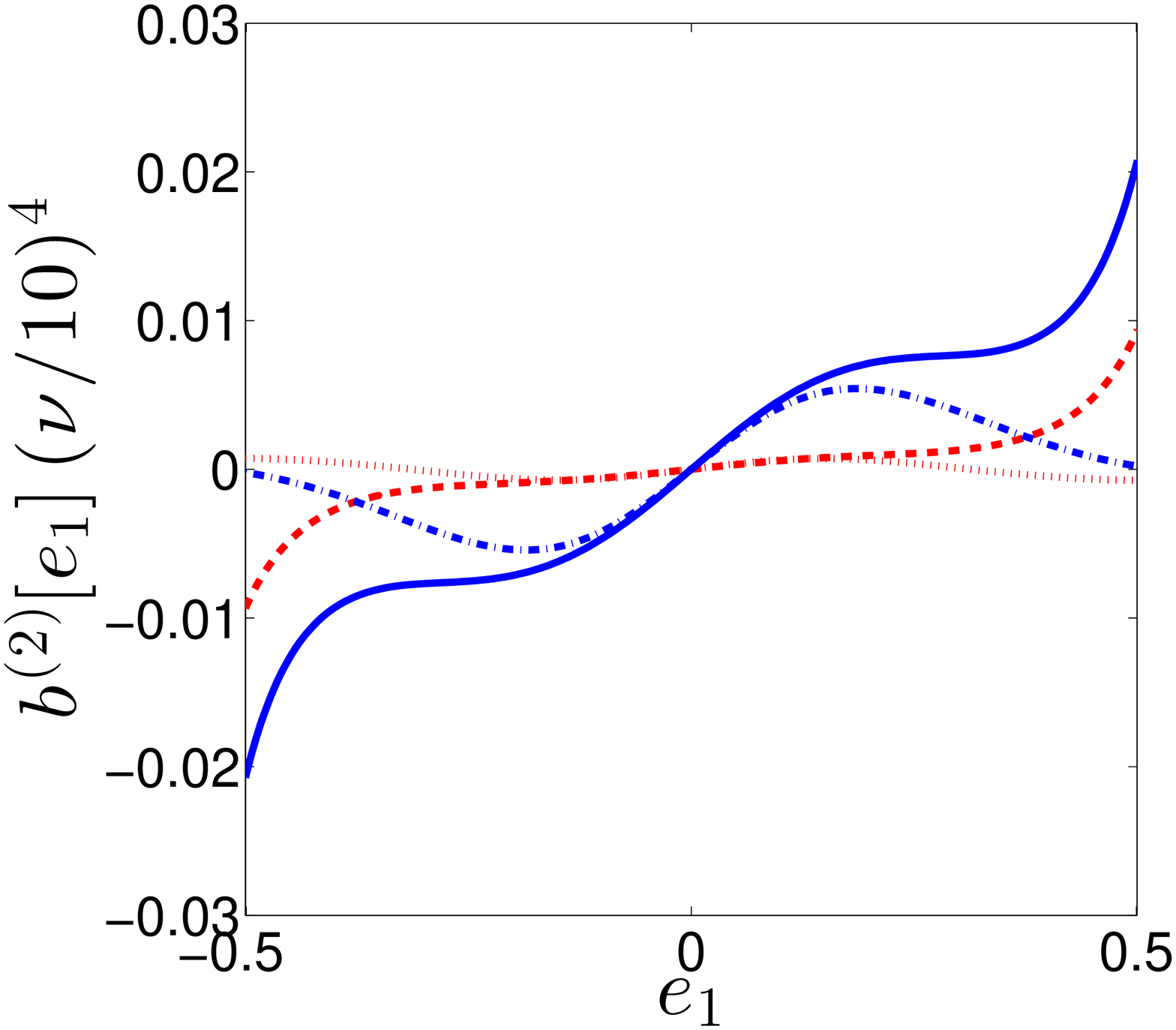}
\includegraphics[width=0.33\textwidth]{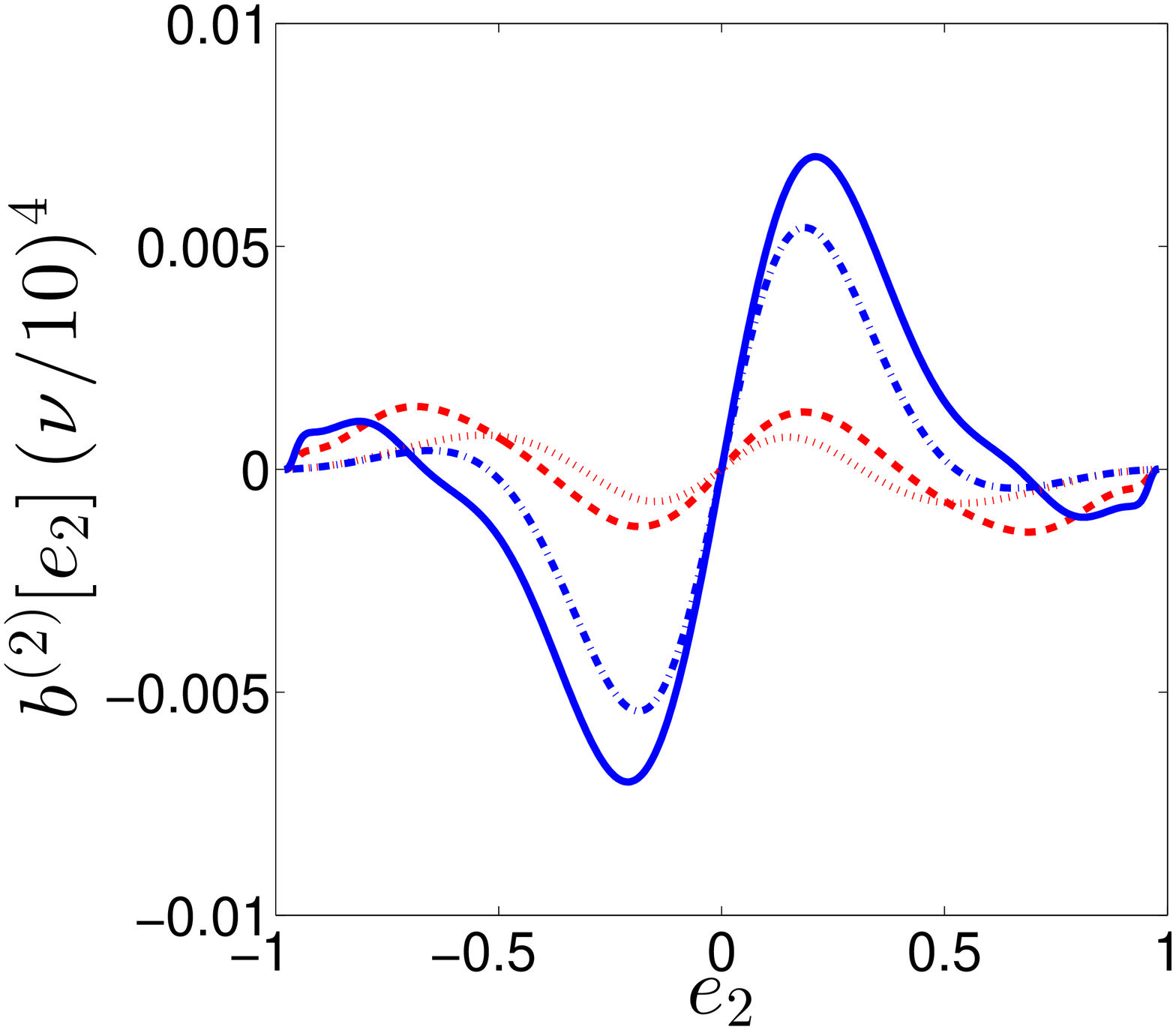}
\caption{\emph{Left panel:} Normalized second-order MLE biases on $e_1$, showing the bias-on-bias term (blue solid) and the intrinsic term (red dashed) arising from second-order terms in the Maximum Likelihood equation. \emph{Middle panel:} Normalized second-order MLE biases on $e_1$, showing the bias-on-bias term (blue solid), its continuum-limit result (blue dot-dashed), the intrinsic term (red dashed) and its continuum-limit result (red dotted). \emph{Right panel:} Same as middle panel for $e_2$.}
\label{fig:b2np}
\end{figure*}

The bias-on-bias term arising from use of the MLE instead of the true parameter value in $b^{(1)}$ is more straightforward to derive, and can be found either by Taylor expanding $b^{(1)}(\hat{\beta})$ around the true parameter value or by considering $b^{(1)}$ as a parameter and finding the first-order MLE bias using Equation~\eqref{eq:MLbias}. We find the result to be
\begin{equation}
b^{(1)}[b^{(1)}] =  \frac{1}{n^2 \bar{F}^3}\left[-\frac{13\bar{K}\bar{P}}{48\bar{F}} + \frac{7\bar{K}^3}{27\bar{F}^2} + \frac{13\bar{K}\bar{M}}{48\bar{F}} - \frac{\bar{R}}{4} + \frac{\bar{Q}}{20}\right].
\end{equation}
In the left-hand panel of Figure~\ref{fig:b2np} we plot the second-order biases on $e_1$ as a function of the true value. As in the case of the first-order bias, both second-order terms become large for highly elliptical galaxies. Furthermore, the two contributions are never guaranteed to be of the same sign, meaning that for different values of $e_1$ the total second-order bias, $b^{(2)} - b^{(1)}[b^{(1)}]$, may be enhanced or reduced through cancellations. For $|e_1|\lesssim 0.75$ however the terms are of the same sign and the bias is reduced.

The second-order bias is much more sensitive to finite-pixellization effects than its first-order counterparts. We find that both second-order terms match their continuum-limit forms well for $|e_1|\lesssim 0.2$, but deviate for more elliptical images. In the middle panel of Figure~\ref{fig:b2np} we zoom in on this region, which demonstrates that for typical galaxy ellipticities the bias-on-bias term is larger than the intrinsic second-order bias. The increased sensitivity to pixellization of second-order terms is due to the higher-order derivatives they require for calculation, coupled with the steepness of the Gaussian surface brightness profile at large $|e_1|$.

In the right-hand panel of Figure~\ref{fig:b2np} we show the second-order biases on $e_2$ as a function of the true parameter value. As in the case of the first-order bias, the second-order bias on $e_2$ is less sensitive to the pixellization and matches its continuum-limit value much more closely even for high $|e_2|$. In contrast to the $e_1$ bias, both second-order terms for $e_2$ are of the same sign, allowing for cancellation in the total bias on $\hat{\beta}^{(1)}$. Similarly to $e_1$, the bias-on-bias tends to be larger than the intrinsic term.


With the second-order bias now calculable, we can define a second-order bias-corrected MLE by
\begin{equation}
\hat{\beta}^{(2)} \equiv \hat{\beta} - b^{(1)}(\hat{\beta}) - b^{(2)}(\hat{\beta}) + b^{(1)}[b^{(1)}](\hat{\beta}).
\end{equation}
This estimator now has a leading-order bias from $\mathcal{O}(\nu^{-6})$ terms, and since the MLE is both third and fourth-order efficient~\citep{Kano} we might speculate that $\hat{\beta}^{(2)}$ is still optimal over other second-order bias-corrected point estimators in the sense of having a smaller mean-squared error.

\begin{figure*}
\centering
\includegraphics[width=0.33\textwidth]{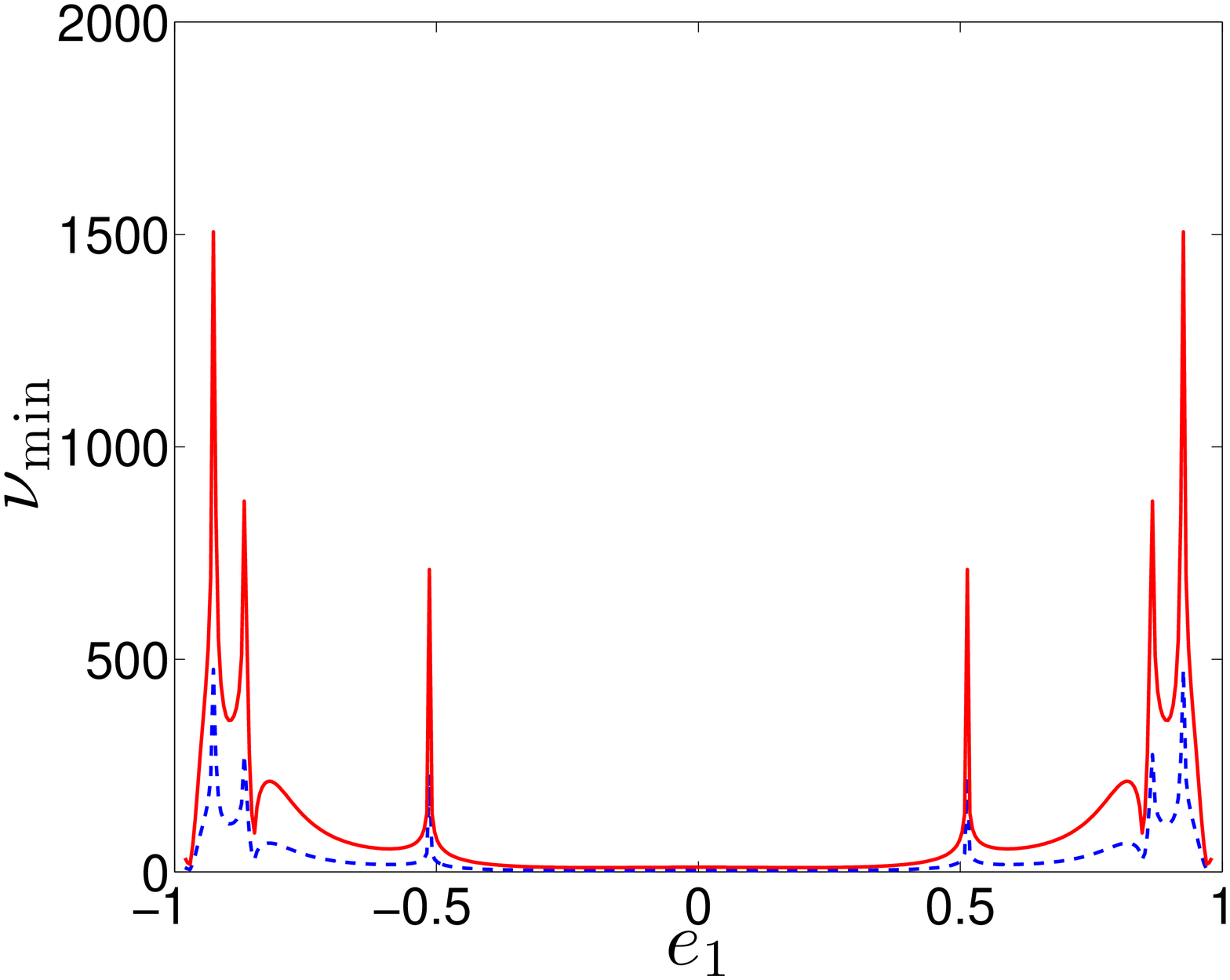}
\includegraphics[width=0.33\textwidth]{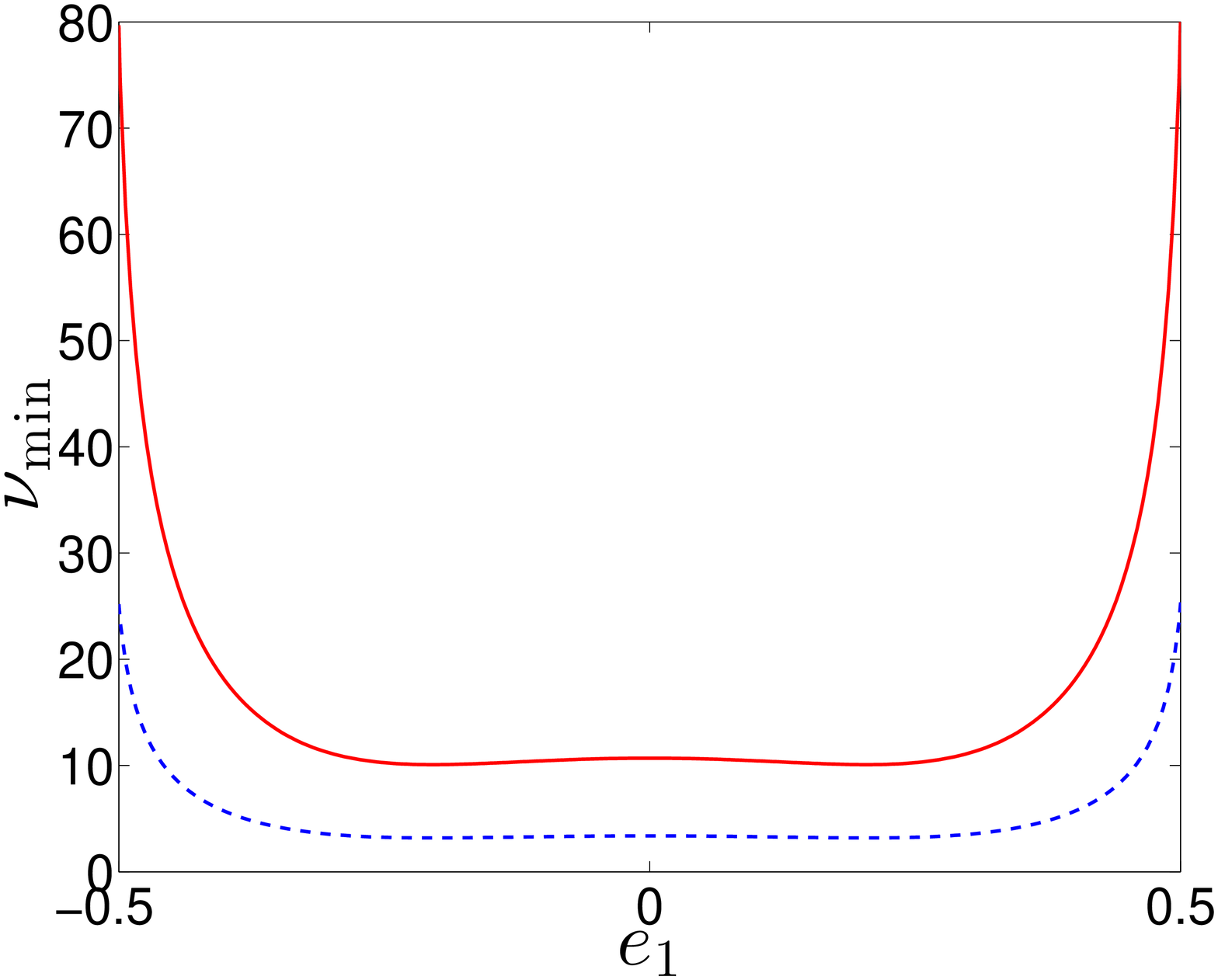}
\includegraphics[width=0.33\textwidth]{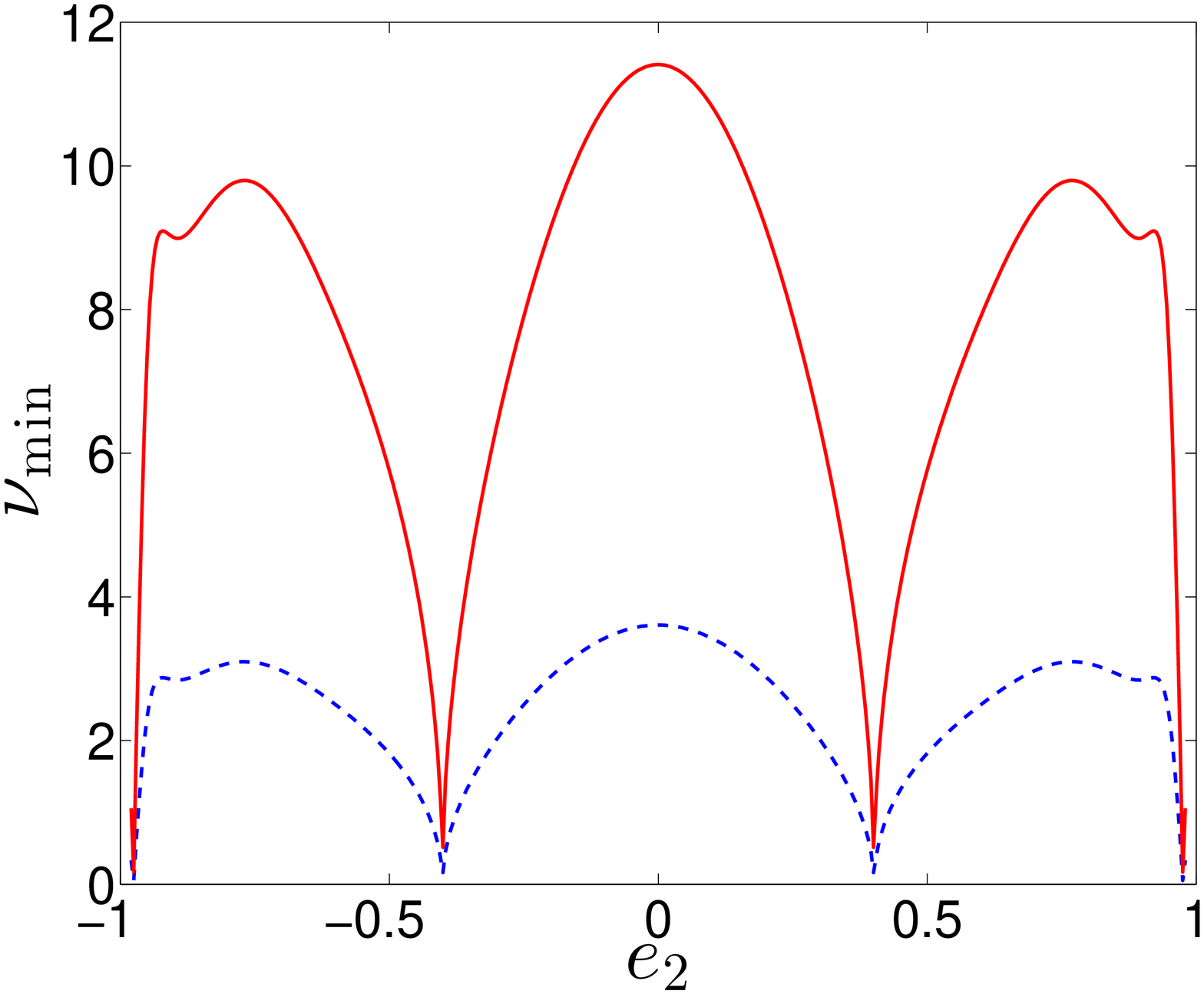}
\caption{\emph{Left panel:} Value of the $S/N$ at which the intrinsic second-order bias on $e_1$ equals the first-order MLE bias as a function of the true parameter value (blue dashed) and the $S/N$ at which it is 10\% of the first-order bias (red solid). \emph{Middle panel:} Same as left panel for small $|e_1|$. \emph{Right panel:} Same as left panel for $e_2$.}
\label{fig:numinnp}
\end{figure*}

Aside from defining higher-order bias-corrected estimators, the advantage of being able to calculate the second-order bias is that it allows us to approximate the minimum $S/N$ for which the perturbative bias predictions are expected to be valid. We find this by setting the intrinsic second-order bias equal to the first-order bias for each true ellipticity value and solving for the $S/N$, denoted $\nu_{\mathrm{min}}$. In the left-hand panel of Figure~\ref{fig:numinnp} we plot this quantity, as well as the $S/N$ at which the second-order bias is 10\% of the first-order bias. The spikes in this plot arise from points where the first-order bias goes to zero, at which point higher-order terms necessarily become significant. It is interesting to note that this can happen even for mildly elliptical galaxies with $e_1 \approx 0.5$. Zooming in on small $|e_1|$ in the middle panel of Figure~\ref{fig:numinnp}, we see that the minimum useable $S/N$ is quite flat in $e_1$, with second-order terms becoming significant at $S/N \approx 10$. The situation is even better for $e_2$, with the right-hand panel of Figure~\ref{fig:numinnp} suggesting that $S/N \approx 10$ is a reasonable approximation to the minimum $S/N$ for which our perturbative bias formalism will be accurate.

\section{Ellipticity biases from simulated images}
\label{sec:eresults}

Now we have the machinery to predict the MLE bias directly from the likelihood, we can test the performance of the bias-corrected estimates on simulated noisy galaxy images. For this purpose we generate a large number of pixellized galaxy images with random Gaussian noise, and a true ellipticity having either $(e_1,e_2) = (0.3,0.0)$ when using $e_1$ as a parameter or $(e_1,e_2) = (0.0,0.3)$ when using $e_2$. The number of galaxies generated at each $S/N$ was $1000\times(S/N)^2$, where the $S/N$ scaling ensures smaller error bars when the bias is smaller. At this stage we stick to one-dimensional parameter inferences. For each noisy image we find the MLE by gridding up the likelihood, and then compute the first and second-order biases with the MLE as a proxy for the true parameter value. This allows us to construct first and second-order bias corrected estimators. We have checked that our results are insensitive to changes in the likelihood grid-spacing when finding the maximum.

\begin{figure*}
\centering
\includegraphics[width=0.45\textwidth]{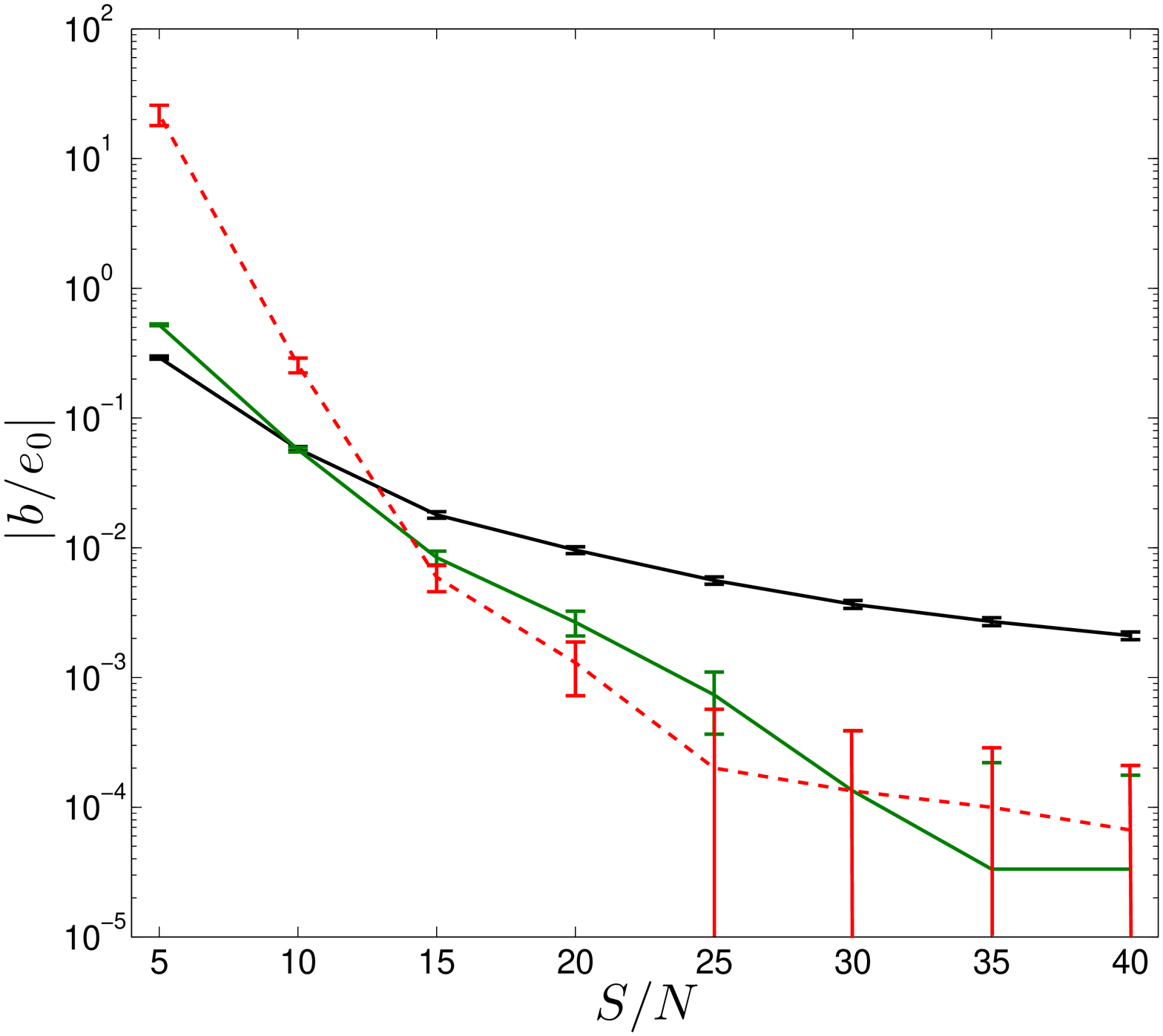}
\includegraphics[width=0.45\textwidth]{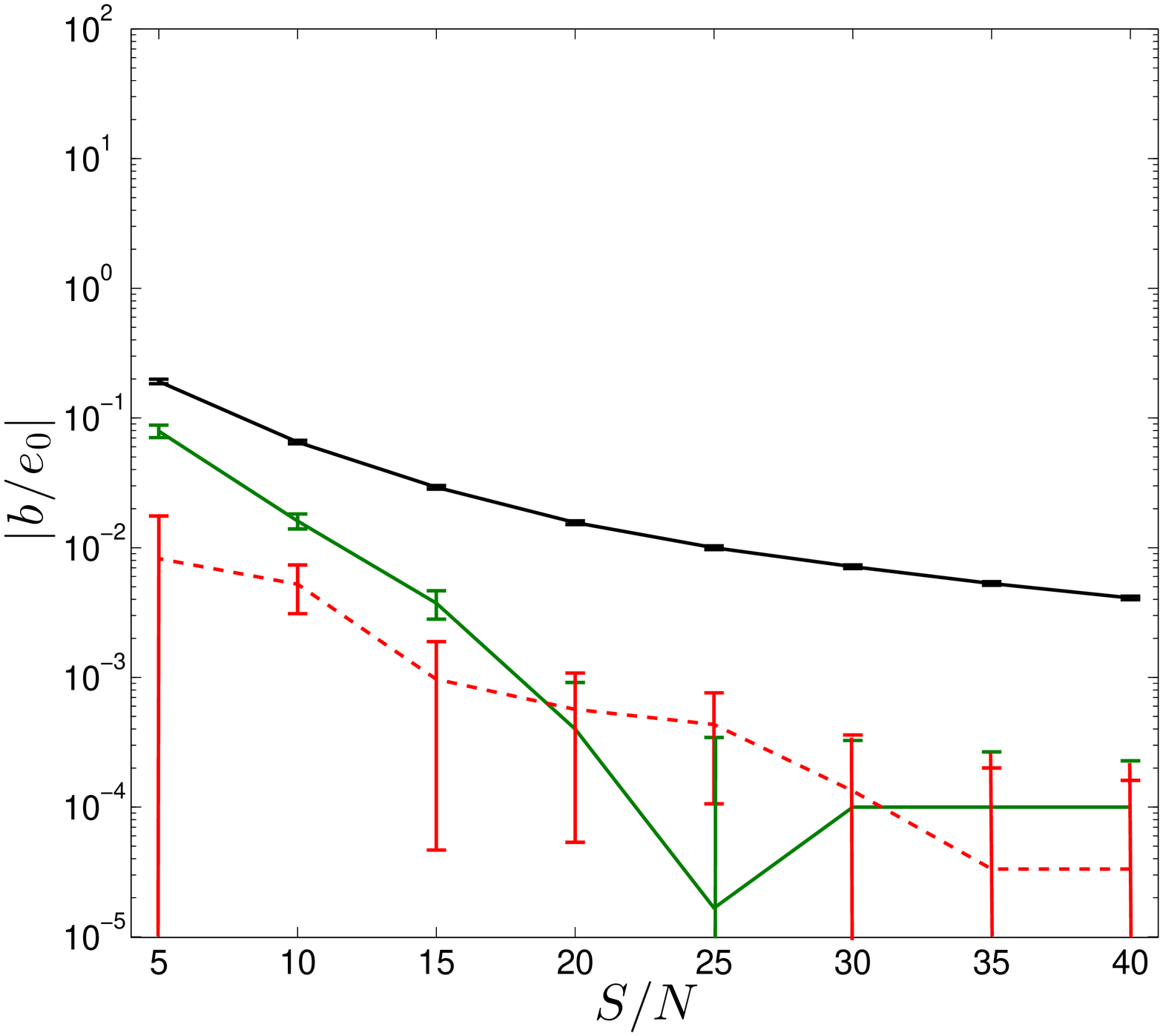}

\caption{\emph{Left panel:} Magnitude of measured bias on the $e_1$ MLE compared to the true value (black solid), the first-order bias-corrected MLE (green solid) and the second-order bias-corrected MLE (red dashed). \emph{Right panel:} Same as left panel for $e_2$.}
\label{fig:simbnp}
\end{figure*}

In the left-hand panel of Figure~\ref{fig:simbnp} we show the resultant biases on $e_1$ for the uncorrected, first-order corrected, and second-order corrected estimators for a range of $S/N$ values. The error bars are calculated from the empirical variance of the MLE for each noise realization at the given $S/N$. We plot the magnitude of the dimensionless quantity $b/e_0$ where $e_0$ is the true ellipticity. Values of unity on the horizontal axis thus represent order unity changes to the measured ellipticity from noise bias.


From this figure we see that a first-order bias correction can reduce the bias on the MLE by several orders of magnitude at high $S/N$, and helps to reduce bias all the way down to $S/N = 15$. At lower $S/N$, higher-order terms become significant and a first-order correction is not sufficient. The second-order correction does even better, clearly reducing the bias further down to $S/N=12$. For noisier images a second-order correction actually increases bias, due to the importance of higher-order terms in the expansion of the ML equation, as expected from Figure~\ref{fig:numinnp}.

\begin{figure*}
\centering
\includegraphics[width=0.45\textwidth]{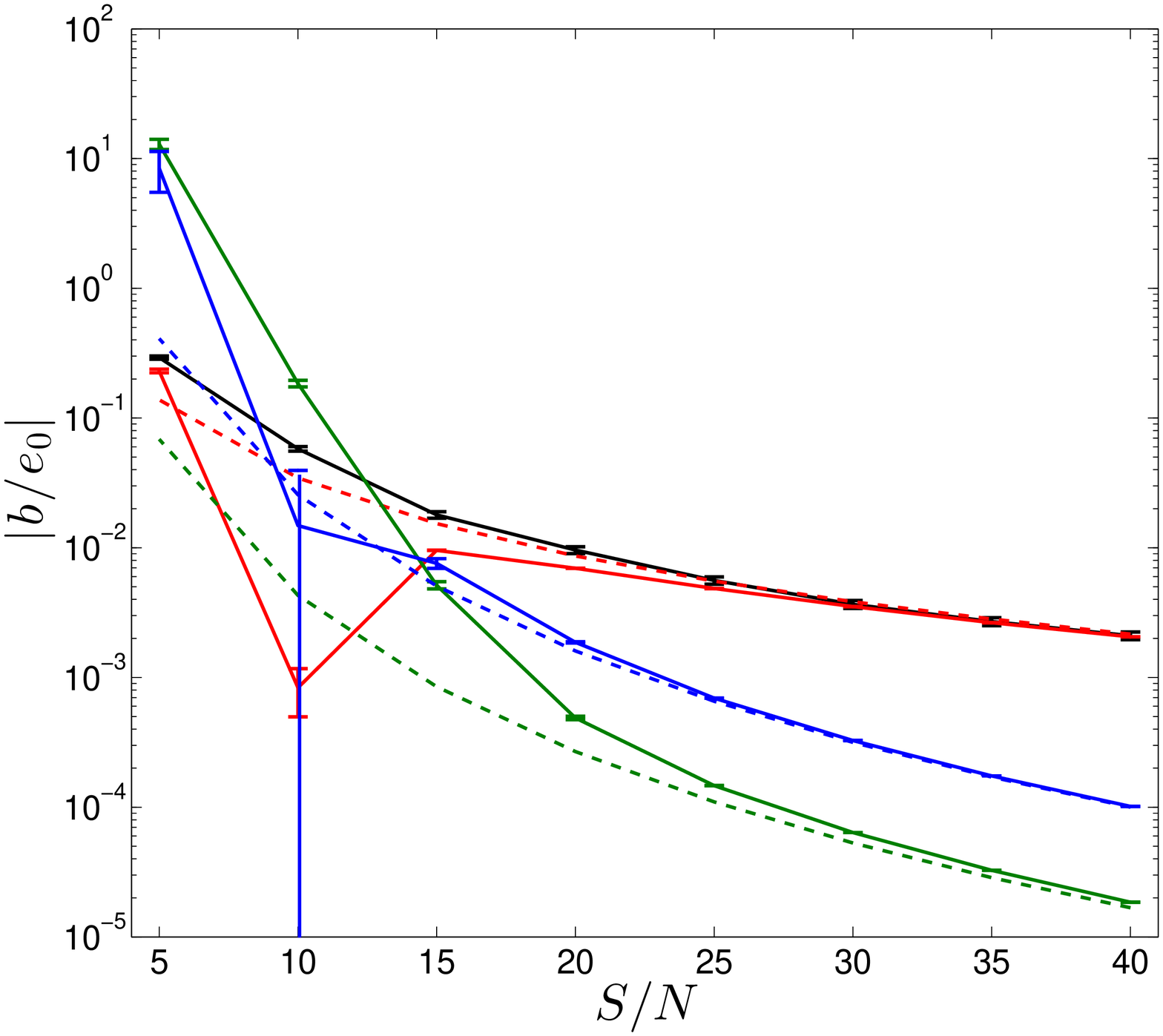}
\includegraphics[width=0.45\textwidth]{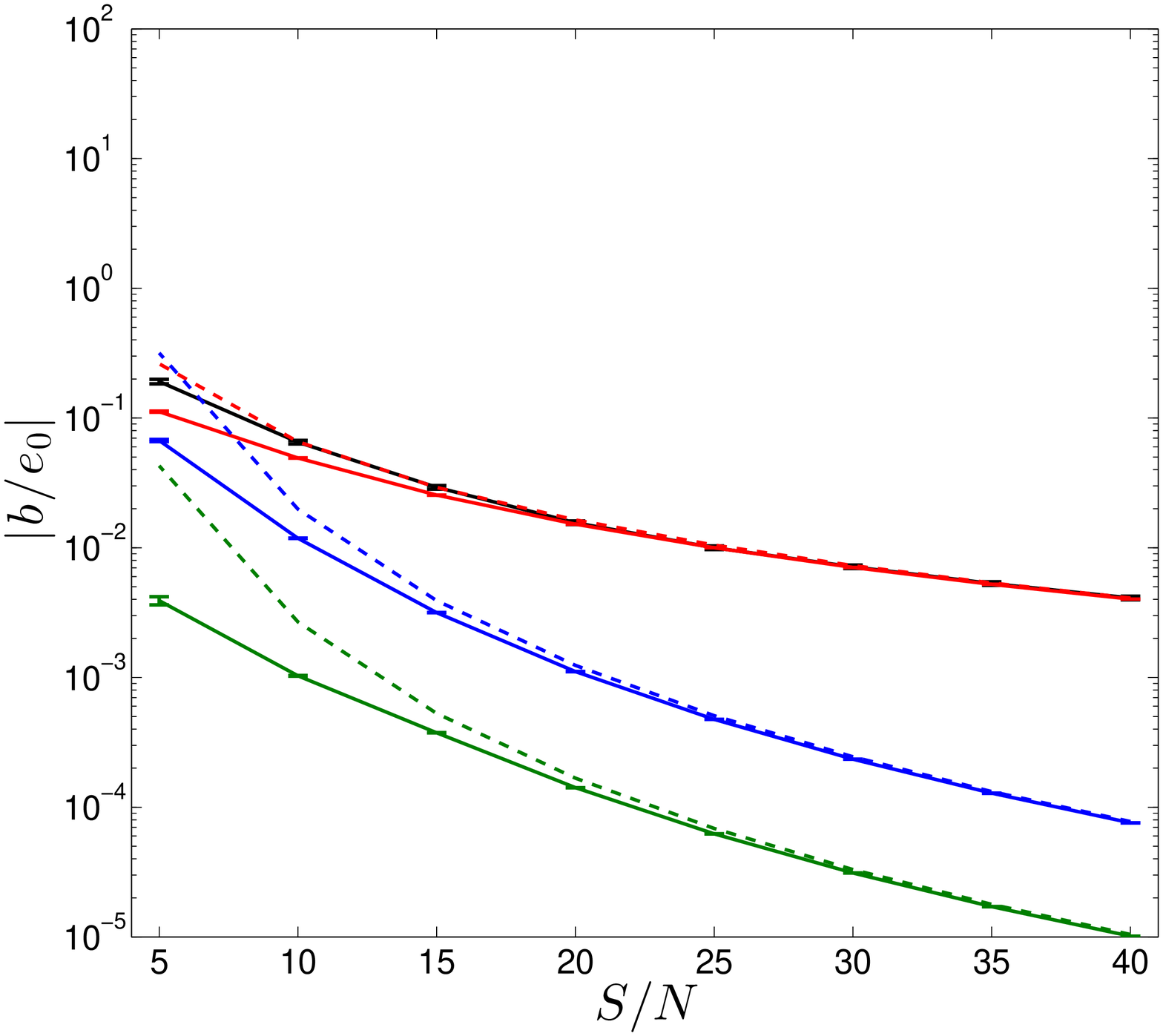}
\caption{\emph{Left panel:} Magnitude of MLE bias compared to true value of $e_1$ (black solid), mean value of measured first-order bias (red solid), true first-order bias prediction (red dashed), mean value of measured intrinsic second-order bias (green solid), true intrinsic second-order bias (green dashed), mean value of measured bias-on-bias (blue solid), true bias-on-bias (blue dashed). \emph{Right panel:} Same as left panel for $e_2$.}
\label{fig:simbnpests}
\end{figure*}

To gain further insight into the performance of the estimators, in the left-hand panel of Figure~\ref{fig:simbnpests} we plot the empirically measured biases from the simulations against their theoretical expectations. The dashed lines in this figure are the predictions from Section~\ref{sec:bias} using the true parameter value, and the solid lines are the empirical estimates of these biases using the MLE as a proxy for truth. Thus the dashed lines scale precisely as $\nu^{-2}$ and $\nu^{-4}$. At high $S/N$ the MLE exhibits small fluctuations about the true parameter value, and hence the measured biases closely match the predictions. As we increase the noise variance, the MLE starts to depart more frequently from the truth in each noise realization. The difference between the mean of the measured first-order biases and the true values is given, to leading order, by the bias-on-bias term introduced above. From Figures~\ref{fig:b1np} and~\ref{fig:b2np} we see that at $e_1=0.3$ this term is positive, whereas the first-order bias is negative. Thus, as the $S/N$ is decreased, the mean first-order bias measured from the simulations becomes less negative than the true value, and so the red points in Figure~\ref{fig:simbnpests} underpredict the true values give by the red dashed line. Around $S/N=10$, the second-order bias-on-bias (blue dashed line) is comparable to the first-order bias, eventually exceeding it and causing a sign-change in the red points at $S/N=5$. This explains the `kink' in the red points at $S/N=10$ in the left-hand panel of Figure~\ref{fig:simbnpests}.

Due to the highly non-linear dependence of the biases on $e_1$ (Figures~\ref{fig:b1np} to \ref{fig:b2np}), the distribution of measured biases becomes highly non-Gaussian, with a mean that departs significantly from the analytic prediction. For very low $S/N$, the MLE can be in the region of $e_1$ parameter space where the biases become very large. These large values of the bias can drag up the mean value significantly, and this non-Gaussianity renders the error bars on these plots unreliable at low $S/N$.


Another way of understanding this behaviour is by recalling that noise bias arises from non-Gaussianities in the likelihood. The extremely non-linear relationship between the biases and the ellipticity induces very large bias-on-bias terms, which explains the behaviour at low $S/N$.

In the right-hand panel of Figure~\ref{fig:simbnp} we plot the measured MLE biases on $e_2$ as a function of $S/N$, for a fiducial model of $(e_1,e_2) = (0.0,0.3)$. At high $S/N$, independent noise realizations give rise to MLEs that exhibit only small fluctuations around the true parameter value. In this regime, the bias is given roughly by the value of curve in the bottom panel of Figure~\ref{fig:b1np} at the true value $e_2 = 0.3$. This is roughly twice the corresponding bias in $e_1$ at the same $S/N$ (middle panel of Figure~\ref{fig:b1np}), so the measured bias at high $S/N$ in the right-hand panel of Figure~\ref{fig:simbnp} is roughly twice as large as the corresponding points in the left-hand panel of Figure~\ref{fig:simbnp}. At low $S/N$ this behaviour is reversed, the bias in $e_1$ now being dominated by the large-$|e_1|$ behaviour seen in the upper panel panel of Figure~\ref{fig:b1np}, such that the bias in $e_2$ is now less than that of $e_1$. 

Furthermore, since neither the first or second-order $e_2$ biases grow very large (Figures~\ref{fig:b1np} and \ref{fig:b2np}), the perturbative expansion of the ML equation is much more accurate than for $e_1$ at low $S/N$, and hence a first-order correction always improves the estimate, reducing the bias by an order of magnitude or so down to $S/N = 15$ and by a factor of a few down to $S/N=5$. A second-order correction does even better, giving more than an order-of-magnitude improvement even at $S/N=5$, with third-order terms expected to be subdominant even here according to Figure~\ref{fig:numinnp}. This point is backed up in the right-hand panel of Figure~\ref{fig:simbnpests}, which clearly demonstrates that the measured biases track their predictions more closely than their $e_1$ counterparts.

The main conclusion of this section is that the MLE noise bias can be reduced by several orders of magnitude for $S/N \gtrsim 30$ with our method, and is reduced for all $S/N \gtrsim 15$. The improvement is better for $e_2$ than $e_1$, due to the more linear relationship of the bias to the ellipticity, and the smaller intrinsic biases caused by the coarse pixellization. Second-order corrections offer further improvement, accounting for much of the bias non-linearity, although these fail too below $S/N \approx 12$ for $e_1$. For $e_2$ second-order corrections can reduce noise bias by almost an order of magnitude even at $S/N \approx 5$. We emphasise that these biases have been reduced without any need for external calibration.


\section{Maximum A Posteriori Estimates}
\label{sec:MAP}

We have seen that the estimation of the MLE bias from a coarsely pixellized noisy galaxy image is made difficult by the extreme behaviour at large values of the ellipticity. However, we know that the galaxies in weak lensing surveys do not exhibit these strong ellipticities, with the r.m.s. ellipticity being roughly 0.3. It may seem strange then that we should allow values of the MLE to take on such extreme values given our prior knowledge of the intrinsic source distribution. It may then be hoped that introducing a prior on intrinsic ellipticity will regularize the bias-estimation procedure as well as more faithfully representing our strong prior expectations on the true noise-free galaxy shape.

With the introduction of a prior, the quantity we now seek to maximize is the \emph{posterior} rather than the likelihood, and the MLE now becomes the MAP. In this section we investigate the performance of the MAP in the problem of shape estimation from noise images.

\subsection{Bias in the MAP}

Introducing a prior to regularize bias-estimation comes at the cost of introducing an extra source of bias into the MLE (now the MAP). This can be straightforwardly accounted for using the formalism of Section~\ref{sec:bias}. Defining the logarithm of the prior probability on $\beta$ as $L_p$, the log-likelihood is now modified as $L \rightarrow L + L_p$. We find the leading-order correction to the first-order noise bias (valid for any noise distribution) to be
\begin{equation}
\label{eq:b1p}
b^{(1)}_p = \frac{L_p'}{n\bar{F}}.
\end{equation}
The second-order intrinsic bias receives an extra contribution given (for Gaussian noise) by
\begin{align}
n^2 \bar{F}^3 b^{(2)}_p &= L_p' \bar{P} + \frac{L_p' \bar{K}^2}{2\bar{F}} + \frac{5L_p''\bar{K}}{6} \nonumber \\ &+\frac{(L_p')^2\bar{K}}{2} + \frac{L_p'\bar{M}}{4} + L_p'L_p''\bar{F} + \frac{L_p'''\bar{F}}{2},
\end{align}
while the bias-on-bias second-order term receives the contribution
\begin{align}
n^2 \bar{F}^3 b^{(1)}[b^{(1)}]_p &= -\frac{3L_p' \bar{P}}{8} + \frac{7L_p' \bar{K}^2}{9\bar{F}} + \frac{5L_p''\bar{K}}{6} \nonumber \\ &+ \frac{2(L_p')^2\bar{K}}{3} + \frac{3L_p'\bar{M}}{8} + L_p'L_p''\bar{F} + \frac{L_p'''\bar{F}}{2}.
\end{align}

We note here that the semi-Bayesian shear measurement method \textsc{Lensfit}~\citep{2007MNRAS.382..315M} also contains a method for removing the effects of prior bias, through a sensitivity correction. Our formalism here is analogous, but is based on a perturbative expansion of the bias of the ellipticity posterior mode rather than a shear measurement derived from a posterior mean.

We choose for the prior the function
\begin{equation}
\label{eq:prior}
p(|e|) \equiv \exp(L_p) \propto (1-|e|^2) \exp{(-|e|^2/2\sigma_e^2)}.
\end{equation}
In this section we fix $\sigma_e = 0.3$, deferring discussion of the sensitivity to the prior until Section~\ref{sec:2Dmresults}. This form of the prior satisfies the requirement of zero weight at the boundary $|e| = 1$.

\begin{figure*}
\centering
\includegraphics[width=0.33\textwidth]{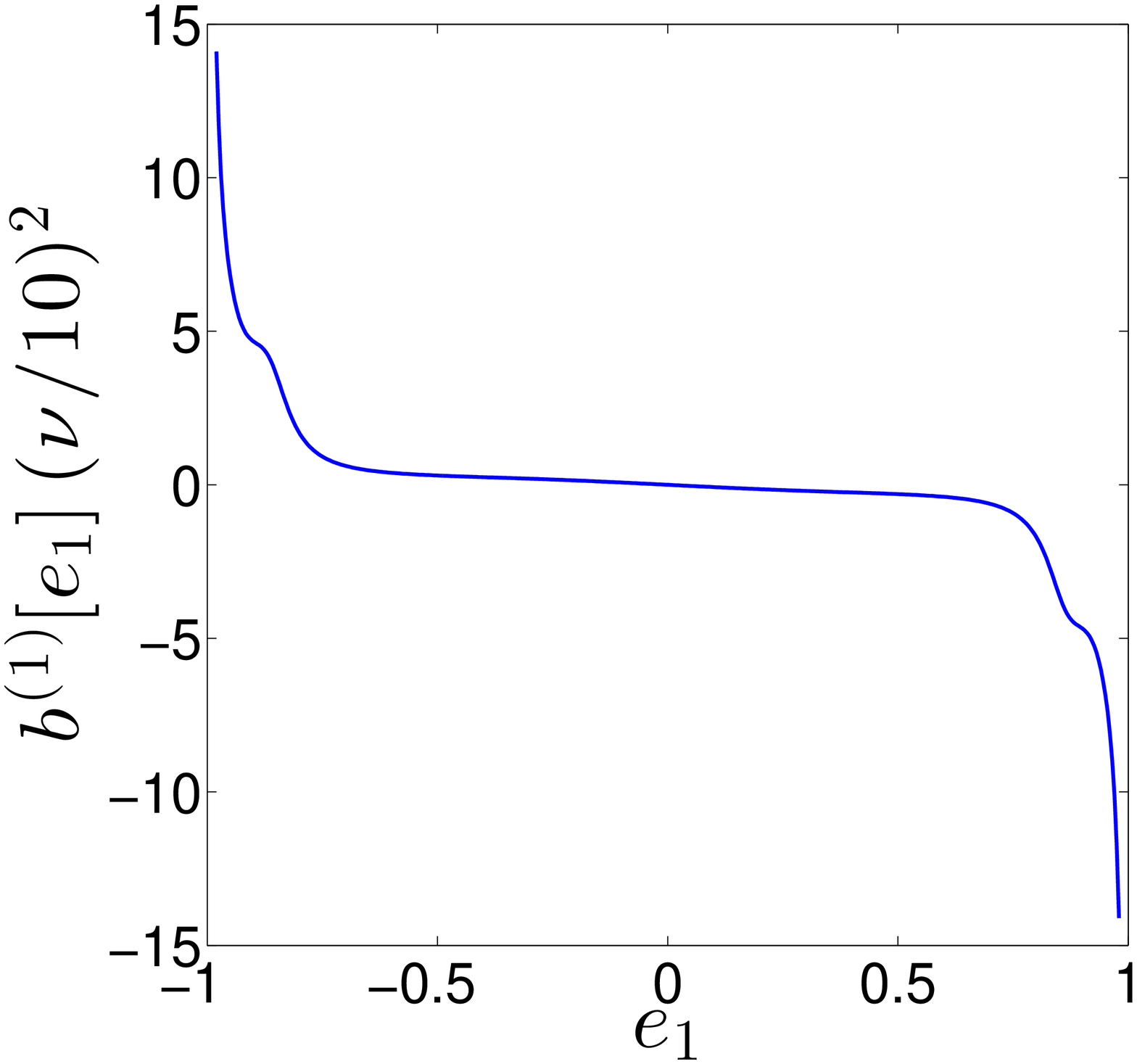}
\includegraphics[width=0.33\textwidth]{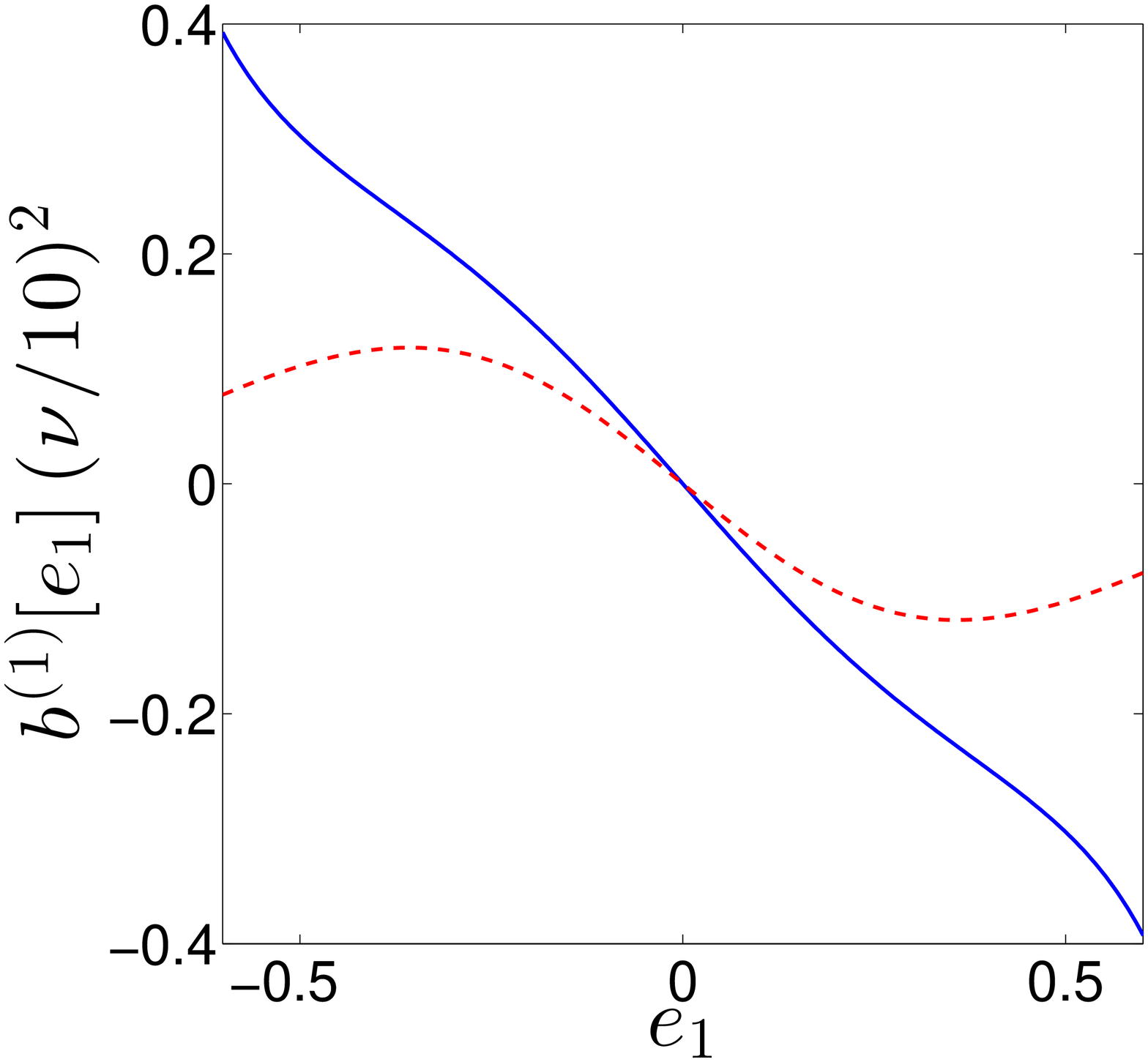}
\includegraphics[width=0.33\textwidth]{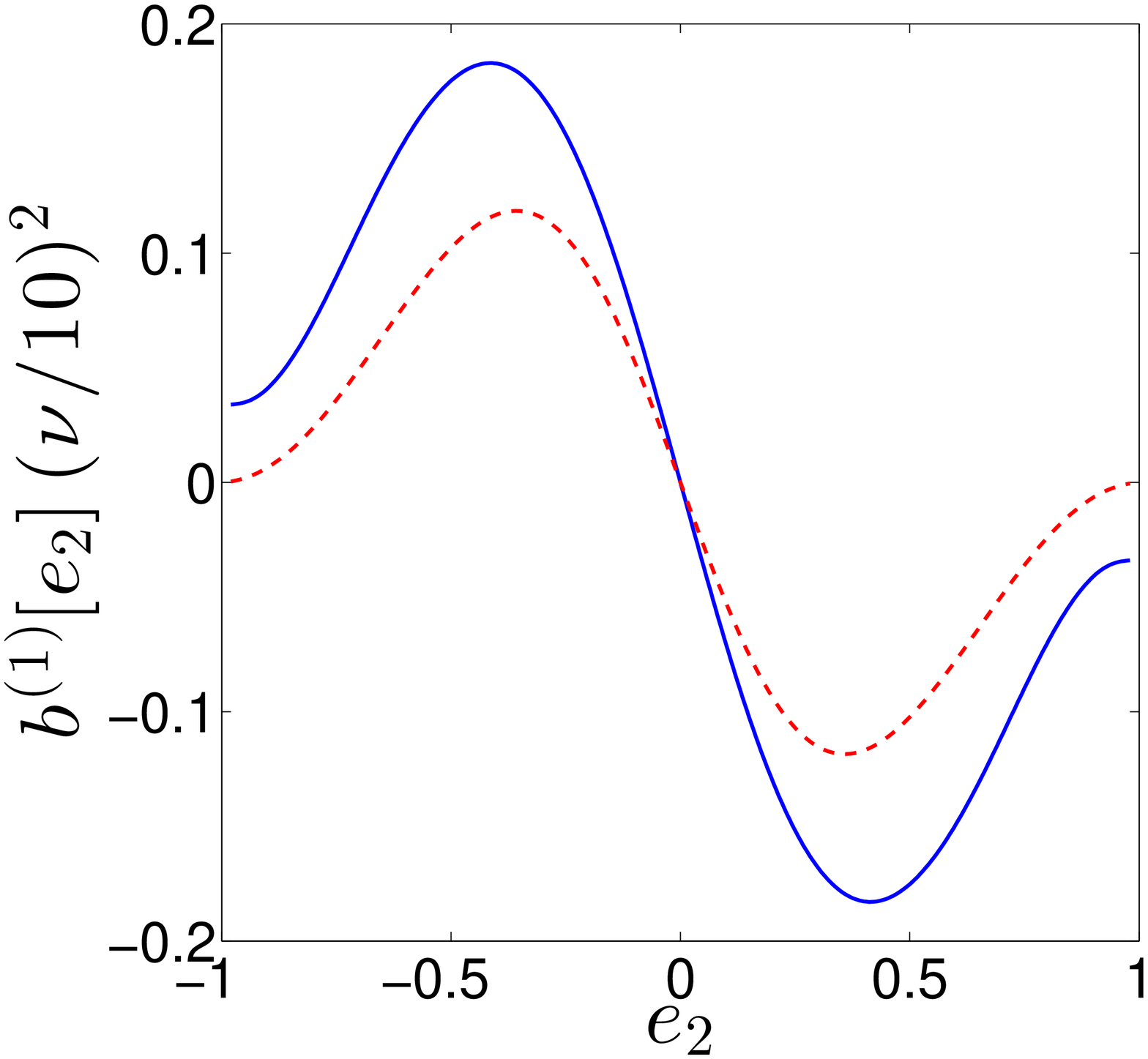}
\caption{\emph{Left panel:} Normalized first-order MAP bias on $e_1$. \emph{Middle panel:} Normalized first-order MAP bias on $e_1$ (blue solid) and its continuum-limit result (red dashed). \emph{Right panel:} Normalized first-order MAP bias on $e_2$ (blue solid) and its continuum-limit result (red dashed).}
\label{fig:b1p}
\end{figure*}

In the left-hand panel of Figure~\ref{fig:b1p} we plot the total (prior + noise) bias on the MAP for $e_1$ as a function of the true value, which should be compared to Figure~\ref{fig:b1np}. In the middle panel of Figure~\ref{fig:b1p} we zoom on the observationally relevant range of small $|e_1|$, which should be compared to Figure~\ref{fig:b1np}. We see from these plots that the prior adds a significant bias to the MLE, and actually diverges at the boundary $|e_1| = 1$. This is partly due to the pre-factor multiplying the Gaussian part of Equation~\eqref{eq:prior}, whose log-derivative diverges, and partly due to the Fisher information in the denominator of Equation~\eqref{eq:b1p} which suffers from the same finite pixellization issues as the MLE. However, since the prior naturally down-weights these extreme ellipticity regions when finding the most likely parameter value, this divergence will not prove problematic.

In the right-hand panel of Figure~\ref{fig:b1p} we plot the total first-order bias on $e_2$. The main effect of the prior here is to increase the bias on all scales, but the shape is rather similar to the prior-free case (Figure~\ref{fig:b1np}). This suggests that the finite-pixellization effect from $\bar{F}$ is more important in determining the high-$|e|$ behaviour of the prior bias than the polynomial pre-factor in Equation~\eqref{eq:b1p}, which gives us some confidence that our results will be broadly insensitive to the precise form of the prior.

\begin{figure*}
\centering
\includegraphics[width=0.33\textwidth]{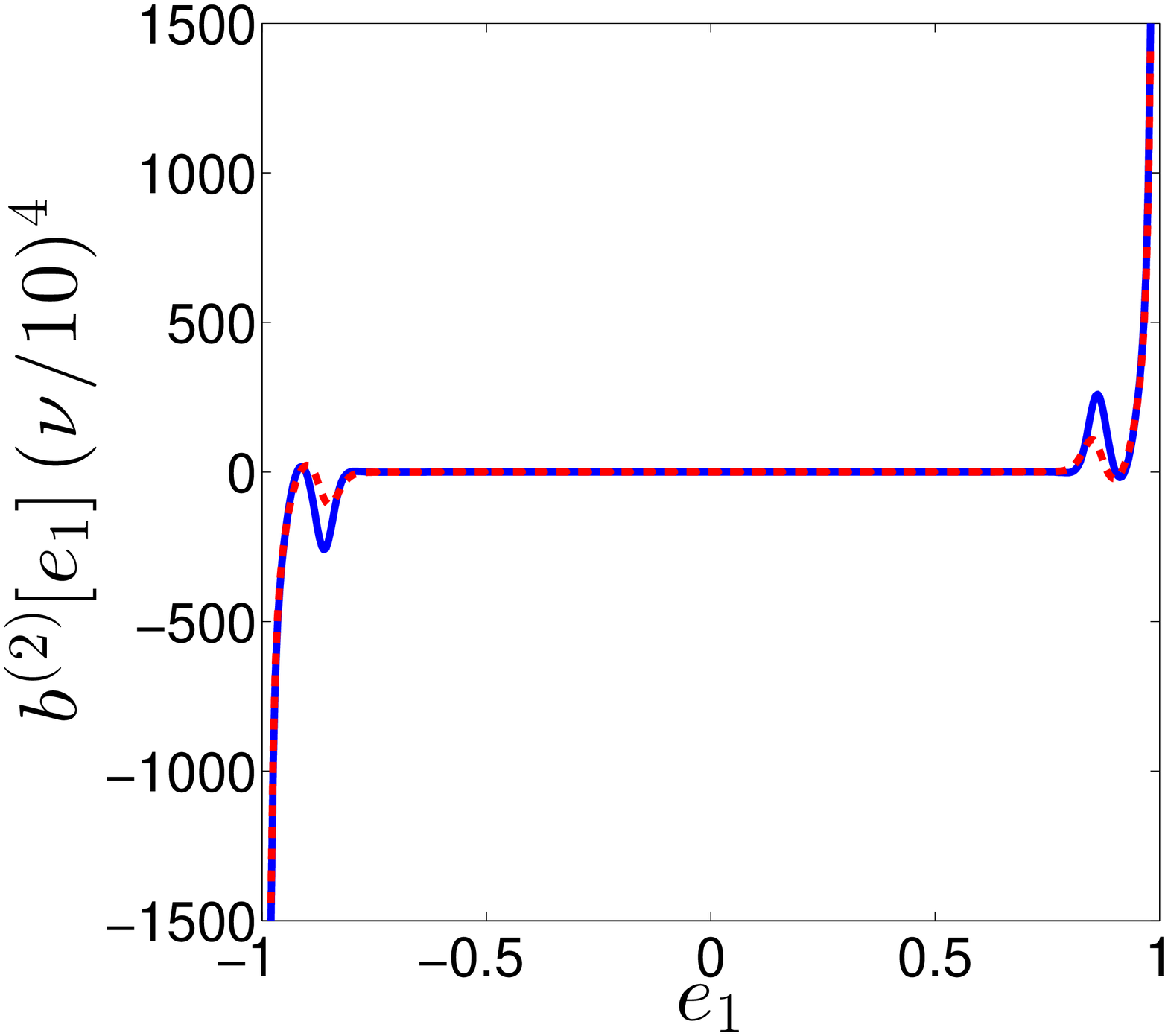}
\includegraphics[width=0.33\textwidth]{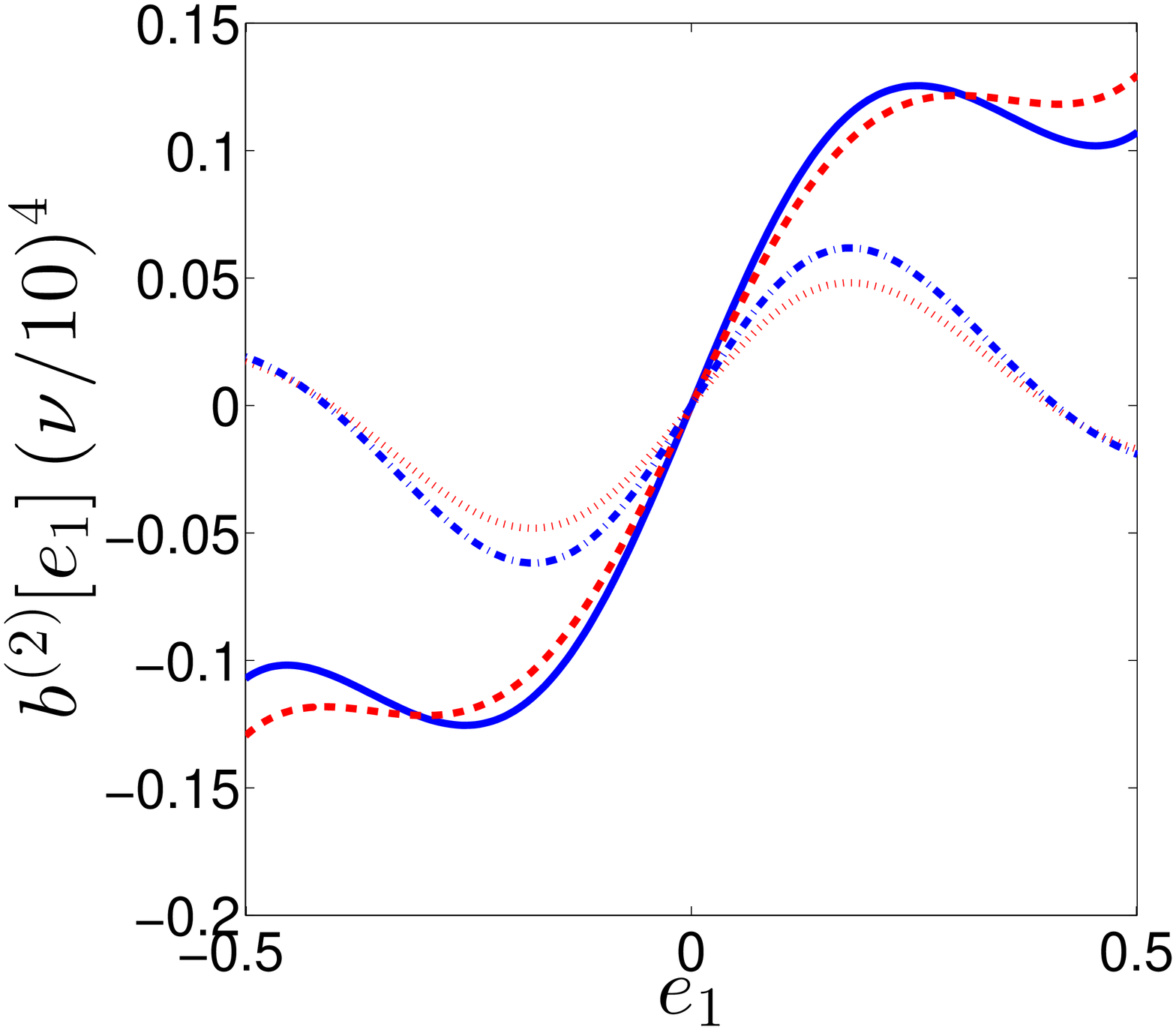}
\includegraphics[width=0.33\textwidth]{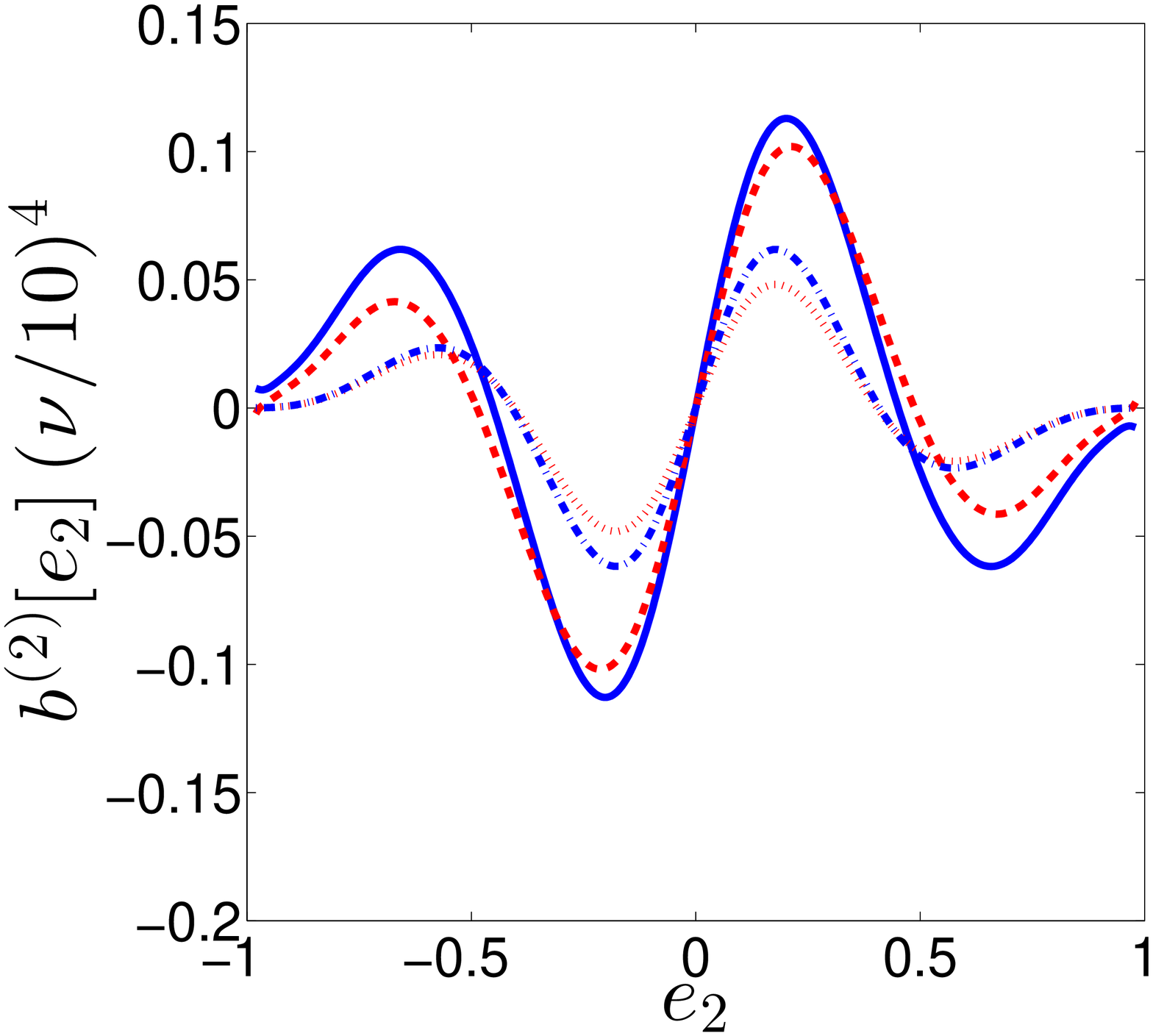}
\caption{\emph{Left panel:} Normalized second-order MAP biases on $e_1$, showing the bias-on-bias term (blue solid) and the intrinsic term (red dashed). \emph{Middle panel:} Normalized second-order MAP biases on $e_1$, showing the bias-on-bias term (blue solid), its continuum-limit result (blue dot-dashed), the intrinsic term (red dashed) and its continuum-limit result (red dotted). \emph{Right panel:} Same as middle panel for $e_2$.}
\label{fig:b2p}
\end{figure*}

In the left-hand panel of Figure~\ref{fig:b2p} we plot the second-order biases induced by the prior on $e_1$, with the middle panel of Figure~\ref{fig:b2p} zooming in on small $|e_1|$. These plots show similar features to their prior-free counterparts, in particular the dramatic increase in bias towards the boundaries of the parameter space. However, a key difference is that the second-order terms are now of much more similar magnitude and have the same sign for all $e_1$. This may be understood by examination of the individual terms in the second-order bias expressions. The dominant terms in both expressions for our choice of prior and galaxy model turn out to be the $L_p'L_p''\bar{F}$ and $5L_p'''\bar{K}/6$ terms, which appear in both second-order bias expressions and hence cancel perfectly. The other important terms have the same sign and similar magnitude, leading to broadly similar values at all $e_1$ and hence a total bias that is much smaller than either individual term. Similar behaviour is observed in the right-hand panel of Figure~\ref{fig:b2p}, where we plot the second-order biases on $e_2$.

\begin{figure*}
\centering
\includegraphics[width=0.33\textwidth]{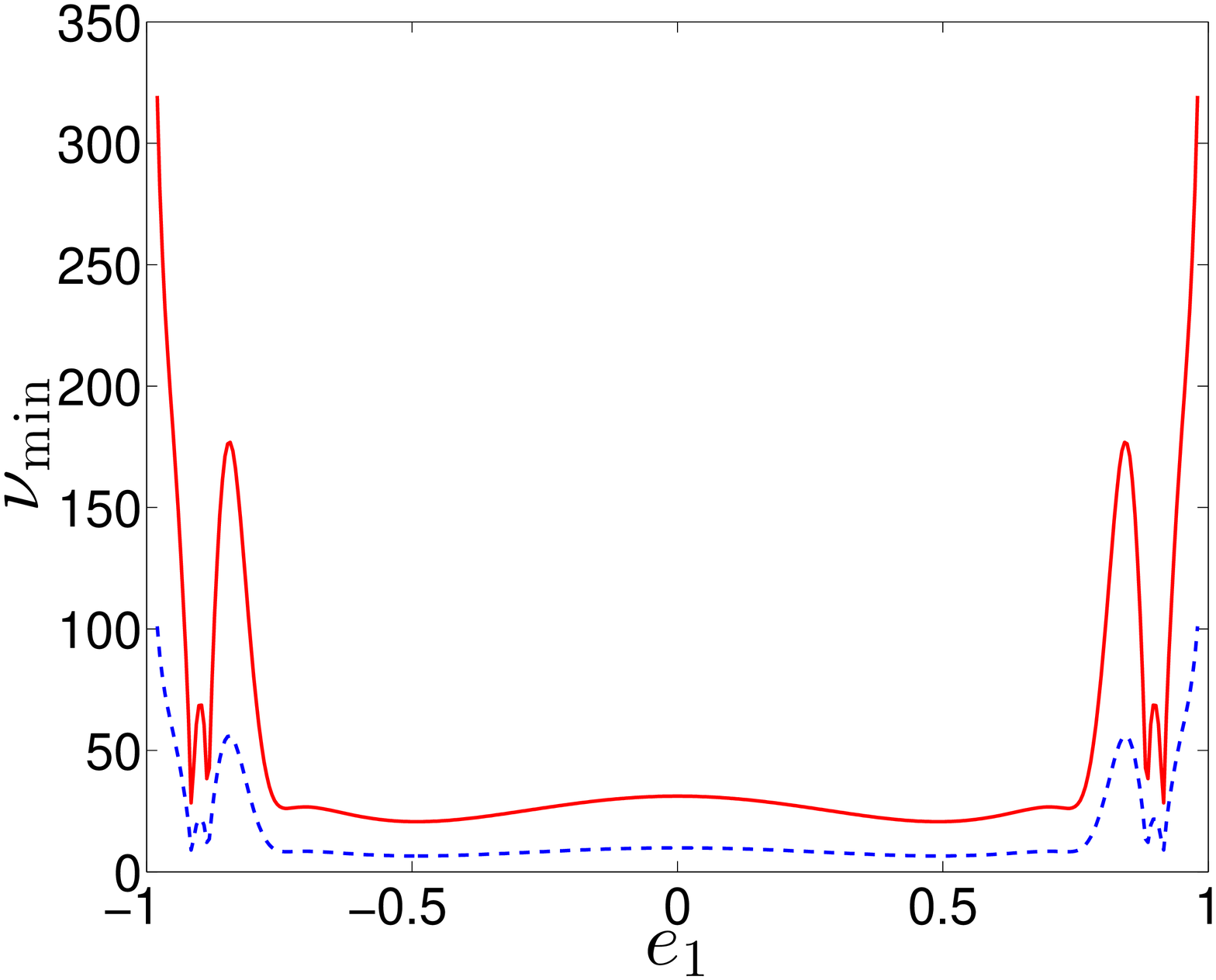}
\includegraphics[width=0.33\textwidth]{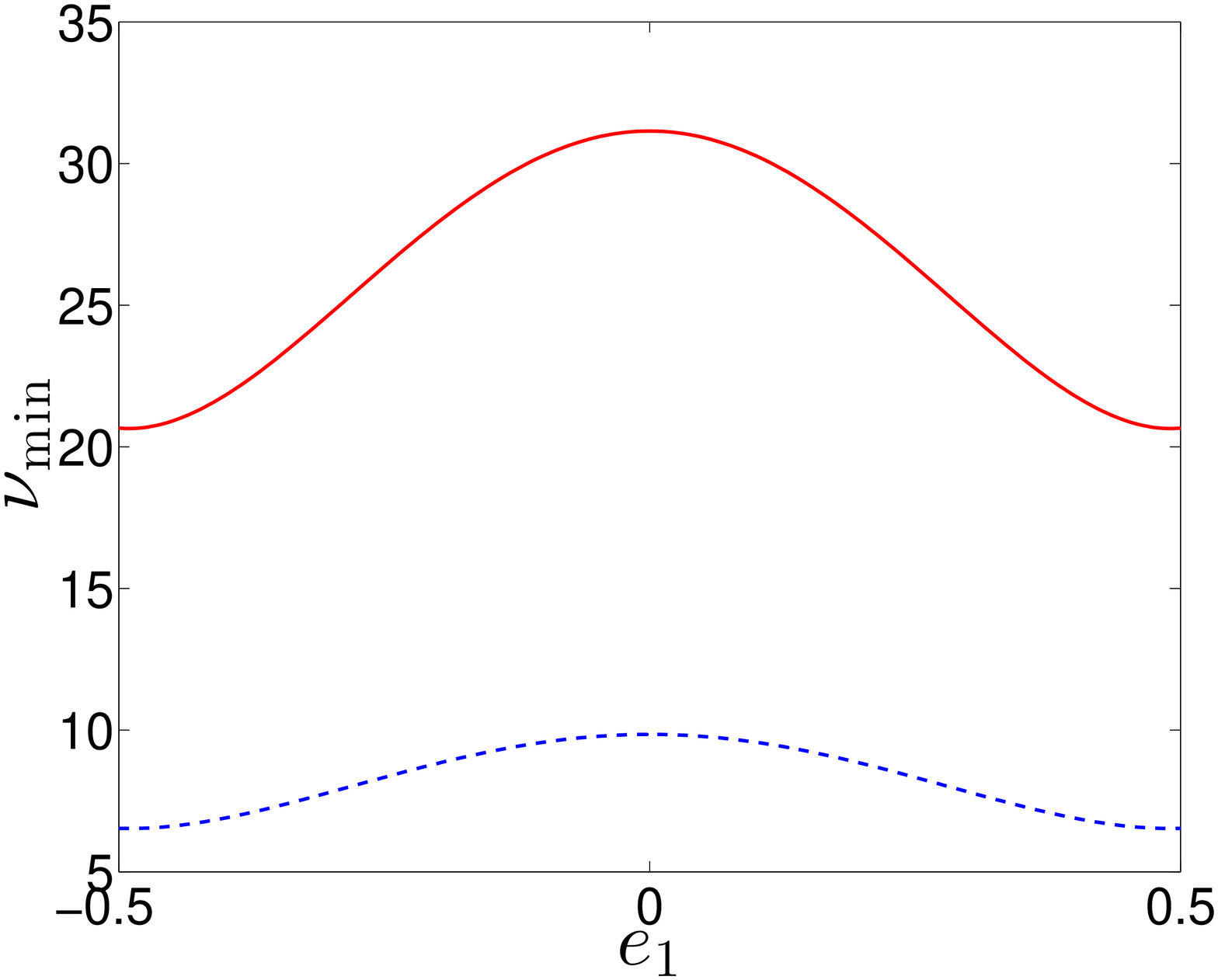}
\includegraphics[width=0.33\textwidth]{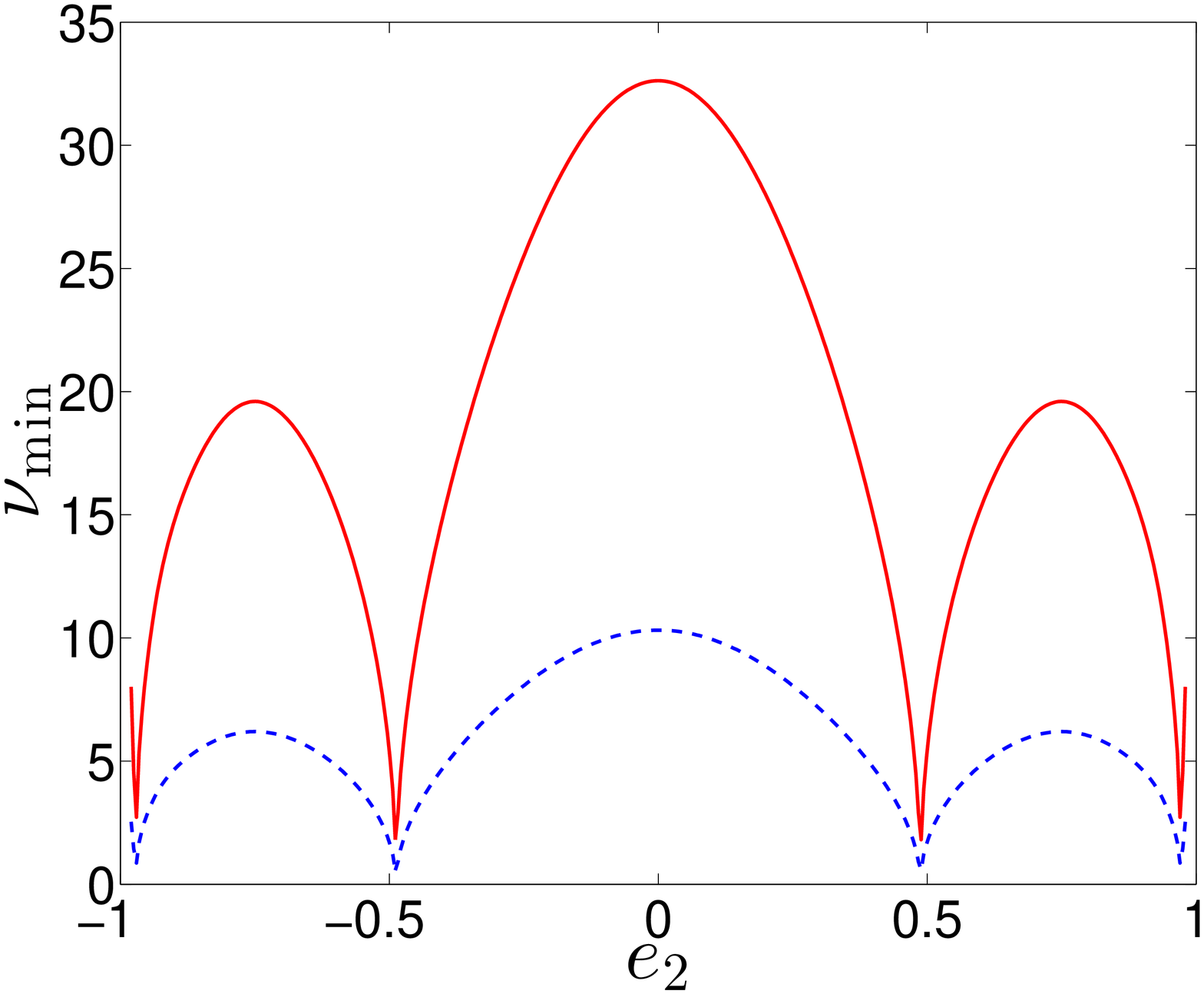}
\caption{\emph{Left panel:} Value of the $S/N$ at which the intrinsic second-order bias on $e_1$ equals the first-order MAP bias as a function of the true parameter value (blue dashed) and the $S/N$ at which it is 10\% of the first-order bias (red solid). \emph{Middle panel:} Same as left panel for small $|e_1|$. \emph{Right panel:} Same as left panel for $e_2$.}
\label{fig:numinp}
\end{figure*}

In the left-hand panel of Figure~\ref{fig:numinp} we plot the minimum $S/N$ that can be used for the $e_1$-MAP with our perturbative bias scheme. The spikes are generally lower in magnitude to their MLE counterpart (Figure~\ref{fig:numinnp}), as the prior bias prevents the first-order term from going through zero. Zooming in on small $e_1$ in the middle panel of Figure~\ref{fig:numinp}, we see that the extra bias induced by the prior has increased $\nu_{\mathrm{min}}$. At the r.m.s. ellipticity of $e_1 = 0.3$ we have $\nu_{\mathrm{min}} \approx 20$. This number is rather similar in the case of the $e_2$-MAP, for which we plot $\nu_{\mathrm{min}}$ in the right-hand panel of Figure~\ref{fig:numinp}.

\subsection{Testing the MAP on image simulations}

With predictions for the extra prior-induced bias now at our disposal, we may proceed with repeating the tests of Section~\ref{sec:eresults} with the MAP as our shape estimator instead of the MLE. 

\begin{figure*}
\centering
\includegraphics[width=0.45\textwidth]{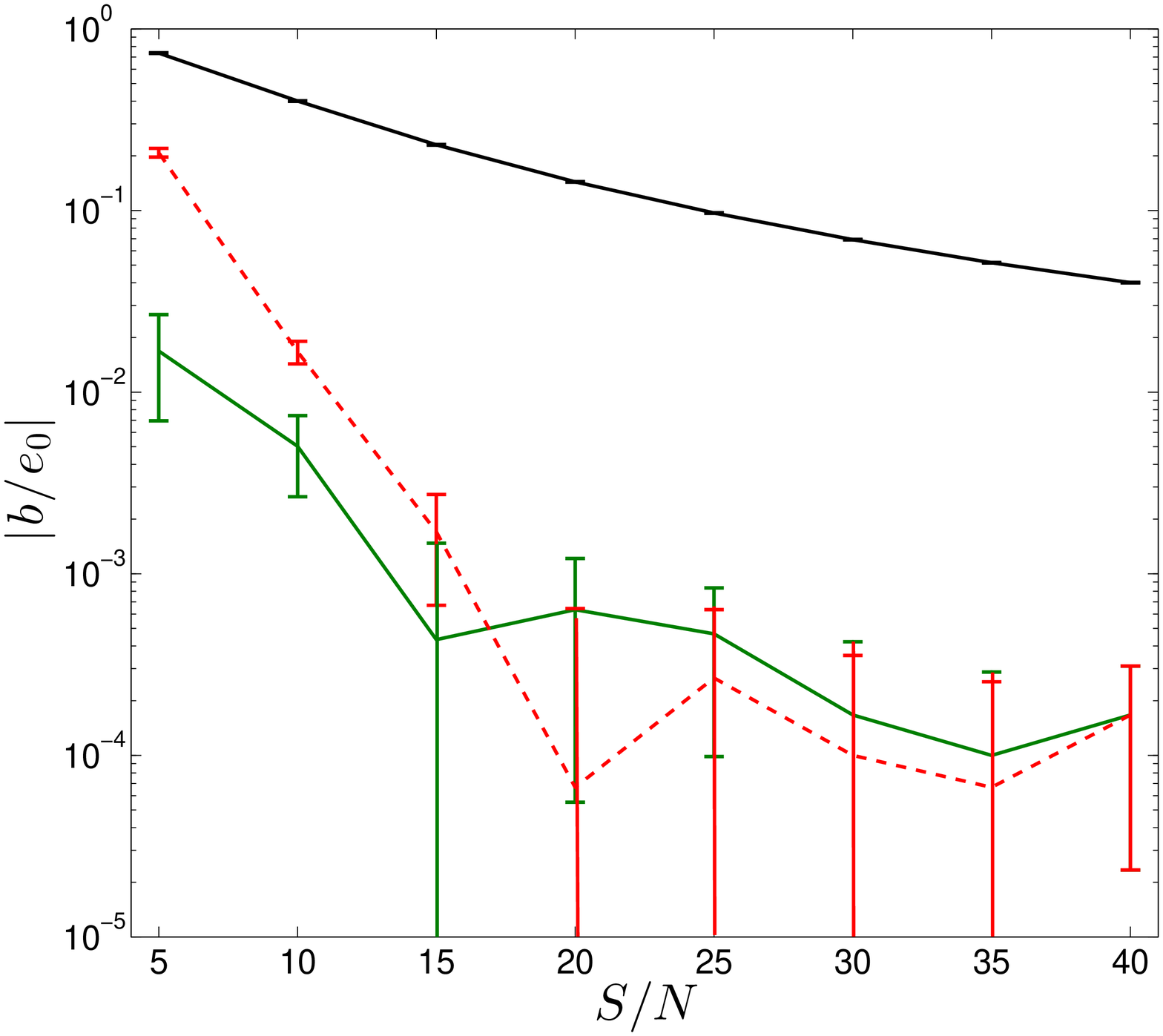}
\includegraphics[width=0.45\textwidth]{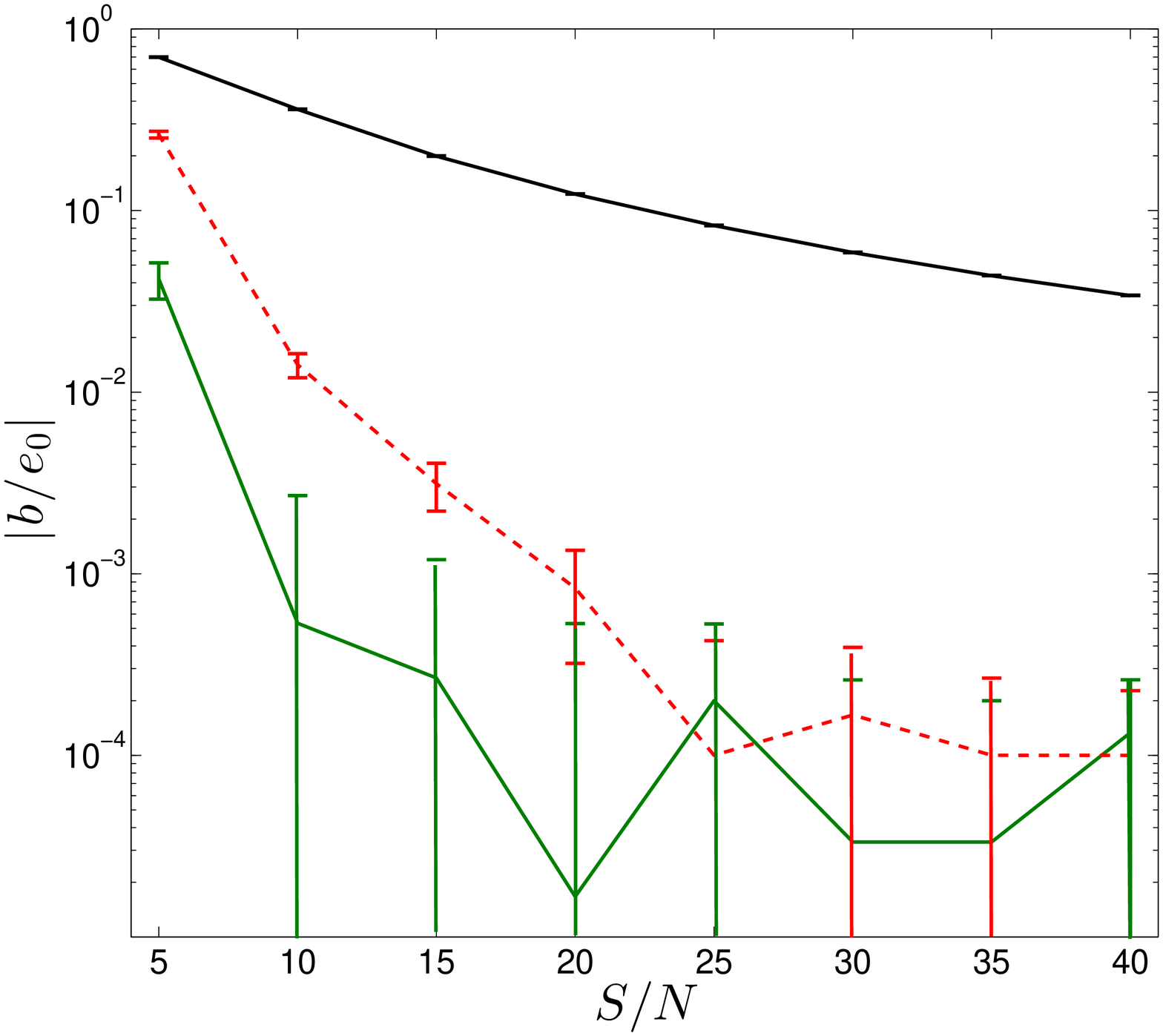}
\caption{\emph{Left panel:} Magnitude of measured bias on the $e_1$ MAP compared to the true value (black solid), the first-order bias-corrected MAP (green solid) and the second-order bias-corrected MAP (red dashed). \emph{Right panel:} Same as left panel for $e_2$.}
\label{fig:simbp}
\end{figure*}

In the left-hand panel of Figure~\ref{fig:simbp} we plot the measured total bias (noise + prior) on the MAP from image simulations. This should be compared to Figure~\ref{fig:simbnp} for the case of the MLE. The black points in this figure show the bias in the MAP as measured from the image simulations, which is now uniformly larger than that of the MLE due to additional bias brought by the prior. However, both the first-order and second-order bias-corrected estimators succeed in removing most of this extra bias, with the resultant residual biases smaller than their MLE counterparts. Note that the error bars in this plot are too large to allow for a comparison between the first and second-order estimators.

The major difference between Figure~\ref{fig:simbp} and Figure~\ref{fig:simbnp} however is the performance at low $S/N$. In this regime the MLE suffers from large biases due to the large ellipticities that are favoured in chance realizations of the noise. When a prior is included these regions of parameter space are downweighted, regularizing the perturbative series and `linearizing' the bias, such that even at $S/N=10$ a first-order correction reduces the bias by roughly two orders of magnitude from $\mathcal{O}(10^{-1})$ to $\mathcal{O}(10^{-3})$. The same correction applied to the MLE actually worsens the bias. This demonstrates that although the prior brings an extra source of bias, its regularizing effects on our perturbative scheme vastly outweigh this. Across the entire $S/N$ range in Figure~\ref{fig:simbp}, the bias-corrected MAP reduces the bias by at least two orders of magnitude for all $S/N$.

\begin{figure*}
\centering
\includegraphics[width=0.45\textwidth]{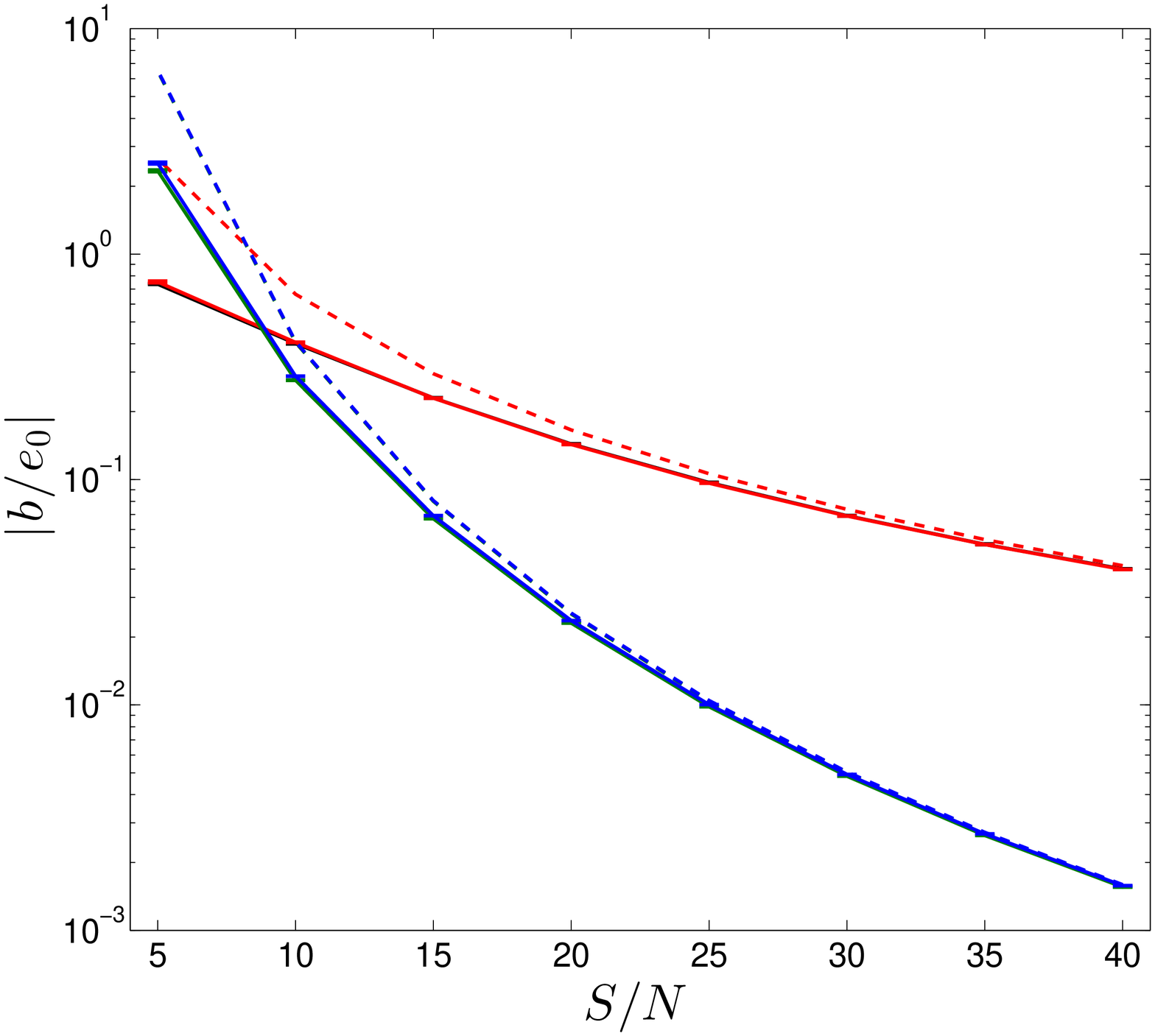}
\includegraphics[width=0.45\textwidth]{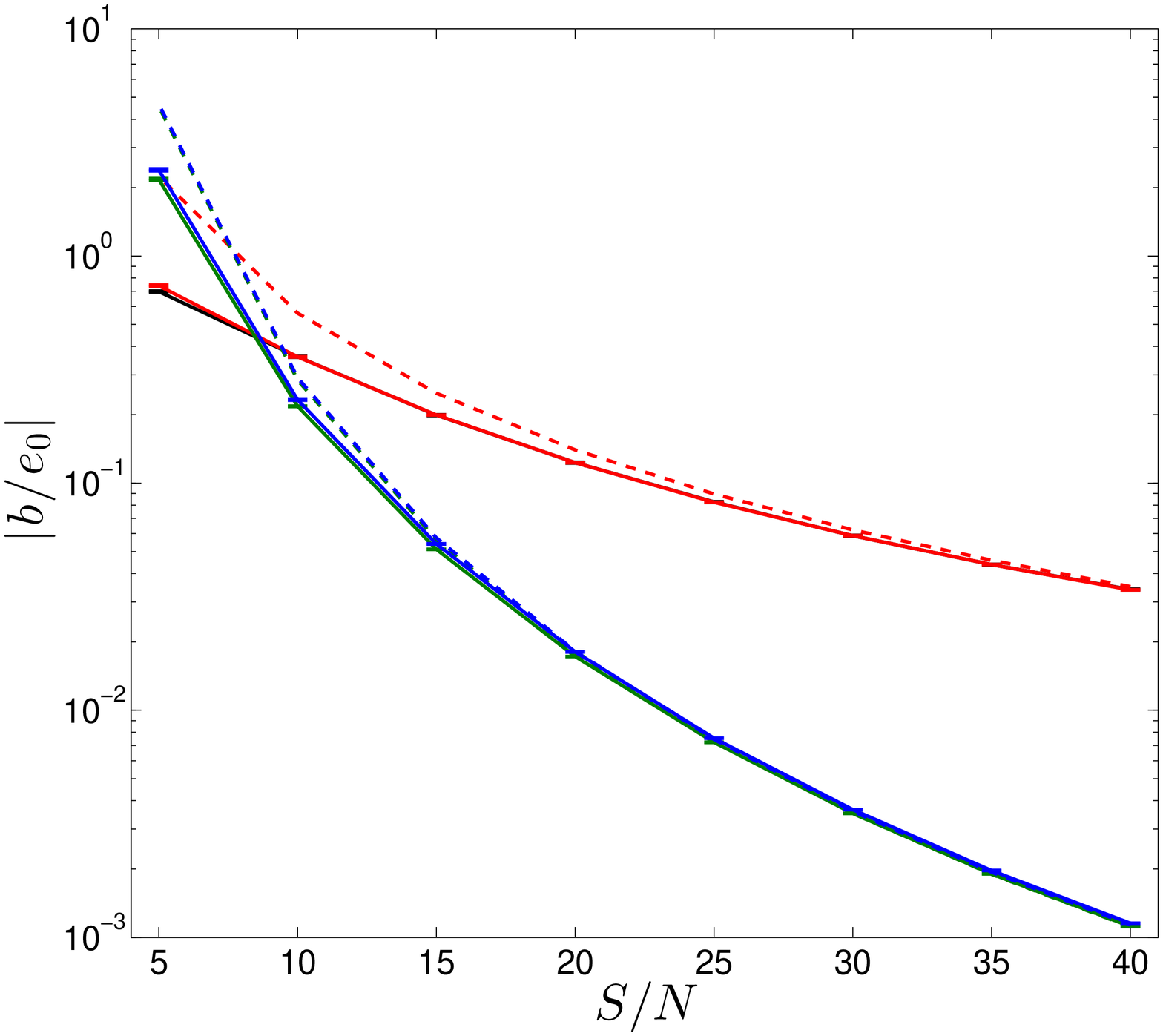}
\caption{\emph{Left panel:} Magnitude of MAP bias compared to true value of $e_1$ (black solid), mean value of measured first-order bias (red solid), noise-free first-order bias prediction (red dashed), mean value of measured intrinsic second-order bias (green solid), noise-free intrinsic second-order bias (green dashed), mean value of measured bias-on-bias (blue solid), noise-free bias-on-bias (blue dashed). Black and red points lie on top of each in this plots, as do blue and green points. \emph{Right panel:} Same as left panel for $e_2$. Black and red points lie on top of each other, as do blue and green points.}
\label{fig:simbpests}
\end{figure*}

This argument is backed up by inspection of the individual biases as estimated from the image simulations, which are displayed in the left-hand panel of Figure~\ref{fig:simbpests}, which should be compared with Figure~\ref{fig:simbnpests} in the MLE case. The solid lines (measured biases) now track the dashed lines (exact biases) much more closely across the $S/N$ range, indicating that the prior is successfully regularizing the perturbative ML expansion. Note that the green and blue lines lie on top of each other in this plot, which is due to the very similar magnitudes of the second-order biases as discussed above. The total second-order bias is the difference in magnitude of these terms, such that the total second-order bias is much less significant in the case of a prior than without. This is exhibited in the close tracking of the first-order bias (red points) by the total measured bias (black points) in Figure~\ref{fig:simbpests}, where the lines are almost indistinguishable. This suggests that second-order effects are subdominant in the presence of a prior, due to both the regularization of the ML expansion and the cancellation between intrinsic and bias-on-bias terms. This cancellation is potentially a consequence of our choice of Gaussian noise, and possibly also due to our Gaussian galaxy model assumption, although further investigation of this would require explicit checking against alternative models. However, our results definitively show that the regularizing effect of the prior brings bias measurements into considerably better agreement with their exact expressions.


We have seen that, unlike $e_1$, the leading-order biases in $e_2$ do not grow large at large $|e|$. This suggests that switching from the MLE to the MAP will not have as much effect on the $e_2$ bias, which is backed up in the right-hand panel of Figure~\ref{fig:simbp} where we plot the measured biases on the $e_2$ MAP. The improvement in the estimator bias is not as dramatic in the first-order-corrected MAP as it is for the first-order-corrected MLE, although there is still an improvement for all $S/N$. Furthermore, the second-order-corrected MAP actually performs worse than the first-order-corrected MAP at $S/N \lesssim 15$, in contrast to the MLE. We attribute this to the fact that the ratio of second-order to first-order bias is larger for $e_2$ with a prior than without (compare Figure~\ref{fig:numinnp} and Figure~\ref{fig:numinp}). This suggests that the perturbative expansion of the ML equation is likely to break down at a higher $S/N$ with a prior than without.


In the right-hand panel of Figure~\ref{fig:simbpests} we plot the individually measured biases on the $e_2$ MAP. We again see that the bias-on-bias and intrinsic second-order biases are of very similar magnitude, suggesting that the total second-order bias is reduced compared to the naive expectation. Of course, this does not affect the argument above that the a first-order correction is not as effective for the $e_2$ MAP, which is based on the ML expansion and hence only on the size of the intrinsic second-order bias. 


The results of this section strongly suggest that introducing a prior into the MLE can reduce its noise bias by orders of magnitude even at low $S/N$. Even though the prior itself is a source of additional bias, this too can be easily accounted for in our formalism. The new estimator, the MAP, is more robust to second-order corrections than the MLE and little is gained by the inclusion of such terms, although they do offer a very useful test of the convergence of the ML equation to its asymptotic form. Although our tests have thus far been on simple galaxy models with Gaussian noise, the formalism we have developed could be extended to more complicated likelihood functions. Correlated noise could in principle also be easily included, although sufficiently many pixels would need to sample the bright regions of the galaxy to ensure the CLT converges efficiently. We defer investigation of this issue to a future work, and instead now apply our formalism to biases on the measured shear field itself.

\section{One-dimensional shear biases}
\label{sec:1Dmresults}

The previous sections have given us an insight into the various factors determining noise bias in galaxy images. However, the relevant observable for gravitational lensing is not the galaxy ellipticity but the shear that acts upon it. In this section we investigate the effects of noise bias on cosmic shear estimates. In this section we will specialise to one-dimension, to facilitate examination of the propagation of the noise bias from the one-dimensional ellipticity parameter spaces discussed in the previous section. We will also largely neglect second-order biases in this section, as the image simulations have shown that they improve biases only modestly over a first-order correction.

Under our definition of ellipticity, the unlensed complex ellipticity $e_s$ is related to the lensed ellipticity $e$ and the shear $g$ by~\citep{1995A&A...294..411S}
\begin{equation}
e = \frac{e_s + 2g + g^2e_s^*}{1+|g|^2 + 2\mathrm{Re}(g e_s^*)}.
\end{equation}
Restricting to weak shear $g \ll 1$ and aligned shear and ellipticity such that $e = \mathrm{Re}(e) = e_1$ and $g = \mathrm{Re}(g) = g_1$, we have
\begin{equation}
e \approx e_s + 2g(1 - e_s^2),
\end{equation}
and hence
\begin{equation}
\label{eq:g}
g = \frac{e - e_s}{2(1-e_s^2)}.
\end{equation}
Therefore, given a noisy estimate of the lensed ellipticity $\hat{e}$, and the true unlensed ellipticity, a natural estimator of shear is given by
\begin{equation}
\hat{g} = \frac{\hat{e} - e_s}{2(1-e_s^2)}.
\end{equation}
It may be wondered whether one could attempt to estimate $g$ directly from an individual galaxy image and hence make use of the bias-correction schemes developed in the preceding sections. Unfortunately, even with a prior on intrinsic ellipticity and the restriction of $|g|<1$~\citep{1995A&A...294..411S} such an estimate would still have infinite noise, i.e. there would be a direction in $(g_1,g_2)$ space such that the posterior probability was constant. To see this, note that an observation of galaxy shape produces two observables ($e_1$ and $e_2$), whereas there are four unknowns (both components of $e_s$ and $g$). A prior on $|e_s|$ reduces this to three unknowns, and so the problem is still unconstrained. Our perturbative scheme will therefore not succeed when applied to such a shear estimator, so we specialise here to the more common approach of averaging over galaxies in a coherence patch of the shear.

%
%
%
%
Since the intrinsic ellipticity is unobservable, we average ellipticity estimates over sources to estimate shear. Equation~\eqref{eq:g} suggests the following form:
\begin{equation}
\label{eq:ghat}
\hat{g} = \frac{\langle \hat{e} \rangle_{\mathrm{S}}}{2(1-\sigma_e^2)},
\end{equation}
where $\sigma_e^2$ is the variance of the intrinsic ellipticity distribution and we have averaged over sources. We have also assumed that $e_s$ has zero mean. The individual ellipticity estimates are biased by noise, which can be written schematically as
\begin{equation}
\hat{e} \approx e + b^{(1)}(e,r^2,S_T)X^2
\end{equation}
where $X$ is a Gaussian random variable with zero mean and unit variance, and we have neglected higher-order noise bias terms. We have made the dependence of the bias on the other galaxy model parameters explicit in the above expression for reasons which will become clear. To average this expression over sources, we need to substitute $g$ and $e_s$ for $e$:
\begin{equation}
\langle \hat{e} \rangle_{\mathrm{S}} = 2 g (1-\sigma_e^2) + \langle b^{(1)}[\mathbf{\Theta}(\mathbf{\Theta}_s,g)]\rangle_{\mathrm{S}} \langle X^2 \rangle_{\mathrm{S}}, 
\end{equation}
where we have assumed that the noise is uncorrelated between galaxies, and $\mathbf{\Theta}$ represents the lensed galaxy model parameters, with $\mathbf{\Theta}_s$ their unlensed values. Now, since we have assumed that the ellipticity biases are separable in $e$ and $X$ (which is true at each order in signal-to-noise), averaging over sources and averaging over noise are commutative operations. So, the shear estimate averaged over noise is
\begin{equation}
\langle \hat{g} \rangle_n = g + \frac{\langle b^{(1)}(\mathbf{\Theta}_s,g)\rangle_{\mathrm{S}}}{2(1-\sigma_e^2)}.
\end{equation}
Finally, we linearize the bias functions around zero shear:
\begin{align}
b[\mathbf{\Theta}(\mathbf{\Theta}_s,g)] &\approx b[\mathbf{\Theta}(\mathbf{\Theta}_s,0)] + g \frac{\partial b[\mathbf{\Theta}(\mathbf{\Theta}_s,g)]}{\partial g} \biggr\rvert_{g=0} \nonumber \\
         &= b(\mathbf{\Theta}_s) + g \frac{\partial b(\mathbf{\Theta})}{\partial \Theta_i}\biggr\rvert_{\mathbf{\Theta} = \mathbf{\Theta}_s} \frac{\partial \Theta_i}{\partial g}\biggr\rvert_{g=0},
\end{align}
%
%
%
where we have implicitly summed over the model parameters $\Theta_i$. Defining the shear bias parameters by $\langle \hat{g} \rangle = (1+m)g + c$ we have
\begin{align}
m &= \frac{1}{2}(1-\sigma_e^2)^{-1}\left\langle \frac{\partial b(\mathbf{\Theta})}{\partial \Theta_i} \frac{\partial \Theta_i}{\partial g} \right\rangle_{\mathrm{S}} \nonumber \\
c &= \frac{1}{2}(1-\sigma_e^2)^{-1}\langle b(\mathbf{\Theta}_s) \rangle_{\mathrm{S}} = 0,
\label{eq:1Dm}
\end{align}
where the final equality follows since both first and second order bias functions are antisymmetric in the unlensed model parameters. This expression can be evaluated using the known relationship of the lensed model parameters to the shear, and by differentiating the biases with respect to the model parameters then averaging the product over one component of the ellipticity. For the intrinsic ellipticity distribution we take the same function used as the ellipticity prior, Equation~\eqref{eq:prior}, although these two functions need not be the same in practice.

Equation~\eqref{eq:1Dm} demonstrates that it is the derivatives of the bias functions with respect to model parameters that determine the shear biases, rather than just the values of the functions themselves. It also shows that noise bias only contributes to the multiplicative bias and not the additive bias, which is a consequence of isotropy, i.e. the fact that the intrinsic distribution depends only on the magnitude of ellipticity and not its direction. We note however that non-circular PSF could provide a source of anisotropy and generate a non-zero $c$-term~\citep{2012MNRAS.427.2711K}.

\begin{figure}
\centering
\includegraphics[width=\columnwidth]{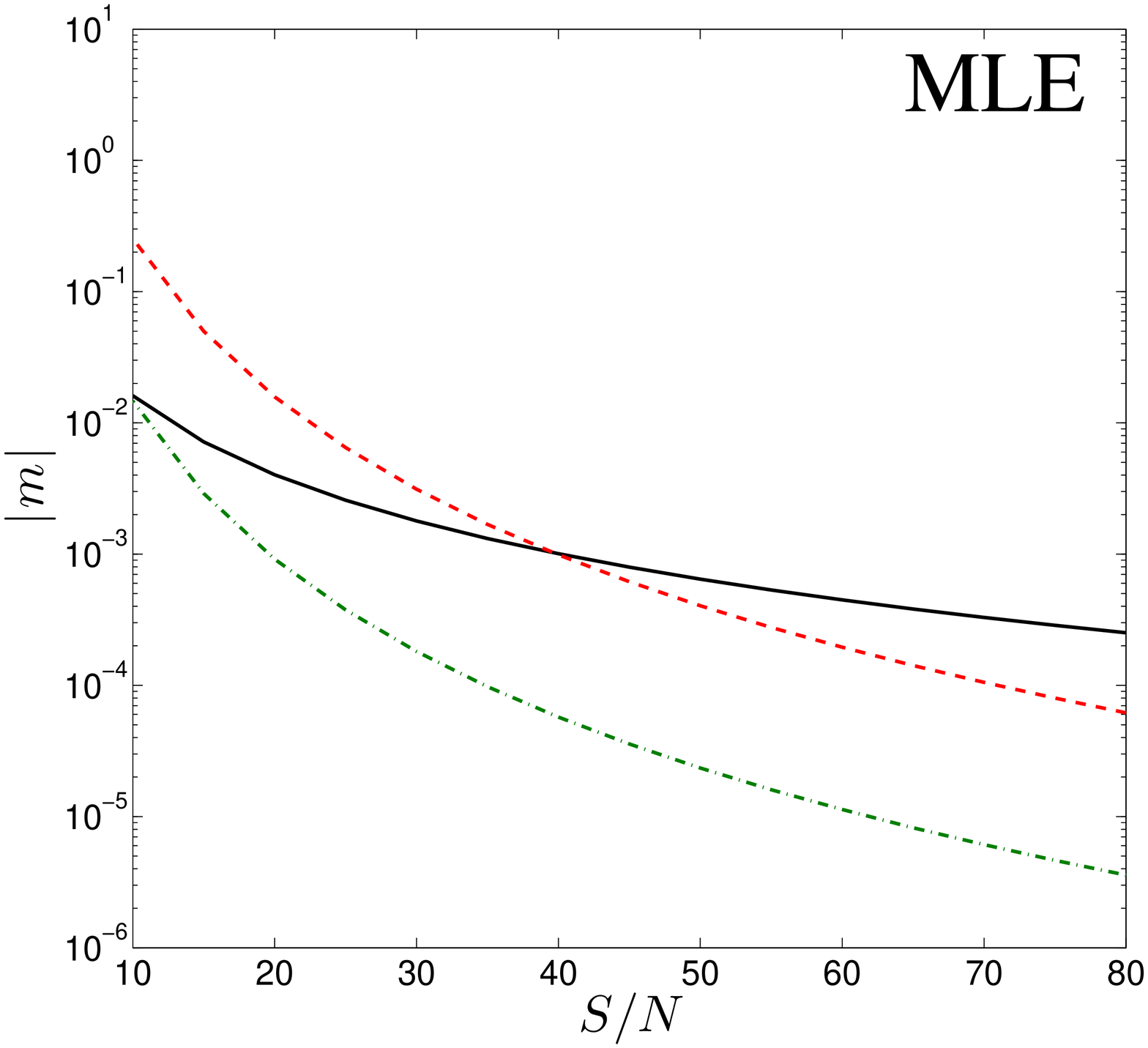}
\includegraphics[width=\columnwidth]{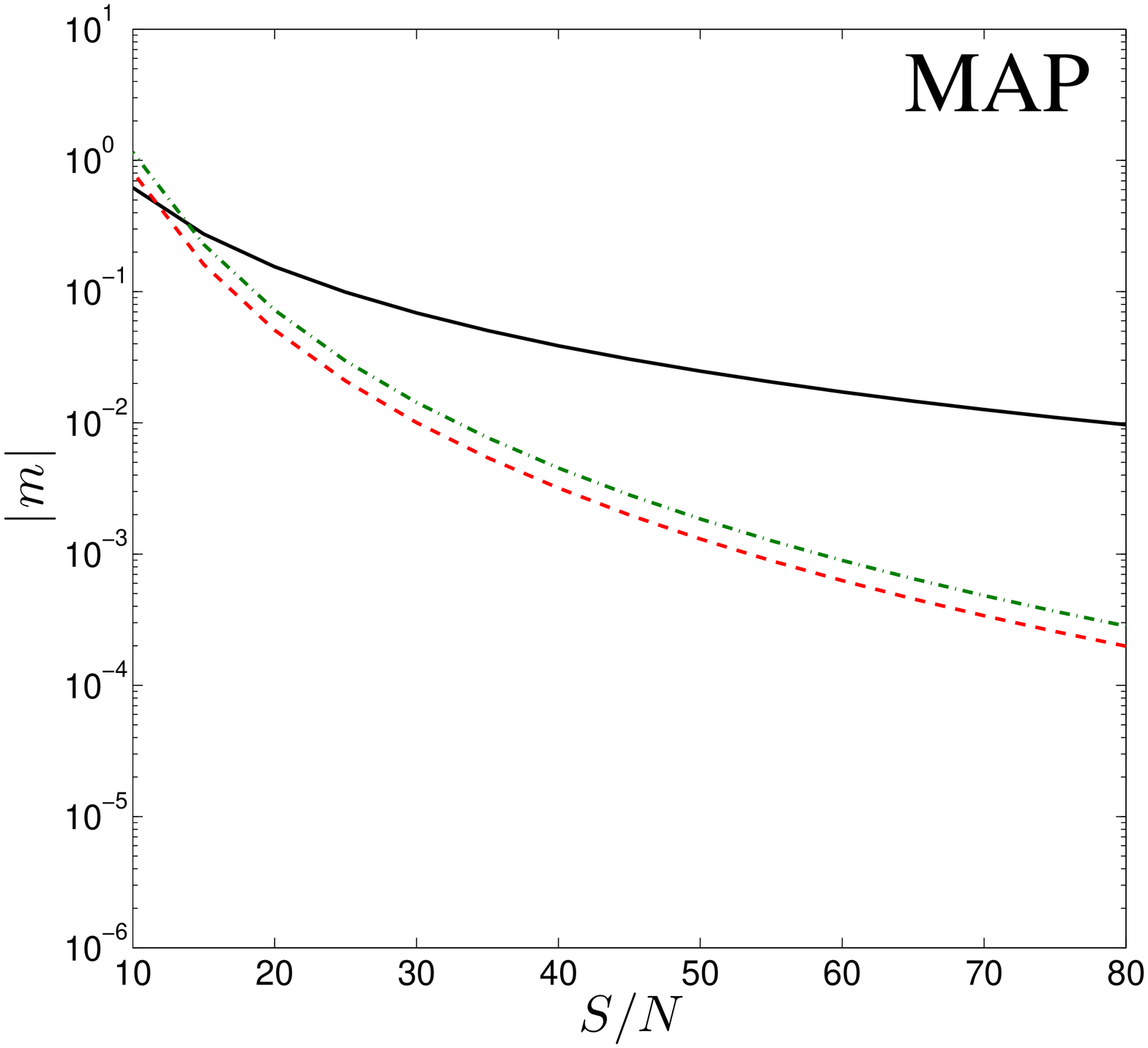}
\caption{\emph{Upper panel:} Theoretical absolute 1D shear multiplicative MLE biases from $e_1$, derived from first-order ellipticity bias (black solid), the second order bias-on-bias term (green dot-dashed), and the intrinsic second-order bias (red dashed). Note that the first-order bias is negative, the intrinsic second-order bias is positive, and the bias-on-bias is negative. \emph{Lower panel:} Same as upper panel for the MAP. The second-order biases now have the same sign.}
\label{fig:1Dm}
\end{figure}

In the upper panel of Figure~\ref{fig:1Dm} we plot the true 1D multiplicative first and second-order shear biases assuming shear has been measured from a population with intrinsic ellipticities given by Equation~\eqref{eq:prior} using only the $e_1$ component of measured ellipticity, assuming that $e_2 = 0$. This plot shows that the intrinsic second-order bias is subdominant to the first-order term only for $S/N \gtrsim 40$. For noisier images, the large $e_1$ ellipticity biases contribute to the average in Equation~\eqref{eq:1Dm}, even after downweighting by the intrinsic distribution of Equation~\eqref{eq:prior}. At these low $S/N$ values the perturbative expansion of the ML equation will start to break down, and so we predict that these bias predictions will not accurately describe the bias measured from image simulations.

The shear multiplicative biases $m$ and $c$ are useful quantities for us to study because weak lensing surveys usually couch the performance of their shear estimators in terms of these numbers. Planned space-based surveys such as the \emph{Euclid} mission will require the total multiplicative shear bias to be no greater than $2 \times 10^{-3}$ in order to gain percent-level accuracy on the dark energy equation of state~\citep{2011arXiv1110.3193L, 2013MNRAS.429..661M}. From Figure~\ref{fig:1Dm} we see that in the regime where our perturbative bias predictions are expected to be accurate ($S/N \gtrsim 40$), the multiplicative bias from noise is $\mathcal{O}(10^{-4})$, so we will have to investigate through image simulations whether our bias-correction schemes might be useful for our fiducial space-based survey at lower $S/N$.

In the lower panel of Figure~\ref{fig:1Dm} we plot the first and second-order bias predictions for the MAP $e_1$ estimator, averaged over sources. Similarly to the ellipticity biases, we see that the biases are all uniformly larger across the $S/N$ range, due to the extra bias induced by the prior. We also see that the individual second-order terms are of similar magnitude, and have the same sign, a consequence of the similar forms of the MAP ellipticity biases. In contrast to the upper panel of Figure~\ref{fig:1Dm}, the second-order terms are subdominant to the first-order bias for almost all $S/N \gtrsim 10$, suggesting that the MAP-based shear estimator might be more amenable to our perturbative bias correction than its MLE counterpart.


Using our image simulations, we can form the shear estimate Equation~\eqref{eq:ghat}, and measure the resulting bias. We apply a uniform weak shear of $g=0.01$ and simulate galaxies with $e_2 = 0.0$ and $e_1$ sampled from the intrinsic distribution Equation~\eqref{eq:prior}. It is straightforward to show that corrections to $m$ from higher-order terms in $g$ enter at $\mathcal{O}(g^2)$, which can also be seen through symmetry considerations~\citep{2015JCAP...01..022M}.

\begin{figure}
\centering
\includegraphics[width=\columnwidth]{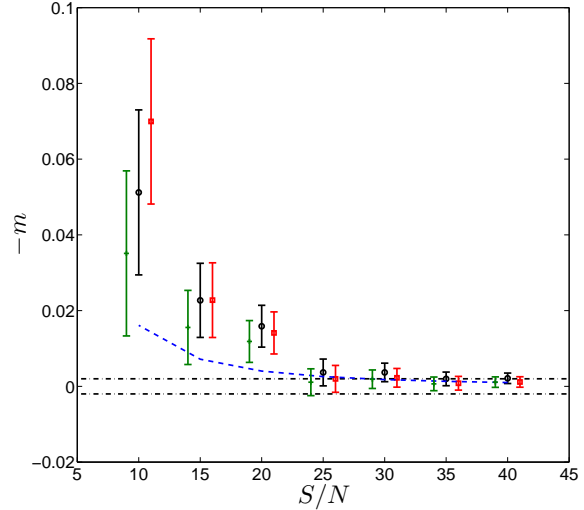}
\caption{Uncorrected shear MLE bias from $e_1$ (black circles), first-order bias prediction (blue dashed), first-order per-galaxy corrected residual bias (red squares), globally-corrected residual bias (green crosses), and the total requirement for our fiducial space-based lensing survey (black dot-dashed). Points have been artificially offset in the horizontal direction for clarity.}
\label{fig:sim1DMLE}
\end{figure}

After finding the MLE (or MAP) for each galaxy, we may apply a first-order (or second-order) correction to that estimate and then proceed to average this `per-galaxy corrected' estimator. The resulting residual bias for the MLE is shown as the red points in Figure~\ref{fig:sim1DMLE}, with the blue points showing the uncorrected bias. Alternatively, if we know the intrinsic ellipticity distribution, we may perform a `global correction' by averaging the uncorrected ellipticity estimates and then subtracting the predicted first-order ellipticity bias averaged over this distribution, which is a known deterministic function, Equation~\eqref{eq:1Dm}. Schematically, the globally corrected estimator is
\begin{equation}
\hat{g}_{\mathrm{GL}} \propto \langle \hat{e} \rangle_S - \langle b(e_s) \rangle_S,
\label{eq:ggl}
\end{equation}
whilst the per-galaxy corrected estimator is
\begin{equation}
\hat{g}_{\mathrm{PG}} \propto \langle \hat{e} - b(\hat{e})\rangle_S.
\end{equation}
At leading order in $S/N$ these procedures are equivalent but at second-order they are not. Subtracting biases on each galaxy leaves two second-order residuals (which we have referred to as the intrinsic and bias-on-bias terms), whilst subtracting a global correction leaves us only with one second-order residual (the intrinsic term). The difference between these two biases is then the second-order bias-on-bias term averaged over source ellipticity. Note that imperfect knowledge of the true distribution will also lead to a difference, and the global correction relies on perfect knowledge of this distribution. In Figure~\ref{fig:sim1DMLE} the globally-corrected estimator bias is given by the green points, and the source-averaged first-order bias by the blue dashed line, such that the green points are the difference between the blue points and the blue dashed line.

At $S/N \gtrsim 20$, both bias correction schemes reduce the mean MLE bias. Note that although the changes are all smaller than the individual error bars, these points are strongly correlated at a given $S/N$ since the data are common to each method. In particular, the uncorrected and globally-corrected estimates are 100\% correlated as evident from Equation~\eqref{eq:ggl}. Thus \emph{differences} between the means of these sets of points are non-stochastic, and are indeed given by the blue-dashed line in Figure~\ref{fig:sim1DMLE}. Therefore, the probability that the correction reduces the \emph{magnitude} of the bias is given by the probability that the true uncorrected bias is greater than half this shift. This is roughly 80\% at high $S/N$ values (assuming Gaussianity), and higher at low $S/N$. Therefore we can say with some confidence that the global correction reduces the biases in this $S/N$ range, by an amount given by the blue-dashed curve in Figure~\ref{fig:sim1DMLE}.

The per-galaxy corrected bias measurements are also correlated with the uncorrected points. At high $S/N$ the correlation is very strong since the difference from the global correction is higher-order in $S/N$. The lack of significant difference between the global (green) points and per-galaxy (red) points between $40 \gtrsim S/N \gtrsim 25$ suggests that in this regime first-order perturbation theory applies, and hence the reduction in bias is statistically significant. At lower $S/N$ the two correction schemes differ significantly, which indicates higher-order terms are important and hence makes firm conclusions difficult due to the large error bars.

Thus from Figure~\ref{fig:sim1DMLE} we can say that although error bars are too large to demonstrate that our method satisfies the fiducial requirement, we can always reduce the MLE bias at $S/N \gtrsim 25$, and at $S/N \gtrsim 10$ for the global correction. At $S/N=10$ the per-galaxy correction is several sigma away from the requirement, so the method seems to fail here. This is indicative of higher order terms in the bias expansion, and hence our results should be treated with caution at this low $S/N$.



\begin{figure}
\centering
\includegraphics[width=\columnwidth]{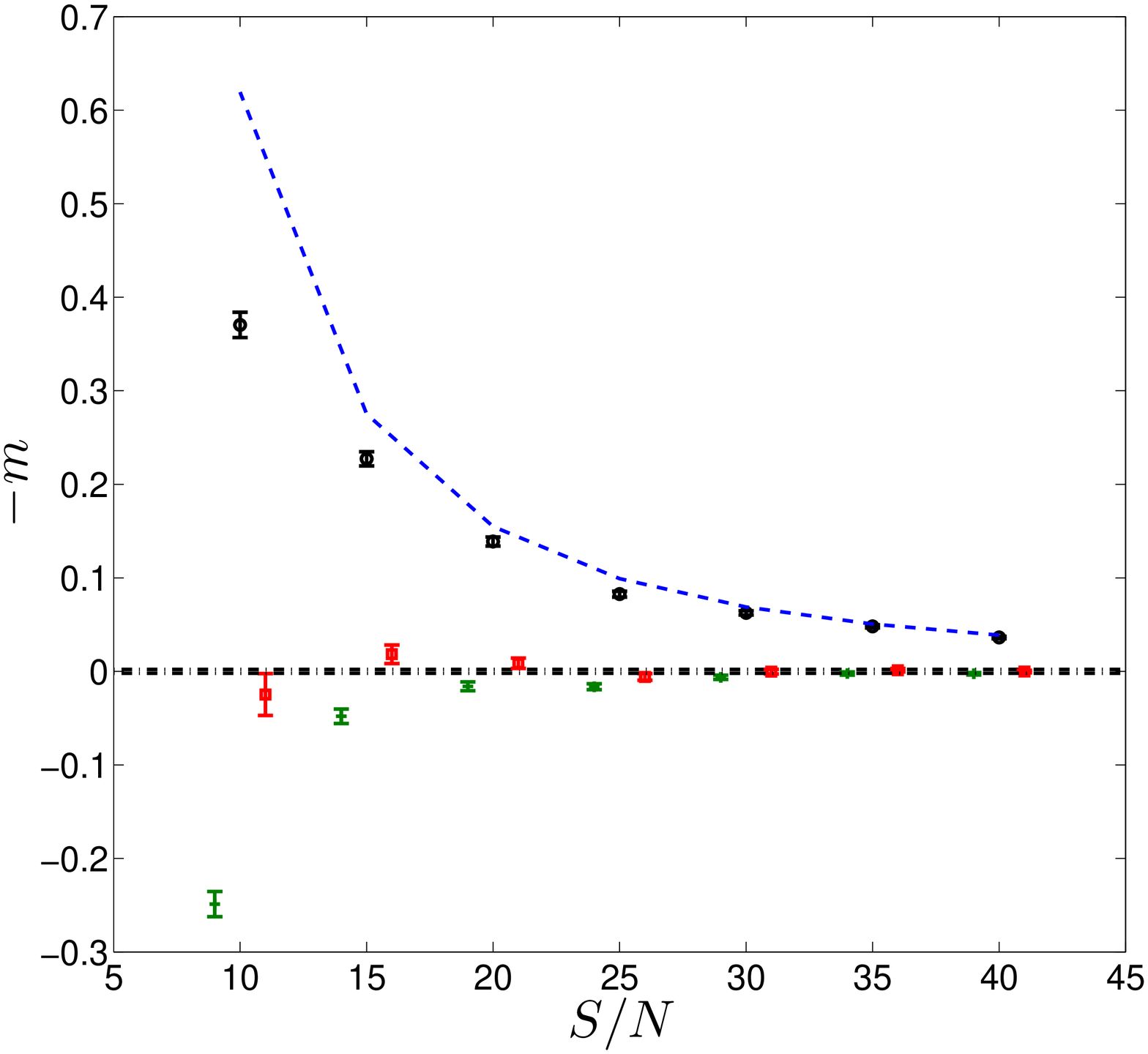}
\includegraphics[width=\columnwidth]{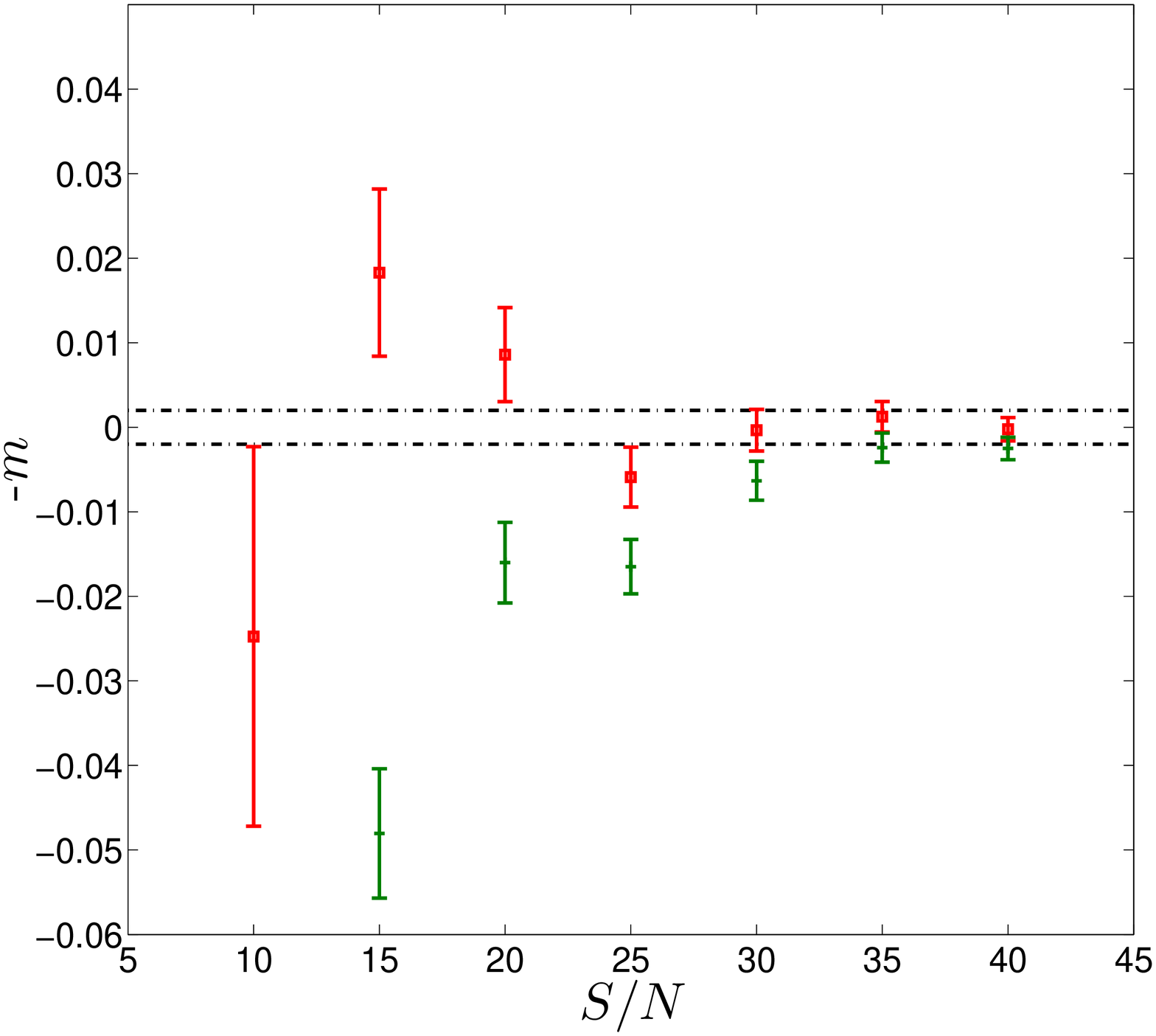}
\caption{\emph{Upper panel:} Same as Fig~\ref{fig:sim1DMLE} for the MAP. \emph{Lower panel:} Same as upper panel zooming in on low bias values and plotting only the residual biases in the bias-corrected MAP.}
\label{fig:sim1DMAP}
\end{figure}

In Figure~\ref{fig:sim1DMAP} we plot the performance of the MAP with the per-galaxy and global first-order corrections. The extra bias induced by the prior now means the uncorrected bias is above the requirement at all values of $S/N$, as seen in the upper panel, but most of this extra bias can be removed with the corrections. The error bars are still too large to determine whether the bias requirement has been met, although there is tentative evidence at the $\sim 1\sigma$ level at $S/N=40$.

At $S/N \gtrsim 30$ the difference between the two first-order correction schemes is given by the source-averaged bias-on-bias, which is higher order in $S/N$ and hence small in this regime. However, below $S/N \approx 30$ this term starts to become significant (see Figure~\ref{fig:1Dm}), and the per-galaxy correction outperforms the global correction, reducing the bias much more effectively. The reason for this should be clear given our previous investigation of second-order biases in the MAP - there is a cancellation between intrinsic and bias-on-bias terms, which suppresses the importance of second-order biases in the MAP. The global correction does not benefit from this cancellation, and hence suffers from large biases at low $S/N$. The lesson here is that higher-order terms not accounted for by a perturbative and predictive bias correction can significantly affect the performance of shear estimators at low $S/N$. We would expect similar conclusions to apply to the MLE/MAP as estimated from $e_2$. Note that even at low $S/N$ the measurements are correlated, so further investigation is required to robustly quantify the improvements.




The results of this section have demonstrated that a first-order bias correction reduces the bias on the MLE for $S/N \gtrsim 25$, and even down to $S/N \gtrsim 10$ for a specific choice of correction. There is little evidence that the MAP improves on this significantly, although we have tentative evidence that requirements are satisfied for $S/N \approx 40$. Further comparison of the MLE and MAP will have to account for their covariance, which we have not attempted here. Per-galaxy bias corrections seem to outperform global shear-bias corrections for the MAP, due to cancellations between second-order terms that become important for $S/N \lesssim 15$. The overall importance of second-order terms is however reduced for the MAP due to the regularizing of high-$|e|$ biases, as was the case for shape measurement.

\section{Two-dimensional shear biases}
\label{sec:2Dmresults}

So far our discussion has focussed on one-dimensional ellipticity and shear estimators. Whilst this permits the essential features of noise bias and bias-correction to be elucidated, it is clearly too simplistic to warrant much further study. In this section we turn our attention to the two-dimensional parameter space of ellipticity.

The multivariate generalisation of the first-order noise bias Equation~\eqref{eq:MLbias} is derived by analogous techniques to the univariate case, the bias on parameter $i$ being given by~\citep{CoxSnell}
\begin{equation}
\label{eq:2DMLbias}
b_i \equiv \langle \hat{\beta_i} - \beta_i \rangle = \frac{1}{2n}\bar{F}^{-1}_{ij}\bar{F}^{-1}_{km}(\bar{K}_{jkm} + 2\bar{J}_{k,jm}),
\end{equation}
where $\bar{F}^{-1}_{ij}$ are elements of the inverse Fisher matrix and
\begin{align}
\bar{K}_{ijk} &= \frac{1}{n} \sum_{m=1}^n \left \langle \frac{\partial^3 \log p_m}{\partial \beta_i \partial \beta_j \partial \beta_k}\right \rangle, \nonumber \\
 \bar{J}_{i,jk} &= \frac{1}{n} \sum_{m=1}^n \left \langle \frac{\partial \log p_m}{\partial \beta_i}\frac{\partial^2 \log p_m}{\partial \beta_j \partial \beta_k} \right \rangle,
\end{align}
where again we assume the expectation values are taken at the true parameter value. Note that the quantities $\bar{J}$ and $\bar{K}$ are not parameter-space tensors as they do not possess the correct transformation properties.

We will repeat our investigation of shear biases via noisy image simulations, this time finding the MLE (or MAP) for both $e_1$ and $e_2$, assuming the true (lensed) size and flux are known. To reduce the computational time, we consider galaxies having a fixed magnitude of intrinsic ellipticity $|e_s|$ of $0.3$ but uniformly distributed orientation. This procedure is sometimes referred to in the literature as a `ring test'~\citep{2007AJ....133.1763N}, and reduces the computational time in our case since it avoids the highly elliptical galaxies which require finer sub-pixel sampling in the convolution step. This setup also avoids the large biases in $e_1$ at high ellipticity. A constant shear of $(g_1,g_2) = (0.01,0.00)$ is applied, and the shear estimate is taken to be
\begin{equation}
\hat{g} = \frac{\langle \hat{e} \rangle_{S}}{2(1-|e_s|^2/2)}.
\end{equation}
The precise form of the denominator in this estimator has been chosen such that in the absence of ellipticity noise bias the above expression is an unbiased estimate of shear.

We define the multiplicative and additive shear biases in the two-dimensional case by $\langle \hat{g_i} \rangle = (\delta^K_{ij} + m_{ij})g_j + c_i$, where $\delta^K_{ij}$ is the Kronecker delta. With the same procedure as Section~\ref{sec:1Dmresults}, we find $c_1 = c_2 = m_{12} = m_{21} =  0$ by virtue of isotropy, $m_{22} = 0$ by virtue of choosing $g_2=0$ and 
\begin{equation}
m_{11} = \frac{1}{2}(1-|e_s|^2/2)^{-1}\left\langle \frac{\partial b_{e_1}(\mathbf{\Theta})}{\partial \Theta_i}\frac{\partial \Theta_i}{\partial g_1} \right\rangle_S,
\end{equation}
where $b_{e_1}(\mathbf{\Theta})$ now refers to the bias on $e_1$ as a function of the intrinsic galaxy parameters $\mathbf{\Theta}$.

\begin{figure*}
\centering

\includegraphics[width=0.7\textwidth]{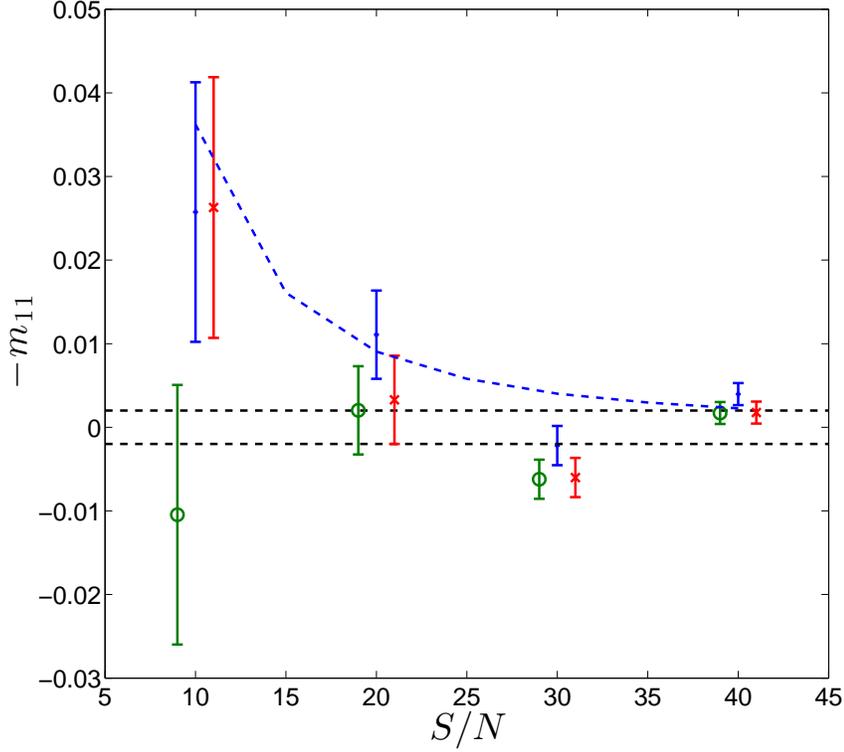}
\caption{2D multiplicative shear biases $m_{11}$ for the MLE, showing uncorrected biases (blue points), the first-order prediction for $m_{11}$ (blue dashed), globally first-order corrected residual biases (green open circles), and per-galaxy first-order corrected residual biases (red crosses). Note that residual biases are artificially offset in the horizontal direction for clarity.}
\label{fig:m11MLE}
\end{figure*}

In Figure~\ref{fig:m11MLE} we plot the multiplicative bias $m_{11}$ on the MLE, including the first-order per-galaxy corrected and globally corrected residual biases measured from the image simulations. The measured bias $m_{21}$ is found to be consistent with zero. As mentioned above, it is only below $S/N \lesssim 40$ that the predicted first-order bias is greater than the total requirement, so we restrict ourselves to this range on the horizontal axis. Due to the limitations of running large numbers of image simulations, the error bars on these points are only smaller than the total requirement for $S/N \sim 40$. At this $S/N$, both the per-galaxy correction and the global correction reduce the bias such that the probability of the mean residual bias being lower than the requirement is about $50\%$, increased from about $16\%$. As in the 1D case the bias we can still say that the MLE bias is reduced here due to the strong correlation of the points. However by the same token, we can say decisively that the bias is \emph{increased} at $S/N=30$. This potentially suggests that higher order terms are important for shear bias in this regime, as suggested by Figure~\ref{fig:1Dm}. 



\begin{figure*}
\centering

\includegraphics[width=0.47\textwidth]{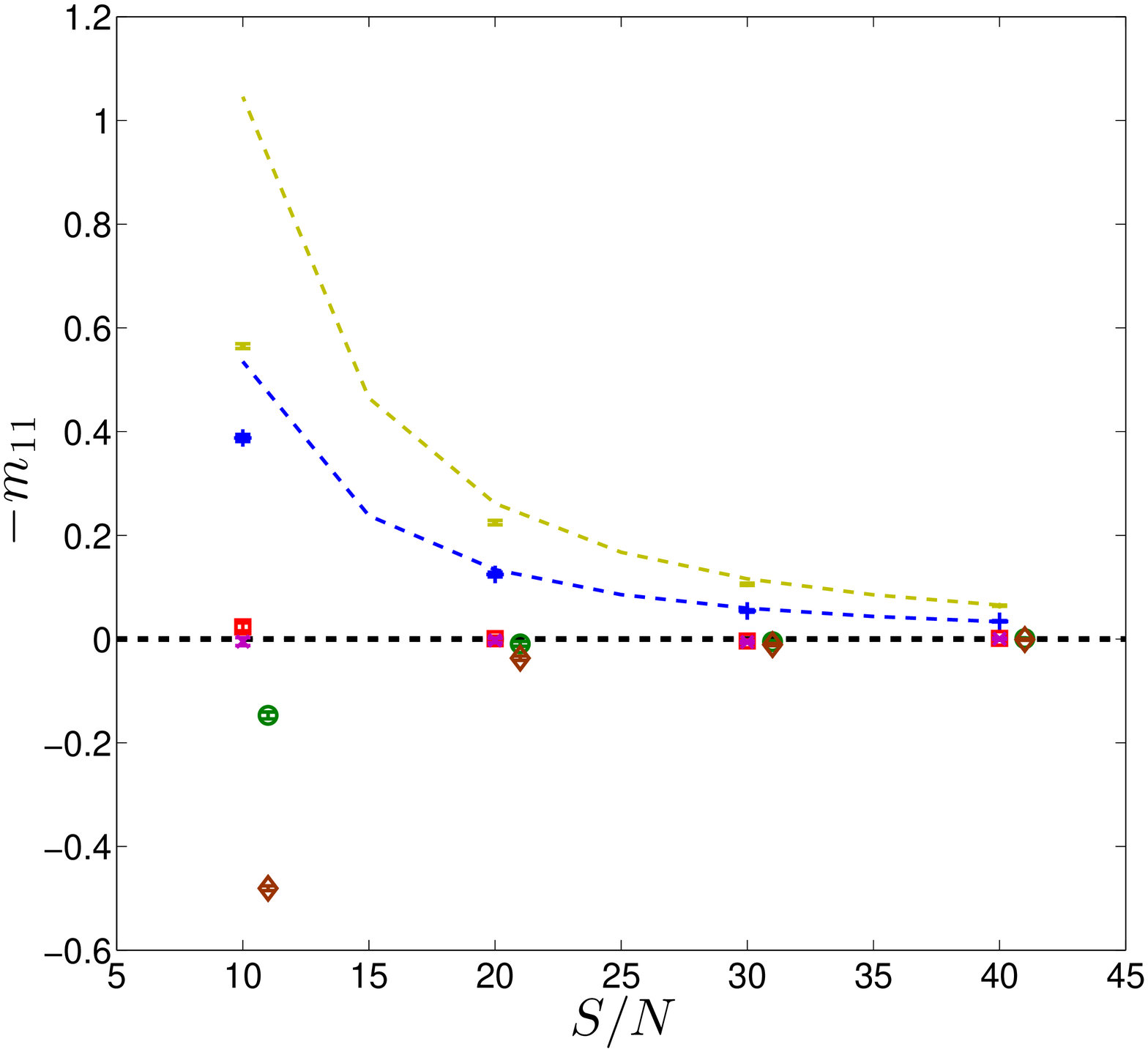}
\includegraphics[width=0.47\textwidth]{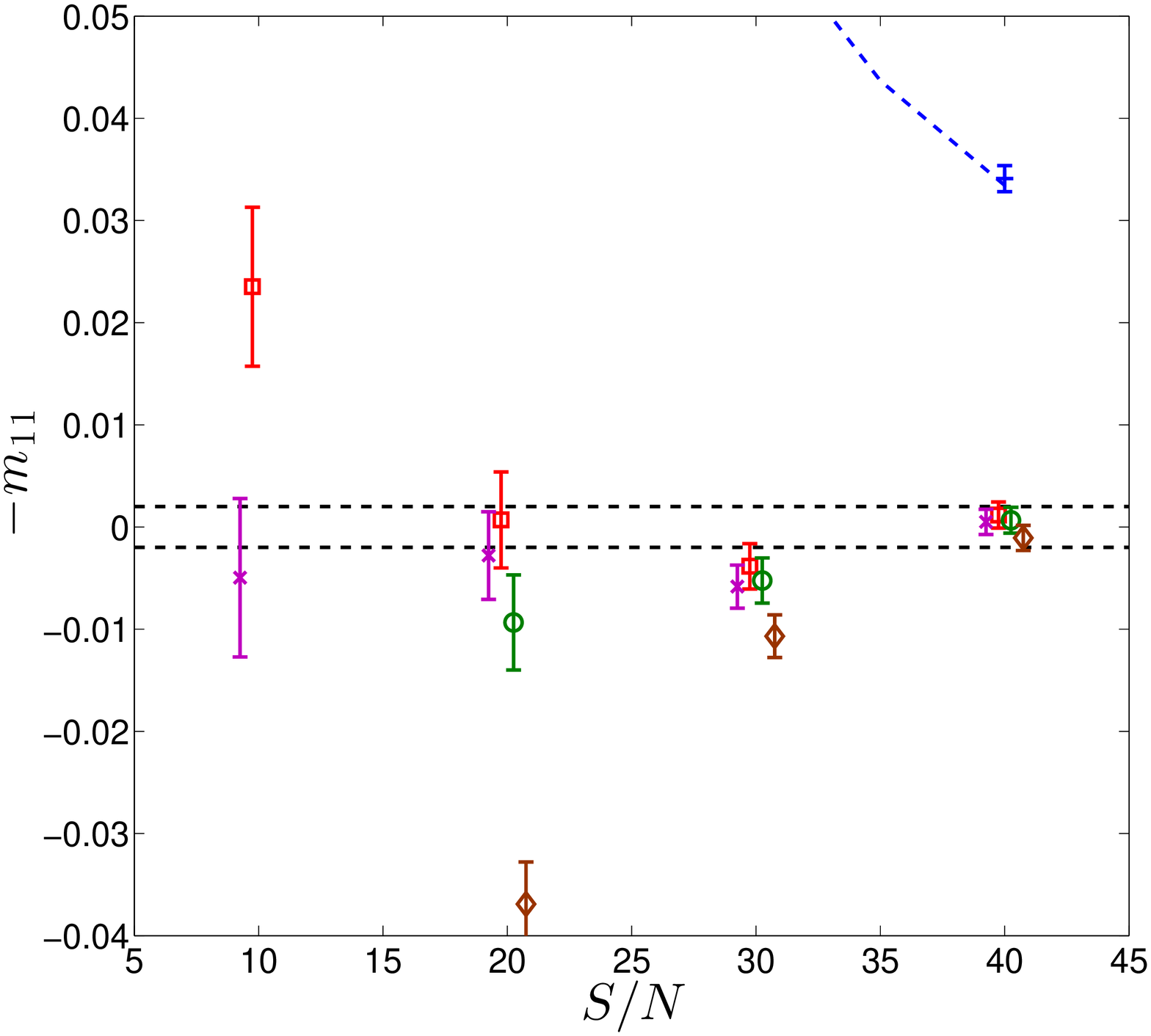}

\caption{\emph{Left}: 2D multiplicative shear biases $m_{11}$ for the MAP, showing uncorrected biases for a prior having $\sigma_e = 0.3$ (prior A, blue points), the first-order prediction for $m_{11}$ with prior A (blue dashed), globally first-order corrected residual biases for prior A (green open circles), per-galaxy first-order corrected residual biases for prior A (red open squares), uncorrected biases for a prior having $\sigma_e = 0.2$ (prior B, yellow filled circles), the first-order prediction for $m_{11}$ with prior B (yellow dashed), globally first-order corrected residual biases for prior B (brown open diamonds), and per-galaxy first-order corrected residual biases for prior B (magenta crosses). Note that residual biases are artificially offset in the horizontal direction for clarity. \emph{Right}: Same as left panel but zooming in on low bias values.}
\label{fig:m11MAP}
\end{figure*}


In Figure~\ref{fig:m11MAP} we plot the multiplicative bias of the MAP measured from the image simulations. The uncorrected biases are higher than those of the MLE due to the prior, but this plot demonstrates that most of this extra bias can be removed with our model, such that the MAP bias is of comparable magnitude to the MLE bias. Some improvement over the MLE is seen at $S/N=40$, where the residual MAP bias for both the per-galaxy and global corrections are smaller in magnitude than the MLE bias. This is likely due to the regularizing influence of the prior, as discussed in the previous section, although it could also be statistically insignificant depending on the degree of correlation between the MLE and MAP at this $S/N$\footnote{Note that at $S/N=10$ a larger set of mock images were used for the MAP compared to the MLE, which complicates this noise-cancellation argument.}.

At $S/N = 30$ the results of the bias-corrected MAP are comparable to those of the bias-corrected MLE, with both per-galaxy and global corrections fairing slightly worse than the uncorrected MLE, although again the error bars are large here. Similar conclusions apply at lower $S/N$, where the global correction now starts to fair considerably worse than the uncorrected MLE. Note that at low $S/N$ the expectation from Figure~\ref{fig:1Dm} is that a first-order correction fails, so any bias-reduction at $S/N=10$ should be treated with some caution. The same should be said for the error bars here, which may not be representative of the true width of the distribution due to non-Gaussianity.

The probability of the total bias requirement being satisfied at $S/N = 40$ is roughly $60\%$, and at lower $S/N$ the error bars are again too large to make conclusive statements about meeting total requirements, except for the globally-corrected MAP which performs poorly for $S/N \lesssim 30$.

It is not surprising that the MAP residual biases are similar to those of the MLE for mild $S/N$. The primary advantage of introducing a prior is to downweight high-ellipticity regions of parameter space which have give a large bias to the MLE. Since the simulated galaxies all have $|e_s| = 0.3$, extremely elliptical images have low likelihood even at the lowest $S/N$ we consider. Thus the MLE in this 2D test suffers less from the problems of high-bias regions than in the 1D test. We predict that a full set of mock images sampled from the full two-dimensional distribution of ellipticity would benefit from a prior more dramatically. What we have conclusively shown in Figures~\ref{fig:m11MLE} and \ref{fig:m11MAP} is that the extra bias from the prior can be almost completely removed for $S/N \gtrsim 20$, and in this regime there is little to choose between the MLE and the MAP as point estimators.

Finally, we investigate the sensitivity of the bias-corrected MAP to the assumed prior on intrinsic ellipticity. We repeat the ring test and bias measurements assuming the same functional form for $p(|e|)$ but setting $\sigma_e = 0.2$, which we refer to as `prior B', with the $\sigma_e = 0.3$ distribution referred to as `prior A'. The new value of $\sigma_e$ tightens the distribution to more circular galaxies\footnote{Note that the true distribution is still a fixed ellipticity magnitude of 0.3 with uniform distribution in angle.}. This has the effect of further down-weighting highly elliptical galaxies in the likelihood which bring with them a large noise bias to the shear estimate, at the cost of increasing the bias brought by the prior. In Figure~\ref{fig:m11MAP} we see that the uncorrected MAP bias has increased by over a factor or 2, which can be understood by inspecting the form of the prior bias in Equation~\eqref{eq:b1p}. For our assumed prior, the bias is roughly proportional to $1/\sigma_e^2$, which explains the increase. Despite this, our per-galaxy bias-correction scheme still brings the MAP bias down to levels similar to the $\sigma_e = 0.3$ case for $S/N \gtrsim 20$. A global correction performs poorly for $S/N \lesssim 30$, similarly to the $\sigma_e = 0.3$ case. At $S/N=10$ the $\sigma_e = 0.2$ prior seems to outperform the $\sigma_e = 0.3$ prior, although as indicated by the difference between the uncorrected MAP bias and the first-order prediction, this is a strongly non-perturbative regime and our results could be due to chance bias cancellations amongst higher order terms.


We have thus collected tentative evidence that the bias-corrected MAP has the capability to reduce shear biases to within the total requirements of planned surveys at $S/N \sim 40$. At higher $S/N$ the MLE is expected to be within this requirement without further correction, and at lower $S/N$ the residual biases are broadly similar, except in the case of a global bias correction which performs poorly for $S/N \lesssim 30$. Thus in 2D a global correction is \emph{not} preferred over a per-galaxy correction, unlike in 1D. For $S/N \lesssim 40$ a first-order bias correction performs poorly in 2D, as higher order terms become important, although further investigation will be required to determine how detrimental this effect is in terms of meeting total bias requirements for a future space-based survey.


\section{Conclusions and comparison to other work}
\label{sec:concs}

In this work we have investigated the properties of two important point estimators for galaxy shapes - the MLE and the MAP. Our analysis has been conducted with the aid of simulated noisy images of toy galaxy models, without the effects of a PSF. Many of our results are, to an extent, predicated on these simplifying assumptions, and some have been arrived at by assuming only a one-dimensional or two-dimensional model-fitting procedure has been implemented. It is therefore legitimate to ask which of our results depend crucially on these simplifications, and which have more general applicability.

The results that should have wider generality are as follows. We have seen that noise bias arises from the non-linear functional relationship between the parameter estimate and the noise~\citep[see also][]{2002AJ....123..583B, 2004MNRAS.353..529H, 2012MNRAS.425.1951R, 2012MNRAS.427.2711K, 2012MNRAS.424.2757M, 2014MNRAS.439.1909V}. Furthermore, given an analytic likelihood function and an analytic galaxy model we can predict the bias on the MLE as a perturbative series in the $S/N$. We have seen that these biases can become large when the galaxy size is comparable to the pixel size, as will be the case for the majority of the galaxies in future space-based weak lensing surveys. The coarseness of the pixel grid can cause a dramatic breaking of isotropy between the biases on the two ellipticity components. Using bias corrections predicted by the model itself (without fitting to external simulations), we have seen that the noise bias can be reduced by orders of magnitude down to some $S/N$ that can only be evaluated through image simulations. At $S/N \gtrsim 40$ the reduction of shape bias carries over to shear estimates that are composed of averaged shape estimates.

We have also seen that the large shape biases that arise at low $S/N$ in the MLE may be mitigated by including a prior to downweight extreme best-fitting models. In this setting the MLE becomes the MAP. The extra bias incurred by including a prior may be predicted, and the MAP still possesses many of the nice asymptotic properties possessed by the MLE~\citep{Lindley,Johnson}, although the property of invariance under general reparametrization of the model is of course lost. We have seen that the prior regularizes the distribution of best-fitting shape parameters, allowing the perturbative bias correction to be pushed to lower $S/N$. We found that second-order bias corrections do offer an estimate with lower bias, but the improvement is only moderate, and the additional CPU time required to compute the second-order terms precludes the significant reduction of the bias error bars that a more quantitative analysis demands.

By propagating ellipticity biases through to shear biases we have been able to directly compare our self-calibration schemes to the requirements of planned space-based weak lensing surveys. We have shown that first-order bias corrections are effective at reducing bias in the MLE and MAP at $S/N \sim 40$, and we have weak evidence that MAP-based shear estimates can satisfy fiducial requirements after a bias-correction. At lower $S/N$ we do not have sufficiently good statistics to draw similar conclusions, although the MLE and MAP perform comparably after a bias correction. We have seen that cancellations amongst higher-order terms resulting from our assumption of Gaussian noise can reduce these biases, although we would not expect this behaviour to persist in a more realistic galaxy model. This clearly serves as a precaution against evaluating shear estimation algorithms with the assumption of Gaussian noise - in reality the noise will not be perfectly Gaussian due to it having multiple sources. Our results should be taken as strong encouragement that these point estimators might be of considerable use in a weak lensing pipeline, and should be incorporated into algorithm testing on more realistic image simulations.

Another question that might be asked of the MLE and MAP estimates is how they compare to previous point estimators in the literature. The most relevant works to mention here are the MLE-based studies of~\citet{2012MNRAS.425.1951R} and~\citet{2012MNRAS.427.2711K} incorporated into the publically available code \textsc{IM3SHAPE}~\citep{2013MNRAS.434.1604Z}. These works cover much of the topics we have discussed here, including perturbative bias predictions, second-order bias terms, and bias-calibration. However, are approach differs to theirs in that we predict biases using the analytic perturbative biases with the MLE as a proxy for the true parameter value, rather than fitting the bias from simulations. We advocate our approach since it has much less sensitivity to the results of image simulations (which may not accurately describe the actual observed galaxy properties), which in our case are only required to find the $S/N$ and parameter regime where the method fails rather than being an integral part of the method per se. Similar comments could be made about other maximum likelihood methods that rely on external calibration, such as \textsc{GFIT}~\citep{2012arXiv1211.4847G}. 

As a caveat to our conclusions, we remind the reader that our method has only been thoroughly tested on shape estimates from highly simplified galaxy modes and shear estimates at fairly high $S/N$. Although in principle internal bias correction seems a promising method for use on real data, more detailed investigation will be required to determine its feasibility for use on realistic images, and we do not exclude the possibility that external calibration will be eventually required. Effects such as PSF subtraction, centroid fitting, model bias and other nuisance parameters can all increase shear bias. Of these, centroid fitting could well make our method \emph{more} effective, since marginalising over the centroid should restore a degree of isotropy and mitigate the large $e_1$ bias found in Section~\ref{sec:bias}, thus making a perturbative description more accurate. Model bias may be more troublesome and is not currently built into our correction scheme, although one could imagine allowing extra freedom in the model through nuisance parameters which could then be incorporated into the bias correction. For a detailed discussion of this issue, see~\citet{2014MNRAS.441.2528K}.

\citet{2014MNRAS.438.1880B} (hereafter BA14) proposed a point estimator based on a shear posterior constructed from an ensemble of galaxies subjected to a constant shear field~\citep[see][for practical implementations]{2014MNRAS.444L..25S, 2015arXiv150805655B}. The method proposed there is not directly comparable to the work here since we are concerned with estimating ellipticities and then averaging over galaxies to compute shear, rather than forming a shear posterior. Despite this difference, there are similarities to our work in that a point estimate is extracted whose bias is then computed. However, rather than studying the expansion of the likelihood around the true parameter value, BA14 expand around zero shear, assuming $g$ is close to zero in some sense. A prior is introduced in their Equation~(5) but then immediately discarded as the data are assumed to be more informative than the prior. In the limit of infinite independent data the shear posterior will become centred on the true parameter value, which need not be zero. With enough galaxies the width of the shear posterior might shrink to such an extent that the $g=0$ point is way out in the tail of the distribution, such that both the second-order and third-order expansion of the posterior around zero shear fail to converge. Use of the MLE (or MAP) as the point estimate guarantees that the perturbative series converges in the limit of infinite independent data. It would be interesting to investigate how the MAP performs compared to the BA14 estimator. Our work also has the advantage that we do not need to assume a constant or parametrized spatially-dependent shear field as in BA14.

Unbiased estimators have also been proposed in~\citet{2015JCAP...01..022M} (hereafter M15). This work proposes to use the fact that certain combinations of likelihood derivatives vanish when averaged over noise realizations and evaluated at the true parameter value. By expanding these quantities around a fiducial model, one can build estimators that are unbiased at a given order in the difference between the fiducial model parameter and the truth. At leading order in this difference the `$E_1$' estimator of M15 is of similar form to the leading order part of the MLE, Equation~\eqref{eq:MLEseries}. The difference is that we have not had to expand around an arbitrary fiducial model. This is clearly a big advantage of the MLE - we do not need to assume anything about the true parameter value and our method does not rely on a fiducial parameter being close to the unknown true value. The leading order M15 estimate also bears some resemblance to the leading order estimator of BA14, the difference being that the BA14 is the leading-order Newton-Raphson approximation to the mode of the posterior at zero shear, whereas the M15 estimator replaces the denominator with the Fisher information and evaluates at a fiducial parameter point.

In this work we have advocated the bias-corrected MLE (or MAP) as a shape/shear estimator. The essential reason we advocate these point estimators over other frequentist methods such as re-Gaussianization~\citep{2003MNRAS.343..459H} or the KSB method~\citep{1995ApJ...449..460K} is that bias calibration can in principle be done internally rather than relying on external simulations. Furthermore, the MLE is an efficient estimator both at leading and next-to-leading order in the $S/N$~\citep{rao}, suggesting that its mean-squared error will be better than other point estimators such as the mean of the ellipticity posterior, or a shear response weighted version of the mean as in \textsc{Lensfit}~\citep{2007MNRAS.382..315M,2008MNRAS.390..149K}, although we have not made the requisite comparison with other methods to show this explicitly.

Finally, we note that although we have focussed on point estimators for galaxy shape and shear, in principle one could correctly propagate the probability distribution functions conditioned on the data, marginalizing over nuisance parameters to recover posteriors on the shear map, power spectra, or cosmological parameters themselves~\citep{2015ApJ...807...87S,2016arXiv160205345H}. This approach provides an optimal way of using all the information present in the data for inference, and is in principle preferable to using biased point estimators of galaxy shapes. However, we believe that the point estimators we have analyzed in this work are still valuable both as a shear measurement technique and as a diagnostic check on the likelihood function. Since the likelihood is a necessary ingredient of the Bayesian approach, we believe its mode could and should be used as a consistency check that the likelihood has been correctly constructed. Thus the bias-correction techniques we have advocated here fit naturally within a hierarchical Bayesian approach. In addition, the MLE and MAP are both very quick to compute, and much of the machinery required to produce shear maps from them already exists~\citep{2015arXiv150705603J}.

To conclude, we have extended the work of~\citet{2012MNRAS.425.1951R} and~\citet{2012MNRAS.427.2711K} to show that a simple first-order bias correction applied to the mode of the likelihood, optionally regularized by a prior, can reduce noise bias in point-estimates of shear.  We have elucidated the origin of noise bias, and investigated the impact of the coarse grid on which galaxies will be imaged by next-generation weak lensing surveys. Our simple toy models for pixel noise and galaxy surface brightnesses have allowed us to make considerable analytic progress, allowing for some more general conclusions to be drawn regarding the performance of bias-correction on galaxy images. We have shown that ellipticity biases can be reduced by orders of magnitude down to low $S/N$, and investigated the improvement gained by correcting for second-order biases, finding it to be small but non-negligible. We propose that these techniques be tested on the more realistic image simulations required to test shear measurement algorithms as part of current and future weak lensing experiments.


\section{Acknowledgments}
AH is supported by a United Kingdom Space Agency \emph{Euclid} grant, and an STFC Consolidated Grant. AT is supported by a Royal Society Wolfson Research Merit Award. We thank Giuseppe Congedo, Chris Duncan, Bryan Gillis, and Lance Miller for helpful conversations. We also thank the anonymous referee for very useful comments which have improved this paper.

\bibliographystyle{mn2e_fix}


\end{document}